\documentclass[twocolumn]{aastex701}
\submitjournal{AJ}

\usepackage{xspace}
\usepackage{amsmath}

\newcommand{\pnAABaryPrecShort}{43.09\xspace}
\newcommand{\pnAEBaryPrecShort}{0.139\xspace}
\newcommand{\pnAIBaryPrecShort}{24.4\xspace}

\newcommand{\pnAAccExposures}{63\xspace}
\newcommand{\pnAAccOnDetector}{51\xspace}
\newcommand{\pnAAccOffDetector}{12\xspace}

\newcommand{\pnAAccImaged}{38\xspace}

\newcommand{\pnAAccImagedUncat}{24\xspace}

\newcommand{\pnAAccAdded}{16\xspace}

\newcommand{\pnAAccRep}{10\xspace}
\newcommand{\pnAAccDet}{5\xspace}

\newcommand{\adamNDet}{57\xspace}
\newcommand{\adamNFrames}{3292\xspace}
\newcommand{\adamNDecam}{28\xspace}
\newcommand{\adamNDecamNights}{6\xspace}
\newcommand{\adamNDecamExp}{16\xspace}
\newcommand{\adamNAtlas}{26\xspace}
\newcommand{\adamNAtlasNights}{24\xspace}
\newcommand{\adamNSkymapper}{3\xspace}
\newcommand{\adamNSkymapperNights}{2\xspace}
\newcommand{\adamAtlasFirst}{2019-07-06\xspace}
\newcommand{\adamAtlasLast}{2026-07-07\xspace}
\newcommand{\adamSkymapperFirst}{2014-08-10\xspace}
\newcommand{\adamSkymapperLast}{2016-08-29\xspace}
\newcommand{\adamDecamFirst}{2016-02-25\xspace}
\newcommand{\adamDecamLast}{2019-05-01\xspace}
\newcommand{\adamSearchStart}{2012-09-01\xspace}
\newcommand{\adamSearchEnd}{2026-10-18\xspace}
\newcommand{\adamRunDate}{2026-10-18\xspace}
\newcommand{\adamJobId}{c742167c-021e-48a8-a8c5-66d86865f227\xspace}
\newcommand{\adamBusiestNight}{2017-05-04\xspace}
\newcommand{\adamBusiestNightN}{17\xspace}

\newcommand{\adamDecamPredSigMin}{33\xspace}
\newcommand{\adamDecamPredSigMax}{74\xspace}

\newcommand{\dcAstNIn}{45\xspace}
\newcommand{\dcAstN}{44\xspace}
\newcommand{\dcAstNNights}{14\xspace}

\newcommand{\dcAstMaxFwhm}{2\xspace}
\newcommand{\dcAstSys}{0.05\xspace}
\newcommand{\dcAstAgree}{40\xspace}
\newcommand{\dcAstNGaiaMin}{79\xspace}
\newcommand{\dcAstNGaiaMax}{278\xspace}
\newcommand{\dcAstHdrOffMax}{107\xspace}
\newcommand{\dcAstHdrRmsMin}{6\xspace}
\newcommand{\dcAstHdrRmsMax}{54\xspace}

\newcommand{\dcAstNscOffMed}{11\xspace}
\newcommand{\dcAstNscRmsMin}{23\xspace}
\newcommand{\dcAstNscRmsMax}{61\xspace}
\newcommand{\dcAstNBiasAgree}{24\xspace}
\newcommand{\dcAstNBiasSingle}{6\xspace}
\newcommand{\dcAstNBiasNone}{14\xspace}
\newcommand{\dcAstBiasMax}{87\xspace}
\newcommand{\dcAstErrMed}{52\xspace}
\newcommand{\dcAstInjGrid}{5\xspace}

\newcommand{\itfNRows}{9,084,064\xspace}

\newcommand{\itfStartYear}{1995\xspace}
\newcommand{\itfNearestDeg}{4.2\xspace}

\newcommand{\itfMatchArcsec}{10\xspace}
\newcommand{\itfPredSigMax}{1.5\xspace}

\newcommand{\pcBetaJoint}{0.14\xspace}
\newcommand{\pcBetaJointErr}{0.03\xspace}
\newcommand{\pcJointOffset}{+0.01\xspace}
\newcommand{\pcJointOffsetErr}{0.01\xspace}

\newcommand{\pcRubXpg}{\ensuremath{-}0.020\xspace}

\newcommand{\pcRubXpStars}{320\xspace}

\newcommand{\pcSameN}{22\xspace}
\newcommand{\pcSameMax}{0.005\xspace}

\newcommand{\pcBudXpg}{\ensuremath{-}0.00\xspace}

\newcommand{\pcBudXpr}{+0.03\xspace}

\newcommand{\pcBudXpi}{\ensuremath{-}0.00\xspace}

\newcommand{\pcBudXpz}{+0.04\xspace}

\newcommand{\pcBudXpY}{\ensuremath{-}0.00\xspace}
\newcommand{\pcNighti}{0.02\xspace}
\newcommand{\pcNightz}{0.00\xspace}
\newcommand{\pcpnAGalB}{\ensuremath{-}12\xspace}
\newcommand{\pcDecamGalBMin}{\ensuremath{-}10\xspace}
\newcommand{\pcDecamGalBMax}{\ensuremath{-}3\xspace}
\newcommand{\pcDecamAlphaMin}{0.7\xspace}
\newcommand{\pcDecamAlphaMax}{1.2\xspace}
\newcommand{\pcRotFloor}{0.03\xspace}
\newcommand{\pcRubMonMaxAbs}{0.007\xspace}
\newcommand{\pcRubMonVisitMaxAbs}{0.03\xspace}
\newcommand{\pcRubXpMaxAbsRIZ}{0.016\xspace}
\newcommand{\pcBudNscObsMin}{0.04\xspace}
\newcommand{\pcBudNscObsMax}{0.21\xspace}
\newcommand{\pcMonNscMaxAbs}{0.03\xspace}

\newcommand{\dcPhotAZpScatter}{0.09\xspace}

\newcommand{\dcPhotCZpScatter}{0.01\xspace}
\newcommand{\dcPhotCNStars}{9\xspace}
\newcommand{\dcPhotCRep}{0.013\xspace}

\newcommand{\dcPhotN}{44\xspace}

\newcommand{\dcDiTplMin}{1\xspace}
\newcommand{\dcDiTplMax}{3\xspace}

\newcommand{\dcDiMinusDirectMaxAbs}{0.16\xspace}
\newcommand{\dcDiMinusDirectSd}{0.02\xspace}
\newcommand{\dcDiFakeBiasMaxAbs}{0.040\xspace}
\newcommand{\dcDiFakeBiasMed}{\ensuremath{-}0.001\xspace}

\newcommand{\dcPhotXpStars}{183\xspace}
\newcommand{\dcPhotYExtraC}{0.02\xspace}
\newcommand{\dcPhotYRedGi}{2.5\xspace}
\newcommand{\dcPhotYNRed}{1\xspace}
\newcommand{\dcPhotYRedExtra}{0.18\xspace}
\newcommand{\dcPhotVRValid}{0.001\xspace}

\newcommand{\dcPhotDHCDirectg}{\ensuremath{-}0.00\xspace}

\newcommand{\dcPhotDHCDirectr}{+0.03\xspace}

\newcommand{\dcPhotDHCDirecti}{+0.06\xspace}

\newcommand{\dcPhotDHCDirectz}{+0.05\xspace}

\newcommand{\dcPhotDHCDirectY}{+0.05\xspace}

\newcommand{\dcPhotNscBiasMax}{0.17\xspace}

\newcommand{\dcNBoxNsc}{1056\xspace}
\newcommand{\dcNBoxArchive}{341\xspace}
\newcommand{\dcNField}{89\xspace}
\newcommand{\dcNFieldNights}{19\xspace}

\newcommand{\dcNNoChip}{37\xspace}
\newcommand{\dcNOnChip}{52\xspace}
\newcommand{\dcNAccepted}{45\xspace}
\newcommand{\dcNAcceptedNights}{14\xspace}

\newcommand{\dcFirstYear}{2016\xspace}
\newcommand{\dcLastYear}{2026\xspace}
\newcommand{\dcFirstAcc}{2016-02-25\xspace}
\newcommand{\dcLastAcc}{2023-04-07\xspace}
\newcommand{\dcNRejEdge}{4\xspace}
\newcommand{\dcNRejOffsetPeak}{3\xspace}

\newcommand{\dcNAccInNsc}{11\xspace}
\newcommand{\dcNAccInNscNights}{5\xspace}
\newcommand{\dcNAccNotNsc}{33\xspace}
\newcommand{\dcNAccNotNscDqSeven}{21\xspace}

\newcommand{\dcSnrMin}{12\xspace}
\newcommand{\dcSnrMax}{87\xspace}
\newcommand{\dcPredSigMax}{0.11\xspace}
\newcommand{\dcMinSnr}{8\xspace}
\newcommand{\dcMinEdge}{5\xspace}
\newcommand{\dcMaxCen}{1\xspace}
\newcommand{\dcMaxTplRatio}{0.5\xspace}
\newcommand{\dcFieldDeg}{1.15\xspace}

\newcommand{\seedNCand}{28\xspace}
\newcommand{\seedNVetted}{11\xspace}
\newcommand{\seedNStatic}{6\xspace}
\newcommand{\seedNRejectedOther}{11\xspace}
\newcommand{\seedNLinked}{11\xspace}
\newcommand{\seedNLinkedNights}{5\xspace}

\newcommand{\seedMinSnr}{8\xspace}
\newcommand{\seedMaxCen}{1\xspace}
\newcommand{\seedMaxTplRatio}{0.5\xspace}
\newcommand{\seedMinTplDays}{253\xspace}
\newcommand{\seedRms}{0.2\xspace}

\newcommand{\seedFirstNight}{2016-02-25\xspace}

\newcommand{\seedFirstRubinOff}{1.8\xspace}
\newcommand{\seedFirstRubinSig}{66\xspace}

\newcommand{\valPsfCorrLooRa}{37\xspace}
\newcommand{\valPsfCorrLooDec}{22\xspace}

\newcommand{\valNscLooRa}{125\xspace}
\newcommand{\valNscLooDec}{48\xspace}
\newcommand{\valABaryMin}{43.0910\xspace}
\newcommand{\valABaryMax}{43.0911\xspace}
\newcommand{\valLooFirstPsfCorrRa}{\ensuremath{-}32\xspace}
\newcommand{\valLooFirstPsfCorrDec}{\ensuremath{-}19\xspace}

\newcommand{\valLooFirstNscRa}{\ensuremath{-}185\xspace}
\newcommand{\valLooFirstNscDec}{93\xspace}
\newcommand{\valLooFirstPredSig}{12\xspace}
\newcommand{\valNscNExp}{11\xspace}
\newcommand{\valNscNNights}{5\xspace}
\newcommand{\valNscWorstNight}{2016-08-11\xspace}
\newcommand{\valNscWorstMin}{171\xspace}
\newcommand{\valNscWorstMax}{669\xspace}
\newcommand{\valNscOtherMax}{90\xspace}

\newcommand{\valFirstGaussBiasRa}{\ensuremath{-}99\xspace}
\newcommand{\valFirstGaussBiasDec}{85\xspace}
\newcommand{\valFirstPsfBiasRa}{\ensuremath{-}43\xspace}
\newcommand{\valFirstPsfBiasDec}{40\xspace}
\newcommand{\valFirstSepBiasRa}{\ensuremath{-}168\xspace}
\newcommand{\valFirstSepBiasDec}{143\xspace}
\newcommand{\valFirstNscMinusFinalRa}{\ensuremath{-}141\xspace}
\newcommand{\valFirstNscMinusFinalDec}{119\xspace}
\newcommand{\valFirstNscMinusSepMax}{37\xspace}
\newcommand{\valShutMin}{\ensuremath{-}0.66\xspace}
\newcommand{\valShutMax}{1.02\xspace}

\newcommand{\valShutterTravel}{1.05\xspace}

\newcommand{\valFlagDate}{2023-04-07\xspace}
\newcommand{\valFlagOpenshut}{\ensuremath{-}24.7\xspace}
\newcommand{\valFlagDarktime}{66.4\xspace}
\newcommand{\valFlagExptime}{90\xspace}

\newcommand{\valAstNRobust}{184\xspace}
\newcommand{\valAstNNights}{13\xspace}
\newcommand{\valAstVmax}{22.5\xspace}

\newcommand{\valAstAlong}{1\xspace}
\newcommand{\valAstAlongErr}{3\xspace}
\newcommand{\valAstCross}{2\xspace}
\newcommand{\valAstCrossErr}{3\xspace}
\newcommand{\valAstDt}{0.4\xspace}
\newcommand{\valAstDtErr}{0.5\xspace}
\newcommand{\valAstTheilSen}{\ensuremath{-}2.9\xspace}
\newcommand{\valAstRateMin}{5\xspace}
\newcommand{\valAstRateMax}{64\xspace}

\newcommand{\valAstDtBound}{3\xspace}
\newcommand{\valObjRateMin}{1.1\xspace}
\newcommand{\valObjRateMax}{2.7\xspace}
\newcommand{\valObjMasPerSecMax}{0.74\xspace}
\newcommand{\valDcrMax}{40\xspace}
\newcommand{\valDcrGMax}{40\xspace}
\newcommand{\valDcrMedian}{4\xspace}
\newcommand{\valDcrNOverTwenty}{4\xspace}
\newcommand{\valAtlasN}{26\xspace}
\newcommand{\valAtlasNCons}{26\xspace}
\newcommand{\valAtlasOcMed}{0.8\xspace}
\newcommand{\valAtlasOcMax}{2.6\xspace}

\newcommand{\valSkymapperN}{3\xspace}
\newcommand{\valSkymapperNCons}{1\xspace}
\newcommand{\valSkymapperOcMed}{0.1\xspace}

\newcommand{\valSkymapperNIncons}{2\xspace}
\newcommand{\valSkymapperInconsOc}{51\xspace}
\newcommand{\valSkymapperInconsDate}{2014-08-10\xspace}
\newcommand{\valSkymapperInconsMag}{17.1\xspace}
\newcommand{\valConsistentArcsec}{3\xspace}

\newcommand{\beKnownHvFour}{33\xspace}

\newcommand{\beRecovered}{8\xspace}
\newcommand{\beDiscoverable}{10\xspace}
\newcommand{\beDiscRecovered}{8\xspace}
\newcommand{\beNew}{1\xspace}
\newcommand{\beBefore}{4\xspace}
\newcommand{\beBeforeLo}{0.2\xspace}
\newcommand{\beBeforeHi}{20\xspace}
\newcommand{\beRemain}{3\xspace}
\newcommand{\bePlaneArea}{2,830\xspace}
\newcommand{\bePlanePct}{14\xspace}
\newcommand{\beBrighterKnown}{9\xspace}

\newcommand{\pnADbUncatalogued}{18\xspace}
\newcommand{\pnADbUncatInTooBig}{15\xspace}

\newcommand{\pnADbNoCatalogue}{6\xspace}

\newcommand{\pnAResNLib}{93\xspace}
\newcommand{\pnAResN}{201\xspace}
\newcommand{\pnAResMyr}{10\xspace}
\newcommand{\pnAResFrac}{0.46\xspace}

\newcommand{\pnAResCenter}{180\xspace}
\newcommand{\pnAResAmp}{95--115\xspace}
\newcommand{\pnAPropI}{25.7\xspace}
\newcommand{\pnAMeanE}{0.139\xspace}
\newcommand{\haumeaPropI}{26.6\xspace}
\newcommand{\famPropILo}{26.8\xspace}
\newcommand{\famPropIHi}{28.1\xspace}
\newcommand{\famMeanELo}{0.087\xspace}
\newcommand{\famMeanEHi}{0.136\xspace}
\newcommand{\famN}{9\xspace}
\newcommand{\pnAResNArgs}{28\xspace}

\newcommand{\pnAResNLibOther}{10\xspace}

\newcommand{\pnATissN}{2.835\xspace}
\newcommand{\haumeaTissN}{2.796\xspace}
\newcommand{\famTissNLo}{2.798\xspace}
\newcommand{\famTissNHi}{2.820\xspace}
\newcommand{\pnADvMin}{214\xspace}
\newcommand{\haumeaDvMin}{390\xspace}
\newcommand{\famDvMinLo}{21\xspace}
\newcommand{\famDvMinHi}{142\xspace}
\newcommand{\famCentreA}{43.10\xspace}
\newcommand{\famCentreE}{0.118\xspace}
\newcommand{\famCentreI}{28.2\xspace}

\newcommand{\knownBoxRadiusArcmin}{2\xspace}
\newcommand{\knownBoxWindowHours}{1\xspace}

\newcommand{\knownOrbitsSelected}{9,003\xspace}

\newcommand{\knownEpochsPerCandidate}{3\xspace}

\newcommand{\pnANearestTno}{2000 JG81\xspace}
\newcommand{\pnANearestTnoSepMin}{3.6\xspace}
\newcommand{\pnANearestTnoSepMax}{3.7\xspace}
\newcommand{\pnANearestTnoH}{8.0\xspace}

\newcommand{\skybotRadiusArcmin}{10\xspace}

\newcommand{\pnASkybotObjects}{5\xspace}

\newcommand{\pnASkybotMaxHelioDist}{3.2\xspace}
\newcommand{\pnASkybotMinSep}{207\xspace}

\newcommand{\knownWideArcmin}{20\xspace}
\newcommand{\knownWideUntil}{2026-10-15\xspace}

\newcommand{\noirlabPrograms}{2014A-0480, PI: R. M. Rich; 2016A-0189, PI: A. Rest; 2016B-0279, PI: D. Finkbeiner; 2017A-0389, PI: D. Tucker; 2018A-0251, PI: D. Finkbeiner; 2018B-0271, PI: D. Finkbeiner; 2019A-0265, PI: D. Finkbeiner; 2019A-0305, PI: A. Drlica-Wagner\xspace}

\newcommand{\pnAChisqFinalFl}{2.2\xspace}

\newcommand{\pnAArcFinalRep}{129\xspace}
\newcommand{\pnAChisqFinalRep}{65.5\xspace}
\newcommand{\pnADofFinalRep}{54\xspace}

\newcommand{\pnAABaryLoOneFinalRep}{43.08\xspace}
\newcommand{\pnAABaryHiOneFinalRep}{43.10\xspace}

\newcommand{\pnAABaryLoThreeOrigRep}{43.0\xspace}
\newcommand{\pnAABaryHiThreeOrigRep}{44.7\xspace}

\newcommand{\pnAChisqPrecFl}{6.2\xspace}

\newcommand{\pnANobsPrecRep}{74\xspace}
\newcommand{\pnAArcPrecRep}{3789\xspace}
\newcommand{\pnAChisqPrecRep}{84.7\xspace}
\newcommand{\pnADofPrecRep}{142\xspace}

\newcommand{\pnAABaryPrecRep}{43.0910\xspace}

\newcommand{\pnAABaryLoOnePrecRep}{43.0909\xspace}
\newcommand{\pnAABaryHiOnePrecRep}{43.0910\xspace}

\newcommand{\pnAQBaryPrecRep}{37.1130\xspace}

\newcommand{\pnAIBaryPrecRep}{24.4339\xspace}

\newcommand{\pnAMaxChiPrecRep}{3.8\xspace}

\newcommand{\rmsQuadArcsec}{0.015\xspace}
\newcommand{\submitQuadArcsec}{0.010\xspace}
\newcommand{\floorArcsec}{0.1\xspace}
\newcommand{\emptySigArcsec}{0.2\xspace}
\newcommand{\pnAPkgNobs}{10\xspace}
\newcommand{\pnAPkgNights}{8\xspace}
\newcommand{\pnAPkgBands}{ugizy\xspace}
\newcommand{\pnAPkgFirst}{2026-05-01\xspace}
\newcommand{\pnAPkgLast}{2026-06-05\xspace}
\newcommand{\pnAPkgArc}{35\xspace}

\newcommand{\pnAPkgMedi}{19.4\xspace}

\newcommand{\pnABeta}{0.13\xspace}
\newcommand{\pnABetaErr}{0.05\xspace}
\newcommand{\pnABetaN}{27\xspace}
\newcommand{\pnABetaChisq}{49.4\xspace}
\newcommand{\pnABetaDof}{21\xspace}
\newcommand{\pnAAlphaMin}{0.42\xspace}
\newcommand{\pnAAlphaMax}{1.21\xspace}

\newcommand{\pnAHg}{3.50\xspace}

\newcommand{\pnAHgErr}{0.05\xspace}
\newcommand{\pnAHi}{2.61\xspace}

\newcommand{\pnAHiErr}{0.04\xspace}
\newcommand{\pnAHr}{2.87\xspace}

\newcommand{\pnAHrErr}{0.04\xspace}

\newcommand{\pnAHy}{2.40\xspace}

\newcommand{\pnAHyErr}{0.06\xspace}
\newcommand{\pnAHz}{2.49\xspace}

\newcommand{\pnAHzErr}{0.05\xspace}

\newcommand{\pnAHVMpcLo}{2.76\xspace}
\newcommand{\pnAHVMpcLoBand}{y\xspace}
\newcommand{\pnAHVMpcHi}{3.22\xspace}
\newcommand{\pnAHVMpcHiBand}{g\xspace}
\newcommand{\pnAHV}{3.19\xspace}
\newcommand{\pnAGR}{0.63\xspace}

\newcommand{\pnAHVErr}{0.04\xspace}
\newcommand{\pnAGRErr}{0.07\xspace}
\newcommand{\pnADiamLo}{612\xspace}
\newcommand{\pnADiamMid}{968\xspace}
\newcommand{\pnADiamHi}{1530\xspace}

\newcommand{\pnAHVObs}{3.27\xspace}

\newcommand{\pnAGRNight}{0.68\xspace}

\newcommand{\tancrediKm}{450\xspace}

\newcommand{\albedoLo}{0.25\xspace}
\newcommand{\albedoMid}{0.10\xspace}
\newcommand{\albedoHi}{0.04\xspace}

\newcommand{\hvTransSig}{0.022\xspace}

\newcommand{\pcSearchDeg}{2.1\xspace}
\newcommand{\pcMagTol}{0.75\xspace}
\newcommand{\pcMinRadius}{2\xspace}

\newcommand{\pcNExp}{63\xspace}

\newcommand{\pnAPcIncompleteCatalog}{18\xspace}

\newcommand{\pnAPcEdge}{9\xspace}
\newcommand{\pnAPcNew}{5\xspace}

\newcommand{\pnAPcNewNights}{4\xspace}

\newcommand{\pnAPcFirst}{2025-05-20\xspace}
\newcommand{\pnAPcLast}{2026-07-12\xspace}

\newcommand{\pnAPcNTwentyFiveAcq}{2\xspace}

\newcommand{\pnAArcCombined}{69\xspace}

\newcommand{\pcNNewNVS}{1\xspace}

\newcommand{\remSnrMin}{5\xspace}
\newcommand{\remChiMax}{1.5\xspace}
\newcommand{\remStampPx}{41\xspace}
\newcommand{\remStampArcsec}{8.2\xspace}
\newcommand{\remNbSnr}{5\xspace}
\newcommand{\remNbFreeSnr}{50\xspace}

\newcommand{\pnAValN}{12\xspace}
\newcommand{\pnAValMedRaMas}{\ensuremath{-}0.1\xspace}
\newcommand{\pnAValMedDecMas}{+0.4\xspace}
\newcommand{\pnAValRmsMas}{2.6\xspace}

\newcommand{\pnAValSigMas}{3\xspace}

\newcommand{\pnAValMagRms}{0.014\xspace}

\newcommand{\pnARemOvrDropped}{1\xspace}
\newcommand{\pnAOvrRepDate}{2026 May 1\xspace}
\newcommand{\pnAOvrRepBand}{z\xspace}
\newcommand{\pnAOvrRepFitCatMas}{18\xspace}
\newcommand{\pnAOvrRepFitParentMas}{4\xspace}
\newcommand{\pnAOvrRepFitMag}{19.33\xspace}
\newcommand{\pnAOvrRepCatMag}{19.39\xspace}
\newcommand{\pnAOvrRepParentMag}{19.32\xspace}
\newcommand{\pnAOvrRepNbSep}{2.0\xspace}
\newcommand{\pnAOvrRepNbRatio}{0.93\xspace}
\newcommand{\pnAOvrDropDate}{2026 May 2\xspace}
\newcommand{\pnAOvrDropBand}{z\xspace}
\newcommand{\pnAOvrDropFitCatMas}{47\xspace}
\newcommand{\pnAOvrDropChi}{2.7\xspace}
\newcommand{\pnAOvrDropStaticSep}{1.4\xspace}
\newcommand{\pnAOvrDropStaticMag}{20.2\xspace}
\newcommand{\pnAOvrDropTwinMin}{8\xspace}

\newcommand{\pnARemNew}{16\xspace}

\newcommand{\pnARemDropSnr}{1\xspace}
\newcommand{\pnARemDropChi}{7\xspace}

\newcommand{\pnARemNewSnrMin}{20\xspace}
\newcommand{\pnARemNewSnrMax}{193\xspace}

\newcommand{\pnARemNewFirst}{2026-03-04\xspace}

\newcommand{\pnAFinalNobs}{30\xspace}
\newcommand{\pnAFinalFirst}{2026-03-04\xspace}
\newcommand{\pnAFinalLast}{2026-07-11\xspace}
\newcommand{\pnAFinalNights}{20\xspace}

\newcommand{\pnAFixedPositions}{15\xspace}

\newcommand{\pnAFixedVisitsTotal}{560\xspace}

\newcommand{\pnAFixedNeighbourPositions}{2\xspace}

\newcommand{\verifyMatchedAPS}{10\xspace}

\newcommand{\verifyMaxSepMas}{0.2\xspace}
\newcommand{\verifyMaxDtMs}{5\xspace}

\newcommand{\verifyRmsWithErr}{10\xspace}

\newcommand{\pnASnrMin}{19.3\xspace}
\newcommand{\pnASnrMax}{163\xspace}

\newcommand{\parentObsTime}{2026-05-01T06:13:30.869Z\xspace}

\newcommand{\parentChildSepMas}{21.9\xspace}

\newcommand{\blendObsTime}{2026-05-23T08:10:34.592Z\xspace}
\newcommand{\blendBand}{i\xspace}
\newcommand{\blendSep}{1.1\xspace}
\newcommand{\blendMag}{22.2\xspace}

\newcommand{\blendObjMag}{19.4\xspace}

\newcommand{\volPvMax}{0.25\xspace}
\newcommand{\volPvMin}{0.03\xspace}
\newcommand{\volRho}{1.8\xspace}
\newcommand{\volDMin}{612\xspace}
\newcommand{\volDMax}{1767\xspace}
\newcommand{\volTperiLo}{44.0\xspace}
\newcommand{\volTperiHi}{45.6\xspace}
\newcommand{\volTeqLo}{41.8\xspace}
\newcommand{\volTeqHi}{44.1\xspace}
\newcommand{\volTeqSpread}{0.2\xspace}

\newcommand{\volPvNtwoAll}{0.033\xspace}
\newcommand{\volDNtwoAll}{1680\xspace}
\newcommand{\volPvCOAll}{0.037\xspace}
\newcommand{\volDCOAll}{1590\xspace}
\newcommand{\volPvCOOne}{0.031\xspace}
\newcommand{\volDCOOne}{1750\xspace}

\newcommand{\pnAWtNRep}{29\xspace}

\newcommand{\pnAWtResRmsMas}{21\xspace}
\newcommand{\pnAWtSigMedMas}{11\xspace}
\newcommand{\pnAWtFloorInflation}{9.0\xspace}

\newcommand{\figDetBoxArcsec}{24\xspace}

\newcommand{\objA}{2026~ED138\xspace}
\newcommand{\fixed}[1]{#1}
\newcommand{\aps}{AP-S\xspace}
\newcommand{\nvs}{NV-S\xspace}

\newcommand{\au}{\,au\xspace}
\newcommand{\kmunit}{\,km\xspace}
\newcommand{\dd}{\,d\xspace}
\newcommand{\mas}{\,mas\xspace}
\newcommand{\arcsecsym}{\ensuremath{^{\prime\prime}}\xspace}
\newcommand{\degsym}{\ensuremath{^{\circ}}\xspace}
\newcommand{\code}[1]{\texttt{#1}}
\xspaceaddexceptions{\arcsecsym \degsym \au \kmunit \mas \dd \arcmin \% \, \sigma}

\shorttitle{A large TNO from early Rubin data}
\shortauthors{Holman, Juri\'c, Napier et al.}

\begin{document}

\title{\objA: A Large Trans-Neptunian Object in Early Rubin Observatory Data,\\
and How Many More May Remain}

\author[0000-0002-1139-4880]{Matthew J. Holman}
\affiliation{Center for Astrophysics $|$ Harvard \& Smithsonian, 60 Garden Street, MS 51, Cambridge, MA 02138, USA}
\email{mholman@cfa.harvard.edu}

\author[0000-0003-1996-9252]{Mario Juri\'c}
\affiliation{DiRAC Institute and Department of Astronomy, University of Washington, Seattle, WA 98195, USA}
\email{mjuric@uw.edu}

\author[0000-0003-4827-5049]{Kevin J. Napier}
\affiliation{Center for Astrophysics $|$ Harvard \& Smithsonian, 60 Garden Street, MS 51, Cambridge, MA 02138, USA}
\email{kevin.napier@cfa.harvard.edu}

\author[0009-0005-5452-0671]{Jacob A. Kurlander}
\affiliation{DiRAC Institute and Department of Astronomy, University of Washington, Seattle, WA 98195, USA}
\email{jkurla@uw.edu}

\author[0000-0003-3313-4921]{Aren Heinze}
\affiliation{DiRAC Institute and Department of Astronomy, University of Washington, Seattle, WA 98195, USA}
\email{aheinze@uw.edu}

\author[0000-0001-5820-3925]{Joachim Moeyens}
\affiliation{DiRAC Institute and Department of Astronomy, University of Washington, Seattle, WA 98195, USA}
\email{moeyensj@uw.edu}

\author[0000-0003-4365-1455]{Megan E. Schwamb}
\affiliation{Astrophysics Research Centre, School of Mathematics and Physics, Queen's University Belfast, Belfast BT7 1NN, UK}
\email{m.schwamb@qub.ac.uk}

\author[0000-0003-0743-9422]{Pedro H. Bernardinelli}
\affiliation{DiRAC Institute and Department of Astronomy, University of Washington, Seattle, WA 98195, USA}
\affiliation{Departamento de Astronomia, Instituto de Astronomia, Geofísica e Ciências Atmosféricas, Universidade de S\~ao Paulo, 05508-090, São Paulo, SP, Brazil}

\email{pedro.bernardinelli@usp.br}

\author[0000-0001-8708-251X]{Ian Sullivan}
\affiliation{DiRAC Institute and Department of Astronomy, University of Washington, Seattle, WA 98195, USA}
\email{sullii@uw.edu}

\author[0000-0001-8018-5348]{Eric C. Bellm}
\affiliation{DiRAC Institute and Department of Astronomy, University of Washington, Seattle, WA 98195, USA}
\email{ecbellm@uw.edu}

\author[0000-0001-8541-8550]{Silvia Protopapa}
\affiliation{Southwest Research Institute, Solar System Science and Exploration Division, Boulder, CO, USA}
\email{silvia.protopapa@swri.org}

\author[0000-0001-7335-1715]{Colin Orion Chandler}
\affiliation{DiRAC Institute and Department of Astronomy, University of Washington, Seattle, WA 98195, USA}
\email{coc123@uw.edu}

\author[0009-0005-9428-9590]{Ian Chow}
\affiliation{DiRAC Institute and Department of Astronomy, University of Washington, Seattle, WA 98195, USA}
\email{chowian@uw.edu}

\author[0009-0001-9424-2291]{Eric J. Christensen}
\affiliation{Vera C. Rubin Observatory, Avenida Juan Cisternas \#1500, La Serena, Chile}
\email{echristensen@lsst.org}

\author[0000-0002-0672-5104]{Samuel Cornwall}
\affiliation{The Grainger College of Engineering, Department of Aerospace Engineering,
    University of Illinois Urbana-Champaign, Urbana, IL, USA
}
\email{samuel41@illinois.edu}

\author[0000-0002-1398-6302]{Siegfried Eggl}
\affiliation{The Grainger College of Engineering, Department of Aerospace Engineering,
    University of Illinois Urbana-Champaign, Urbana, IL, USA
}
\email{eggl@illinois.edu}

\author[0000-0002-4439-1539]{Sarah Greenstreet}
\affiliation{NSF-DOE Vera C. Rubin Observatory/NSF NOIRLab, 950 N.\ Cherry Avenue, Tucson, AZ 85719, USA}
\affiliation{DiRAC Institute and Department of Astronomy, University of Washington, Seattle, WA 98195, USA}
\email{sarah.greenstreet@noirlab.edu}

\author[0000-0003-4285-4453]{Chester Li}
\affiliation{DiRAC Institute and Department of Astronomy, University of Washington, Seattle, WA 98195, USA}
\email{zhuofu@uw.edu}

\author[0000-0003-3827-4691]{Erfan Nourbakhsh}
\affiliation{Department of Astrophysical Sciences, Princeton University, Princeton, NJ 08544, USA}
\email{erfan@astro.princeton.edu}

\author[0000-0002-2121-6375]{James E. Robinson}
\affiliation{Institute for Astronomy, University of Edinburgh, Royal Observatory, Blackford Hill, Edinburgh EH9 3HJ, UK}
\email{james.robinson@ed.ac.uk}

\author[0009-0004-0059-3229]{Devanshi Singh}
\affiliation{DiRAC Institute and Department of Astronomy, University of Washington, Seattle, WA 98195, USA}
\email{ds2004@uw.edu}

\author[0000-0002-5726-3640]{Danica Zilkova}
\affiliation{NSF NOIRLab, 950 N.\ Cherry Avenue, Tucson, AZ 85719, USA}
\email{danica.zilkova@noirlab.edu}

\collaboration{22}{(Rubin Observatory LSST builders)}

\begin{abstract}
We characterize a large trans-Neptunian object (TNO), \objA, discovered in data taken by the
NSF-DOE Vera C.\ Rubin Observatory with the linking code \code{pointdexter}, and validate it independently. \objA has absolute magnitude
$H_V=\pnAHV$, implying a diameter of \pnADiamLo--\pnADiamHi\kmunit for
geometric albedos between \albedoLo and \albedoHi
and making it one of the ten intrinsically brightest known TNOs and a dwarf-planet candidate. At
$i\approx\pnAPkgMedi$ it is bright enough in the sky for ground-based spectroscopy and light curves
and for detailed study with HST and JWST. Archival DECam detections extend its arc to
more than ten years (\pnAArcPrecRep days). Its orbit, with barycentric
$a=\pnAABaryPrecShort$\au, $e=\pnAEBaryPrecShort$ and $i=\pnAIBaryPrecShort\degsym$,
lies at the edge of the \fixed{12:7} mean-motion resonance with Neptune, like
that of the dwarf planet Haumea, but with $g-r=\pnAGR$ it is too red to be a
member of Haumea's collisional family. \objA was missed by earlier surveys because
it lies far south and near the crowded regions of the Galactic plane. Our recoveries of the known TNOs with
$H_V<\fixed{4}$ suggest that about \beRemain more objects like them, and probably more,
remain undiscovered away from the Galactic plane. More such objects probably lie within
\fixed{10}\degsym of the plane, which no search has yet covered to this depth.
\end{abstract}


\section{Introduction} \label{sec:intro}

The largest trans-Neptunian objects (TNOs), the dwarf planets and their near
relatives, formed in the disk that also formed the giant planets. Their number
and orbits constrain how the outer solar system formed and how Neptune migrated.
They are also the easiest TNOs to find.  The wide-field searches of recent
decades \citep{Larsen2001,Larsen2007,Trujillo2003,Brown2008,Schwamb2010,Sheppard2011,Rabinowitz2012,Brown2015}
found most of them in searches over much of the sky. Most of those searches avoided the crowded star fields near the
Galactic plane, however, and reached at most $R\approx\fixed{21.6}$
(Section~\ref{sec:earlier}), about two magnitudes brighter than the Rubin search that found
the object reported here \citep{HolmanRubin}.
The Legacy Survey of Space and Time \citep[LSST;][]{Ivezic2019} at the
NSF-DOE Vera C.\ Rubin Observatory (Rubin) will increase the number of known TNOs several-fold
\citep{Schwamb2023,Kurlander2025}, and it observes those fields routinely.

This paper reports a large TNO, \objA (PN01109 in the search), found in early Rubin data
by the search pipeline described by \citet{HolmanRubin}, which found \fixed{2,440} previously unknown
TNOs; \objA has the lowest absolute magnitude of them. The search uses
\code{pointdexter} software (M.~J.~Holman \& K.~J.~Napier, in preparation)
running as a part of Rubin Solar System Pipelines (M.~Juri\'c et al., in preparation) that reads Rubin's single-visit source catalogs and
groups detections from different nights that belong to one slow-moving
distant object. The search reported \objA as a candidate.

This paper has three aims:
\begin{enumerate}
  \item to describe the independent checks that show \objA is real and new,
        and to measure its orbit, size and colors from every Rubin and archival
        detection;
  \item to understand why it had not been found before, including why
        Rubin's own catalogs missed it in most of the visits whose images
        show it;
  \item to estimate how many more TNOs with $H_V<\fixed{4}$ may remain
        undiscovered, and where.
\end{enumerate}
Section~\ref{sec:data} describes the Rubin data, the initial identification, and our orbit fitting.
Sections~\ref{sec:validation}--\ref{sec:decam} describe the checks, and
Sections~\ref{sec:orbits}--\ref{sec:physical} give the orbit, colors and size, and
whether \objA can have kept volatile ices.
Section~\ref{sec:missed} explains why earlier surveys and Rubin's single-visit
catalogs missed it. Section~\ref{sec:howmany} uses the search's recoveries of
the known TNOs with $H_V<\fixed{4}$ to estimate how many such
objects remain, and Section~\ref{sec:conclusions} summarizes.

\section{Data} \label{sec:data}

\subsection{Rubin data} \label{sec:catalogs}

All data are from Rubin's \objA data release.\footnote{Announced at \url{https://www.rubin.community/t/2026-ed138-a-dwarf-planet-candidate-found-in-early-rubin-data-and-its-data-on-the-rsp/12877}; a DOI has not yet been assigned. The release is described in the Data availability section.} They consist of images taken with LSSTCam \citep{LSSTCam} at Rubin Observatory's Simonyi Survey Telescope (MPC code \code{X05}) between
\pnAPcFirst and \pnAPcLast. We use Rubin's calibrated single-visit images \citep{RubinPromptProducts} and two
source catalogs measured on them: \aps, from Daytime Alert Production (Daytime
AP; \citealt{RubinAP}), which detects sources at $\fixed{5}\sigma$ and gives a position error for each, and \nvs, from the nightly
validation pipeline, which gives no position errors. Both are catalogs of detections in direct images. None of the visits of the observations that the search reported, which we call the
\emph{delivered observations} (Section~\ref{sec:package}), has difference-image sources.

Both catalogs are made automatically from the calibrated images, by two different
runs of the LSST Science Pipelines \citep{PSTN019}: \aps by Daytime Alert
Production, which reprocesses each night's exposures the following day, and \nvs by
the nightly validation run, which has been utilized to provide quick reductions for situational awareness and has since been retired. Neither was tuned for crowded fields, so both miss many detections
of \objA (Section~\ref{sec:deblend}). Where a
catalog has no entry for an object, we measure it ourselves on the calibrated image
(Section~\ref{sec:remeasure}).

Positions are calibrated to Gaia DR3 \citep{GaiaDR3}. For each \aps position the
delivered observations give the catalog error plus \rmsQuadArcsec\arcsecsym in quadrature;
we report the catalog error plus \submitQuadArcsec\arcsecsym, the convention of Rubin's
submissions to the MPC. \nvs gives no position errors, and we leave those uncertainties empty.
The orbit fits use the observations as submitted to the MPC, weighted by these uncertainties
and their RA--Dec correlations (Appendix~\ref{sec:weights}). Appendix~\ref{sec:products} gives further details of
the products.

\subsection{The initial identification} \label{sec:package}

\objA was first detected by {\tt pointdexter} as a cluster of \pnAPkgNobs observations on
\pnAPkgNights nights, spanning \pnAPkgArc days from \pnAPkgFirst to
\pnAPkgLast in the \pnAPkgBands bands. These delivered observations were the starting point for orbital arc extension and all further studies.
The median $i$-band magnitude of \pnAPkgMedi pointed to the lowest absolute magnitude of the new objects found by the search of early operations Rubin data. The object lies far south,
at declination about $-\fixed{43}^\circ$ and ecliptic latitude about $-\fixed{20}^\circ$,
\fixed{12}\degsym south of the Galactic mid-plane (Section~\ref{sec:earlier}).
Table~\ref{tab:summary} summarizes the observations of \objA by data set and band, and
Table~\ref{tab:astrometry} lists every position used in this paper.

\begin{deluxetable*}{lrrrrrrrrrccr}
\tablecaption{Observations of \objA used in this paper\label{tab:summary}}
\tablehead{\colhead{Data set} & \colhead{$u$} & \colhead{$g$} & \colhead{$r$} & \colhead{$i$} & \colhead{$z$} & \colhead{$y$} & \colhead{$VR$} & \colhead{Total} & \colhead{Nights} & \colhead{First} & \colhead{Last} & \colhead{Arc (d)}}
\startdata
Delivered & 2 & 2 & \nodata & 1 & 4 & 1 & \nodata & 10 & 8 & 2026-05-01 & 2026-06-05 & 35 \\
Rubin, all & 3 & 3 & 4 & 6 & 9 & 5 & \nodata & 30 & 20 & 2026-03-04 & 2026-07-11 & 129 \\
DECam & \nodata & 7 & 6 & 6 & 7 & 6 & 12 & 44 & 14 & 2016-02-25 & 2023-04-07 & 2598 \\
Adopted fit & 3 & 10 & 10 & 12 & 16 & 11 & 12 & 74 & 34 & 2016-02-25 & 2026-07-11 & 3789
\enddata
\tablecomments{Number of observations in each band. Rubin bands are $ugrizy$; the DECam data are in $grizY$ (listed under $y$) and the wide $VR$ band. ``Delivered'' are the observations the search reported (Section~\ref{sec:package}); ``Rubin, all'' adds the cataloged and remeasured detections of Section~\ref{sec:precovery}; ``DECam'' are the precoveries of Section~\ref{sec:decam}; ``Adopted fit'' is their union. Nights are local nights in Chile.}
\end{deluxetable*}

\subsection{Orbit fitting} \label{sec:orbitmethod}

We iteratively refitted the orbit of \objA, beginning with the delivered observations and then each time
we added detections in the arc extension process (Sections~\ref{sec:precovery}--\ref{sec:decam}), first with Rubin and then with DECam observations. Two fits
recur in the paper: the \emph{Rubin-only} fit, to all the Rubin observations, and
the \emph{adopted} fit, which adds precovery detections in images from the Dark Energy Camera \citep[DECam;][]{Flaugher2015}. We fit
with our own least-squares code (Levenberg--Marquardt), which propagates orbits with
\code{ASSIST} \citep{Holman2023}, built on \code{REBOUND} \citep{Rein2012} and its
IAS15 integrator \citep{Rein2015}, with the DE440 planetary ephemeris
\citep{Park2021}. The model includes the \fixed{16} most massive asteroids as
perturbers and the light-travel time, and computes each observation from its
own observing site. The state is parametrized in
topocentric spherical coordinates at a mid-arc epoch \citep{Bernstein2000}; we
quote $1\sigma$ and $3\sigma$ ranges from $\chi^2$ profiles along the poorly
constrained radial velocity ($\Delta\chi^2=\fixed{1}$ and \fixed{9}) and derive
posterior probabilities by statistical ranging \citep{Virtanen2001}. We weight
the observations by their reported uncertainties; repeating every fit with a
\floorArcsec\arcsecsym floor added gives the same conclusions
(Appendix~\ref{sec:weights}).

\section{Validation} \label{sec:validation}

\begin{figure*}
\centering
\includegraphics[width=\textwidth]{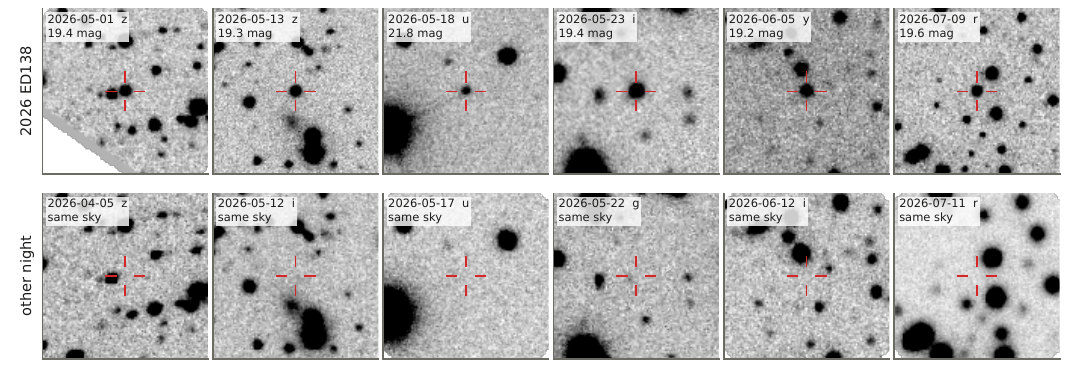}
\caption{Single-visit cutouts, \figDetBoxArcsec\arcsecsym on a side, north
up, east left, with the zscale display range. Top row: \objA
in six visits, centered on the detection and labeled with the date,
band and magnitude. Bottom row: the same sky on another night,
after the object has moved away; the red cross marks the same position.
White areas lie off the detector.}
\label{fig:detections}
\end{figure*}

We checked that \objA is real and that
it was not already known. No observation in the MPC database lies within
\knownBoxRadiusArcmin\arcmin and $\pm$\knownBoxWindowHours\,h of any reported
detection, no known orbit passes near it, and no tracklet in the MPC's
Isolated Tracklet File (ITF), which holds observations not linked to any object,
matches its track (Appendix~\ref{sec:known}).
Every reported row matches a source in the Rubin
catalogs, with signal-to-noise ratios (S/N) of \pnASnrMin--\pnASnrMax, and
Figure~\ref{fig:detections} shows \objA
in single-visit images, absent from the same sky on another night
(Appendix~\ref{sec:real}).

\section{Additional detections in Rubin data} \label{sec:precovery} \label{sec:remeasure}

We call any Rubin detection that is not among the delivered observations an
\emph{additional detection}, and keep \emph{precovery} for the archival DECam
detections (Section~\ref{sec:decam}). Searching the catalogs of every LSSTCam
visit near the predicted track added \pnAPcNew additional detections of \objA,
which extend its arc to \pnAArcCombined days (Appendix~\ref{sec:precoverydetail}, Table~\ref{tab:visits}). In most
other visits whose images show \objA, the catalogs are incomplete around it
(Section~\ref{sec:deblend}).

\begin{figure*}
\centering
\includegraphics[width=\textwidth]{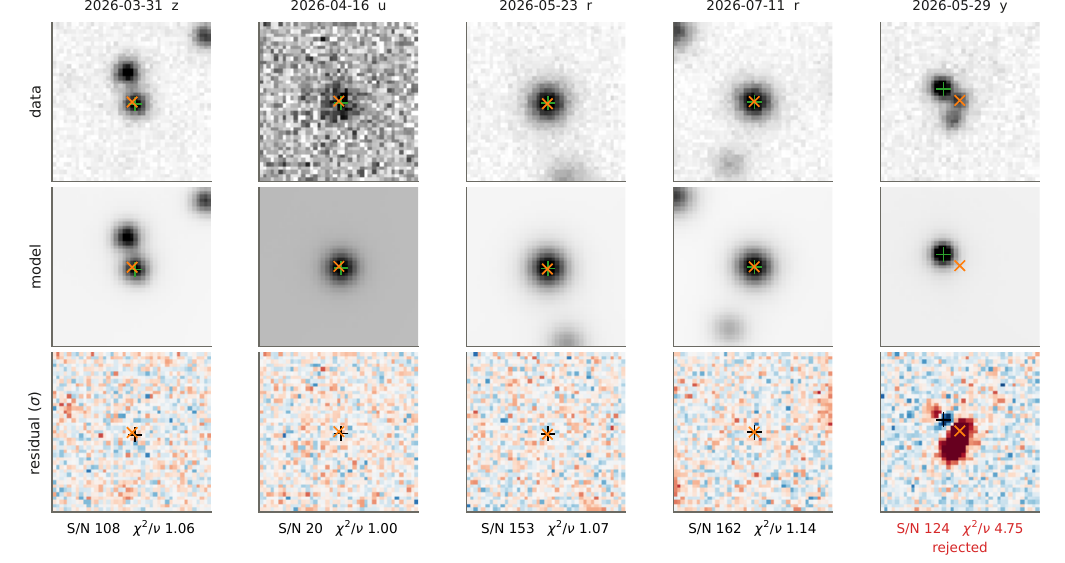}
\caption{Local PSF fits at the predicted positions in visits with no catalog
detection. Each column is one visit, labeled with the date and band;
the rows show the data, the fitted model and the residual in units of the
pixel noise. Orange $\times$: predicted position; $+$: fitted position. The
S/N and reduced $\chi^2$ of each fit are given below it. The last column
shows a fit rejected by the $\chi^2$ cut, where the object is blended with
a star.}
\label{fig:remeasure}
\end{figure*}

We then measured \objA directly at its predicted positions with a
\emph{local PSF fit} (Appendix~\ref{sec:remeasuredetail}): the point-spread function (PSF) model that the pipeline
stored with the image, placed at the object and at each detected neighbor, and a background plane, fitted to a small
cutout of the calibrated image, with no
catalog and no deblending step. Where a clean catalog detection exists, the fit
agrees with it to \pnAValRmsMas\mas rms. It adds \pnARemNew measurements of \objA, the earliest from
\pnARemNewFirst, in visits where no catalog has the object
(Figure~\ref{fig:remeasure}). After
inspection, we replaced one cataloged row of \objA with its local PSF fit and
dropped another (Appendix~\ref{sec:astrometry}). The complete
Rubin data set has \pnAFinalNobs observations of \objA on \pnAFinalNights nights,
from \pnAFinalFirst to \pnAFinalLast.

\section{Precovery in DECam images} \label{sec:decam} \label{sec:decamval}

The Rubin data span only \pnAArcFinalRep days. Earlier detections of \objA
would constrain its orbit far better, and at an $i$-band magnitude of about
\pnAPkgMedi, \objA is within reach of several earlier surveys of the southern sky.

\begin{figure*}
\centering
\includegraphics[width=\textwidth]{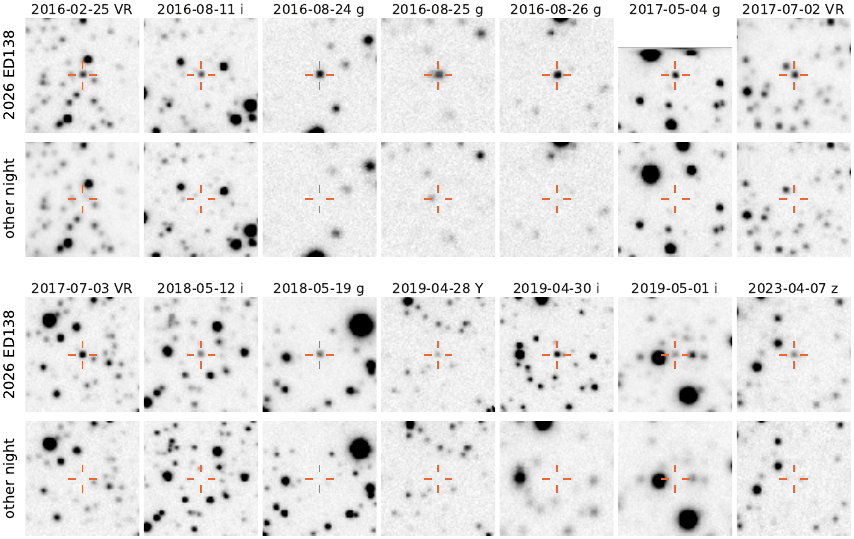}
\caption{\objA in DECam images, one exposure per night (the one with the
highest S/N), each above an image of the same sky taken on another night.
Each cutout is \fixed{24}\arcsecsym on a side, north up and east left; the
marks show the predicted position.}
\label{fig:decam}
\end{figure*}

We first searched with the precovery service of the Asteroid Institute's ADAM
platform \citep{Kiker2023}, which returns the cataloged sources near an orbit's predictions in the
surveys it indexes. Given the Rubin-only orbit, it returned candidates from DECam, in the NOIRLab Source Catalog \citep[NSC DR2;][]{Nidever2021},
and from SkyMapper \citep{Keller2007} and ATLAS \citep{Tonry2018ATLAS}. After we checked each DECam candidate in the
images, \seedNLinked observations on \seedNLinkedNights nights link to the Rubin orbit; the
earliest lay \seedFirstRubinOff\arcsecsym from the Rubin-only prediction, whose
formal uncertainty there was \seedFirstRubinSig\arcsecsym. With the orbit refit to include these observations, we
searched every DECam image in the NOIRLab archive that contains the ephemeris, measuring the image at that position rather than relying on the
catalogs. Following this process, we find that \objA is detected in \dcNAccepted exposures on \dcNAcceptedNights
nights, from \dcFirstAcc to \dcLastAcc (Figure~\ref{fig:decam}).

We tied each position locally to the Gaia DR3 stars \citep{GaiaDR3} in a cutout
around \objA, fitted it with an empirical PSF together with its known static
neighbors, and corrected the bias from uncataloged faint stars, which we measured
by injecting \objA into images of the same sky taken on other nights. We kept
\dcAstN detections on \dcAstNNights nights, with a median uncertainty of
\dcAstErrMed\mas per axis. With each night left out in turn, the fit predicts
that night's positions with an rms of \valPsfCorrLooRa and \valPsfCorrLooDec\mas
in RA and Dec. Timing errors are far smaller. Differential chromatic refraction
is \valDcrMedian\mas in the median and exceeds \fixed{20}\mas in only
\valDcrNOverTwenty exposures, reaching \valDcrMax\mas in $g$ at high airmass
(Appendices~\ref{sec:decamdetail} and \ref{sec:decamchecks}).
The positions are given in Table~\ref{tab:astrometry}.

\section{Orbits} \label{sec:orbits} \label{sec:orbitresults}

Table~\ref{tab:elements} gives the elements of the adopted fit (Section~\ref{sec:orbitmethod}), heliocentric and
barycentric. We quote and classify with barycentric elements, which change
little between the different epochs of the fits; heliocentric elements oscillate
with the Sun's motion about the barycenter. Appendix~\ref{sec:orbitdetail} shows how
the orbit evolved as detections were added (Table~\ref{tab:orbits},
Figure~\ref{fig:arange}) and the residuals of the Rubin observations
(Figure~\ref{fig:residuals}).

The adopted fit of \objA spans \pnAArcPrecRep days, from
\seedFirstNight to \pnAFinalLast, with \pnANobsPrecRep observations; no
residual exceeds \pnAMaxChiPrecRep$\sigma$. It gives barycentric
$a=\pnAABaryPrecRep$\au ($1\sigma$ range \pnAABaryLoOnePrecRep--\pnAABaryHiOnePrecRep\au).
With the Rubin data alone (\pnAArcFinalRep days) the $1\sigma$ range was
\pnAABaryLoOneFinalRep--\pnAABaryHiOneFinalRep\au, and on the delivered
observations the $3\sigma$ range was
\pnAABaryLoThreeOrigRep--\pnAABaryHiThreeOrigRep\au (Figure~\ref{fig:arange}).
The DECam detections fit the Rubin-only orbit's predictions
(Appendix~\ref{sec:adam}), and the two fits agree. With barycentric
$a=\pnAABaryPrecRep$\au, $i=\pnAIBaryPrecRep\degsym$ and
$q=\pnAQBaryPrecRep$\au, \objA lies near the \fixed{12:7} mean-motion resonance
with Neptune. We integrated the nominal orbit and clones
drawn from its covariance, \pnAResN particles in all, for \pnAResMyr\,Myr. The
model has the Sun, with the mass of the inner planets added to it, and the four
giant planets. We work in the invariable plane, the plane perpendicular to the
total angular momentum of the solar system, and measure the elements from the
system's center of mass. In this integration the resonant angle
\fixed{$\phi=12\lambda-7\lambda_N-3\varpi-2\Omega$}, where $\lambda$ and $\varpi$
are the mean longitude and the longitude of perihelion, $\Omega$ is the longitude
of the ascending node and $\lambda_N$ is the mean longitude of Neptune, librates about
\pnAResCenter\degsym with a large amplitude (\pnAResAmp\degsym, the median over
\fixed{200}-kyr windows for the resonant particles), so the orbit lies at the
edge of the resonance. By the criteria of \citet{Khain2020}, \pnAResNLib of the
\pnAResN particles are resonant over \pnAResMyr\,Myr and the rest are not. That
fraction, \pnAResFrac, cannot be distinguished from the \fixed{50}\% at which
\citet{Khain2020} call an object a resonant candidate, and it is far below the
\fixed{80}\% for a secure class. In earlier integrations with \fixed{31} particles, it ranged from about two
fifths to about three quarters between models and between refits of the orbit
that differ by a hundredth of its uncertainty. About half of the orbits that fit
the data are therefore resonant. The
orbit is now so well determined that the clones differ only by chaotic spreading
at the resonance's edge, so a longer arc will not settle the class.
We tested all \pnAResNArgs arguments of this fifth-order resonance allowed by the
d'Alembert rules, including those that involve Neptune's longitude of perihelion
$\varpi_N$ and node $\Omega_N$. Only two librate: $\phi$, and
\fixed{$12\lambda-7\lambda_N-\varpi_N-4\Omega$} in \pnAResNLibOther of the same
particles. No particle librates in any argument without also librating in $\phi$.

The orbit of \objA is close to that of the collisional family of the dwarf
planet Haumea \citep{Brown2007Family,Ragozzine2007,Proudfoot2019,Proudfoot2022},
and Haumea is in the \fixed{12:7} resonance. Family members share two
properties \citep{Ragozzine2007}: a surface of nearly pure water ice, with
neutral visible colors \citep{Brown2007Family,Schaller2008,Snodgrass2010}, and
an orbit that the collision could have produced. The color of \objA rules out
membership: with $g-r=\pnAGR\pm\pnAGRErr$ (Section~\ref{sec:colors}), it is
redder than the Sun ($g-r=\fixed{0.41}$ in Rubin's bands; \citealt{HolmanRubin}) by
about three times its uncertainty, whereas Haumea, whose visible reflectance is
neutral \citep{Rabinowitz2006}, and the confirmed members have nearly solar colors.
The orbit is consistent with this, though on its own it would not decide
membership. We integrated Haumea and its \famN members
confirmed by their spectra for \pnAResMyr\,Myr together with \objA, and computed
proper inclinations (inclinations with the periodic perturbations by the planets
averaged out) relative to the invariable plane. Reaching the orbit of \objA from the
center of the family (proper $a=\famCentreA$\au, $e=\famCentreE$,
$i=\famCentreI\degsym$, as computed by \citealt{Ragozzine2007} with their own
method) needs an ejection velocity of at
least \pnADvMin\,m\,s$^{-1}$, against \famDvMinLo--\famDvMinHi\,m\,s$^{-1}$ for the
members and about \fixed{150}\,m\,s$^{-1}$ for the family as a whole
\citep{Ragozzine2007}; Haumea itself, which has drifted in the resonance, needs
\haumeaDvMin\,m\,s$^{-1}$. Because of such drift, the
family is also compared through the Tisserand parameter with respect to Neptune, a
combination of $a$, $e$ and $i$ that such drift nearly conserves: it is
\pnATissN for \objA, above the family's \famTissNLo--\famTissNHi (Haumea
\haumeaTissN). The proper inclination of \objA, \pnAPropI\degsym, is lower than
Haumea's (\haumeaPropI\degsym) and than that of every member
(\famPropILo--\famPropIHi\degsym), and its mean eccentricity, \pnAMeanE, is just
above theirs (\famMeanELo--\famMeanEHi). \objA is therefore not a
family member. It is one of the unrelated objects expected in this part of the
belt \citep{Ragozzine2007}. A near-infrared
spectrum without water-ice absorption would add direct evidence.

\centerwidetable  %
\begin{deluxetable*}{lc}
\tablecaption{Adopted orbits (reported uncertainties).\label{tab:elements}}
\tablehead{\colhead{} & \colhead{2026 ED138}}
\startdata
\cutinhead{Heliocentric}
Epoch (MJD TDB) & 59337.8171 \\
$a$ (au) & $43.269536^{+0.000055}_{-0.000059}$ \\
$e$ & $0.1339372^{+0.0000075}_{-0.0000075}$ \\
$q$ (au) & $37.47414^{+0.00027}_{-0.00028}$ \\
$i$ (deg) & $24.3846266^{+0.0000018}_{-0.0000018}$ \\
$\Omega$ (deg) & $322.797458 \pm 0.000029$ \\
$\omega$ (deg) & $169.4550 \pm 0.0029$ \\
$M$ (deg) & $120.4534 \pm 0.0041$ \\
\cutinhead{Barycentric}
$a$ (au) & $43.090975^{+0.000056}_{-0.000060}$ \\
$e$ & $0.1387280^{+0.0000074}_{-0.0000074}$ \\
$q$ (au) & $37.11305^{+0.00027}_{-0.00027}$ \\
$i$ (deg) & $24.4338512^{+0.0000018}_{-0.0000018}$ \\
$\Omega$ (deg) & $322.606057 \pm 0.000029$ \\
$\omega$ (deg) & $168.6998 \pm 0.0028$ \\
$M$ (deg) & $121.0797 \pm 0.0040$ \\
\cutinhead{Fit}
$r$ at epoch (au) & 46.726 \\
$\Delta$ at epoch (au) & 46.052 \\
Observations & 74 \\
Arc (d) & 3789.0 \\
$\chi^2$/dof & 84.72/142 \\
\cutinhead{Dynamics}
\fixed{12:7} resonant fraction & 0.46 \\
\quad libration amplitude (deg) & 95--115 \\
Tisserand parameter $T_N$ & 2.835 \\
Proper inclination (deg) & 25.7
\enddata
\tablecomments{Rubin and DECam observations (``+ DECam'' in Table~\ref{tab:orbits}). Ecliptic J2000 osculating elements at each fit's epoch; ASSIST with DE440 and sb441-n16. Heliocentric
elements use the Sun's mass; barycentric elements use the mass of the Sun, the planets, the Moon and Pluto
(DE440), about the solar-system barycenter. Uncertainties on $a$, $e$, $q$, $i$ are the $\Delta\chi^2 = 1$
profile range; on $\Omega$, $\omega$, $M$ the linearized 1$\sigma$ from the fit covariance. The barycentric
state vector and its covariance are in the data release. Dynamics: from \fixed{10}\,Myr integrations of the
nominal orbit and clones drawn from its covariance (Section~\ref{sec:orbits}); the resonant fraction is that of the
particles resonant over the integration by the criteria of \citet{Khain2020}, and the libration amplitude is the
median over \fixed{200}-kyr windows, for the resonant particles; the proper inclination is relative to the
invariable plane.}
\end{deluxetable*}

\section{Photometry} \label{sec:phot}

\subsection{Rubin} \label{sec:photrubin}

\begin{figure}
\centering
\includegraphics[width=\columnwidth]{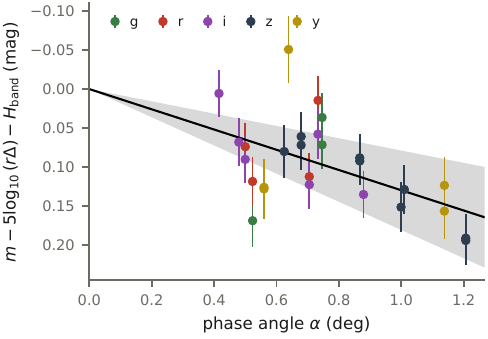}
\caption{Phase curve of \objA: the reduced magnitude of each of the \pnABetaN
Rubin $grizy$ measurements, minus the fitted $H$ of its band, against phase angle.
The error bars include the \pcRotFloor\,mag allowance for rotation. The line is the
fitted slope $\beta=\pnABeta$\,mag\,deg$^{-1}$, and the shaded band its
$1\sigma$ range.}
\label{fig:phase}
\end{figure}

We use the Rubin magnitudes of \objA from the catalogs and the local
PSF fits (Section~\ref{sec:precovery}), as submitted to the MPC. Unlike the DECam magnitudes
(Section~\ref{sec:photdecam}), these are measured on direct images, where unresolved stars could bias them.
The local PSF fits model each detected neighbor and agree with the catalog magnitudes to \pnAValMagRms\,mag rms
where both exist (Appendix~\ref{sec:remeasuredetail}), and the one catalog row blended with a static star involves
a star of $\blendBand\approx\blendMag$, against $\blendBand\approx\blendObjMag$ for \objA
(Appendix~\ref{sec:real}). We compute
reduced magnitudes $m-\fixed{5}\log_{10}(r\Delta)$, where $r$ and $\Delta$ are
the distances from the Sun and from the observer in au, from the adopted orbit
geometry. At the small phase angles $\alpha$ (the Sun--object--observer angle)
of a distant object, the reduced magnitude grows almost linearly with $\alpha$,
so we model it as $H+\beta\alpha$, where the phase slope $\beta$ is in
mag\,deg$^{-1}$ and $H$ is the absolute magnitude, at zero phase.
The phase angles of \objA span \pnAAlphaMin--\pnAAlphaMax\degsym, enough to
measure $\beta$: a fit of one $\beta$ and one $H$ per band to the
\pnABetaN $grizy$ measurements, each weighted with
\pcRotFloor\,mag of rotational variation added in quadrature, gives
$\beta=\pnABeta\pm\pnABetaErr$\,mag\,deg$^{-1}$ ($\chi^2=\pnABetaChisq$ for
\pnABetaDof degrees of freedom; the error is scaled by $\sqrt{\chi^2/\nu}$;
Figure~\ref{fig:phase}).
This is typical of TNOs \citep{Rabinowitz2007,AlvarezCandal2016}. Phase slopes
can differ between bands \citep{AyalaLoera2018,AlvarezCandal2019}, which a single
$\beta$ does not capture; over the small phase angles of our observations the effect
on the colors is small. We reduce
\objA to zero phase with it, which lowers its $H_V$ from \pnAHVObs at the mean
observed phase to \pnAHV at zero phase.

Rubin's calibration in the visits of \objA agrees with The Monster, its reference
catalog \citep{Ferguson2025}, to within \pcRubMonMaxAbs\,mag in every band; against
synthetic photometry from the low-resolution Gaia DR3 XP spectra
\citep[Gaia XP;][]{Montegriffo2023} its $g$ differs by \pcRubXpg\,mag, which we treat as a
systematic uncertainty (Appendix~\ref{sec:photchecks}).

\subsection{DECam} \label{sec:photdecam}

The DECam magnitudes of \objA are PSF fits on difference images, in all \dcPhotN
exposures: each exposure minus PSF-matched templates of the same sky from other
nights (Appendix~\ref{sec:photdecamcal}). The difference removes the static
stars of this crowded field, cataloged or not; at its own position \objA is
blended with stars that bias a direct PSF fit by up to \dcDiMinusDirectMaxAbs\,mag
in single exposures. We tied each exposure to Gaia XP synthetic
photometry of stars in the same cutout
(Appendix~\ref{sec:photchecks}; Table~\ref{tab:decamphot}). We do not use the
NSC DR2 zero points: most reference stars in these fields are reddened, which
biases those zero points by up to \dcPhotNscBiasMax\,mag. Reduced to zero phase and converted to the LSST
bandpasses, the DECam magnitudes of \objA differ from Rubin's by \pcBudXpg,
\pcBudXpr, \pcBudXpi, \pcBudXpz and \pcBudXpY\,mag in $g$, $r$, $i$, $z$ and $Y$
(Table~\ref{tab:photbudget}). A joint fit of the
Rubin and DECam magnitudes, with one $\beta$ and one DECam offset, gives
$\beta=\pcBetaJoint\pm\pcBetaJointErr$\,mag\,deg$^{-1}$, consistent with the
Rubin value of Section~\ref{sec:photrubin}, and an offset of
$\pcJointOffset\pm\pcJointOffsetErr$\,mag.

\section{Physical properties} \label{sec:physical}

Table~\ref{tab:physical} collects the physical properties of \objA derived in
this and the previous sections.

\begin{deluxetable}{llc}
\tablecaption{Physical properties of \objA.\label{tab:physical}}
\tablehead{\colhead{Quantity} & \colhead{Value} & \colhead{Section}}
\startdata
\cutinhead{Photometry, zero phase}
$H_u$ (mag) & $5.05\pm0.07$ & \ref{sec:colors} \\
$H_g$ (mag) & $3.50\pm0.05$ & \ref{sec:colors} \\
$H_r$ (mag) & $2.87\pm0.04$ & \ref{sec:colors} \\
$H_i$ (mag) & $2.61\pm0.04$ & \ref{sec:colors} \\
$H_z$ (mag) & $2.49\pm0.05$ & \ref{sec:colors} \\
$H_y$ (mag) & $2.40\pm0.06$ & \ref{sec:colors} \\
$H_V$ (mag) & $3.19\pm0.04$ & \ref{sec:colors} \\
$H_V$, MPC band offsets (mag) & 2.76--3.22 & \ref{sec:colors} \\
$u-g$ (mag) & $1.55\pm0.09$ & \ref{sec:colors} \\
$g-r$ (mag) & $0.63\pm0.07$ & \ref{sec:colors} \\
$r-i$ (mag) & $0.26\pm0.05$ & \ref{sec:colors} \\
$i-z$ (mag) & $0.12\pm0.06$ & \ref{sec:colors} \\
$z-y$ (mag) & $0.08\pm0.07$ & \ref{sec:colors} \\
Phase slope $\beta$ (mag\,deg$^{-1}$)\tablenotemark{a} & $0.14\pm0.03$ & \ref{sec:photdecam} \\
Phase angles (deg) & 0.42--1.21 & \ref{sec:photrubin} \\
Median $i$, \fixed{2026} (mag) & 19.4 & \ref{sec:package} \\
Rotational variation & not measured\tablenotemark{b} & \ref{sec:photrubin} \\
\cutinhead{Size}
$D$, $p_V=0.25$ (km) & 612 & \ref{sec:size} \\
$D$, $p_V=0.10$ (km) & 968 & \ref{sec:size} \\
$D$, $p_V=0.04$ (km) & 1530 & \ref{sec:size} \\
Albedo, rotation period, mass & not measured & \nodata \\
\cutinhead{Temperature and volatiles, $p_V=0.25$--0.03}
Perihelion temperature (K) & 44.0--45.6 & \ref{sec:volatiles} \\
Equivalent temperature (K) & 41.8--44.1 & \ref{sec:volatiles} \\
CH$_4$ & lost at every $p_V$ & \ref{sec:volatiles} \\
CO & kept if $p_V<0.037$ & \ref{sec:volatiles} \\
N$_2$ & kept if $p_V<0.033$ & \ref{sec:volatiles}
\enddata
\tablecomments{Absolute magnitudes are at zero phase, with the measured phase slope; colors are differences of
them. $H_V$ is from $g$ and $i$ with the relation of \citet{Porter2025}; the MPC-style range applies the MPC's
fixed band-to-$V$ offsets, without a color term, to each band. Diameters are
$D=\fixed{1329}\,\mathrm{km}\,p_V^{-1/2}\,10^{-H_V/\fixed{5}}$. Temperatures and volatile retention are for an
isothermal body with the albedo range shown (Figure~\ref{fig:volatiles}).}
\tablenotetext{a}{Joint fit of the Rubin and DECam magnitudes, with one DECam offset. The absolute magnitudes above use the Rubin data alone, $\beta=\pnABeta\pm\pnABetaErr$.}
\tablenotetext{b}{The photometric fits allow \pcRotFloor\,mag of rotational variation.}
\end{deluxetable}

\subsection{Colors and absolute magnitude} \label{sec:colors}

The absolute magnitudes below are at zero phase (Section~\ref{sec:photrubin}).
Per band, we average the reduced magnitudes: $H_g=\pnAHg\pm\pnAHgErr$, $H_r=\pnAHr\pm\pnAHrErr$,
$H_i=\pnAHi\pm\pnAHiErr$, $H_z=\pnAHz\pm\pnAHzErr$ and
$H_y=\pnAHy\pm\pnAHyErr$.
The errors are standard errors of the mean, from the scatter within each
band, and include the uncertainty of $\beta$.
\objA is redder than the Sun: $g-r=\pnAGR\pm\pnAGRErr$ (\pnAGRNight from
a same-night pair).

We convert to $V$ with the Rubin relation of \citet{Porter2025},
$V=g-\fixed{0.302}\,(g-i)-\fixed{0.041}$, fitted to stars in the bands of ComCam,
Rubin's commissioning camera; its rms scatter is \hvTransSig\,mag. Propagating the $g$ and $i$ errors and adding
that scatter in quadrature gives $H_V=\pnAHV\pm\pnAHVErr$. These errors do not include the
difference between the ComCam and LSSTCam bandpasses or the photometric
zero-point uncertainty; Rubin's $g$ is \pcRubXpg\,mag offset from Gaia XP in the
field of \objA (Section~\ref{sec:photrubin}), which would change $g-r$ by as much.  Applying a transformation
fitted to stars to the spectrum of a TNO is itself uncertain: for bright TNOs with
published colors, \citet{HolmanRubin} find Rubin's $g-r$ about a tenth of a
magnitude bluer than the literature values transformed to Rubin's bands. If that
holds here, \objA is redder still, and the color argument of
Section~\ref{sec:orbits} is stronger.
The Minor Planet Center computes $H_V$ differently: it converts each band to $V$
with a fixed offset (for example, $+\fixed{0.37}$\,mag for $z$), with no color
term.\footnote{\url{https://data.minorplanetcenter.net/api/mag-band}} For an object
as red as \objA, that makes the red bands too bright in $V$. Applied to the $H$
values above, the offsets give $H_V$ from \pnAHVMpcLo (from $\pnAHVMpcLoBand$) to
\pnAHVMpcHi (from $\pnAHVMpcHiBand$). An $H_V$ computed mostly from the $i$, $z$ and
$y$ bands, as the MPC's is likely to be, will therefore be brighter than ours.

\subsection{Size and albedo} \label{sec:size}

With $D=\fixed{1329}\,\mathrm{km}\,p_V^{-1/2}\,10^{-H_V/\fixed{5}}$ \citep{Pravec2007},
geometric albedos of \albedoLo, \albedoMid and \albedoHi give diameters of
\pnADiamLo, \pnADiamMid, and \pnADiamHi\kmunit.
Measured TNO albedos span this range \citep{Lellouch2013}, but they depend
on size: the largest TNOs tend to have higher albedos than smaller ones
\citep{Brown2008,Lellouch2013}. For an object with an absolute magnitude as low as that of \objA, an albedo
at the low end of the range is therefore less likely, though low albedos do
occur among large TNOs: Salacia, about one magnitude fainter in absolute
magnitude than \objA, has a geometric albedo of only about \fixed{0.04}
\citep{Fornasier2013}. Several large TNOs whose sizes have been measured by stellar
occultation or thermal emission have albedos near \fixed{0.1}: Quaoar
$\fixed{0.109}\pm\fixed{0.007}$ \citep{BragaRibas2013}, \fixed{2002~MS4}
$\fixed{0.10}\pm\fixed{0.025}$ \citep{Rommel2023} and Gonggong \fixed{0.089}
\citep{Pal2016}; at that albedo \objA would be about \pnADiamMid\kmunit across.
Albedos can also be much higher: Pluto, Eris and Makemake, whose surfaces are rich in
CH$_4$ ice, have geometric albedos of about \fixed{0.6} to \fixed{0.9} \citep{Brown2008}. With such an
albedo, \objA would be only about \fixed{320}--\fixed{400}\kmunit across. This is less likely, but
its surface composition is still unknown. At that size it could not have kept CH$_4$
(Section~\ref{sec:volatiles}), so so bright a surface would need CH$_4$ supplied from its interior.
A diameter between \pnADiamLo and \pnADiamMid\kmunit (albedos between \albedoLo and
\albedoMid) is more probable. Even the smaller of these sizes is above the
size at which \citet{Tancredi2008} expect icy bodies to be close to hydrostatic
equilibrium, about \tancrediKm\kmunit, so \objA is a dwarf-planet candidate.
Size is only the first of the
tests of \citet{Tancredi2008}; a lightcurve or a shape measurement is needed to
apply the others. Whether \objA is compact and round is less certain: \citet{Grundy2019} suggest that icy bodies of
\fixed{400}--\fixed{1000}\kmunit are the transition between small, porous objects
and larger ones that have collapsed, so the answer depends on the albedo.
\objA is one of the ten intrinsically brightest known TNOs (Section~\ref{sec:howmany}). Other large TNOs
found in recent years, such as \fixed{(532037)} Chiminigagua (\fixed{2013~FY27}; \citealt{Sheppard2018}),
\fixed{2017~OF201} \citep{Cheng2026}, \fixed{2010~JO179}
\citep{Holman2018b} and \fixed{2015~RR245} \citep{Bannister2016}, are comparable or
fainter.
Thermal photometry or a stellar occultation is needed to decide between
these sizes.

\subsection{Volatile retention} \label{sec:volatiles}

\begin{figure*}
\centering
\includegraphics[width=0.72\textwidth]{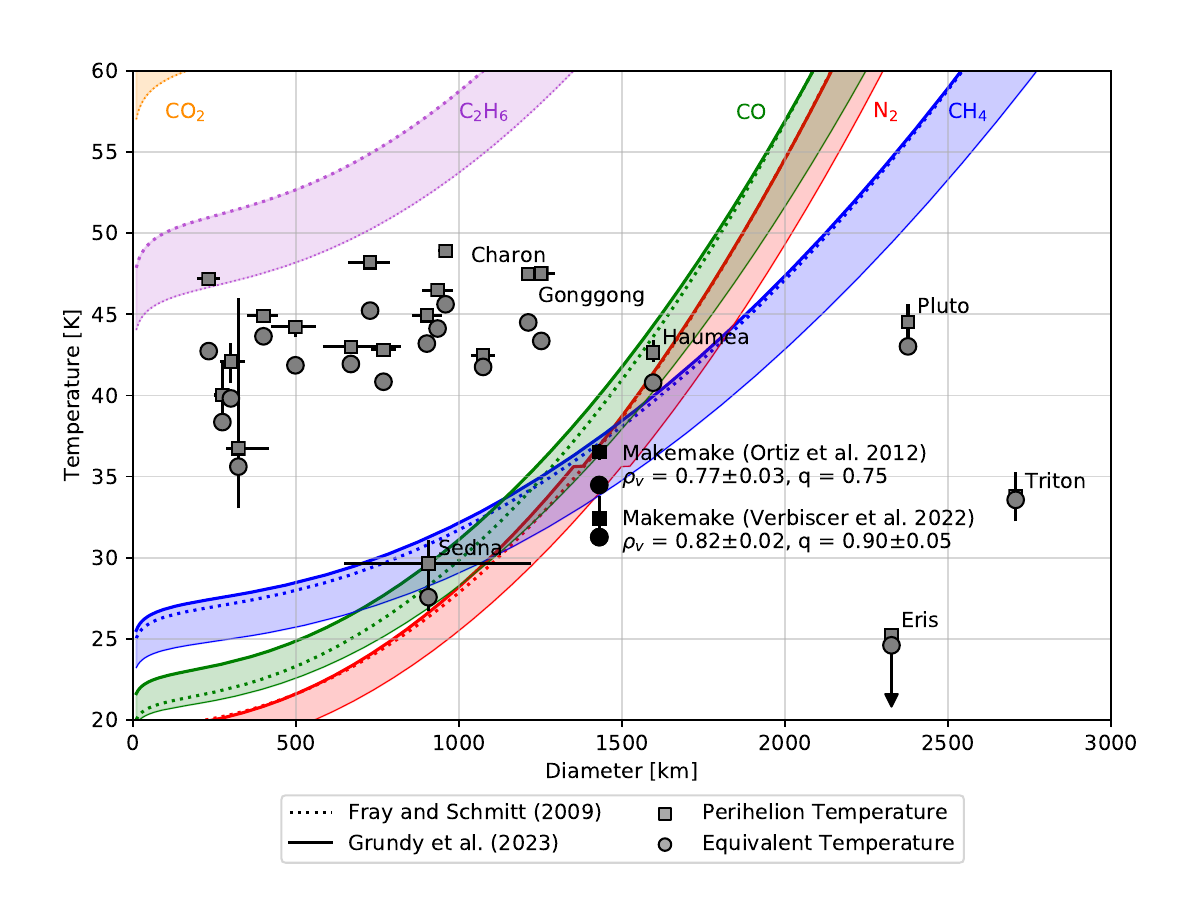}\llap{\includegraphics[width=0.72\textwidth]{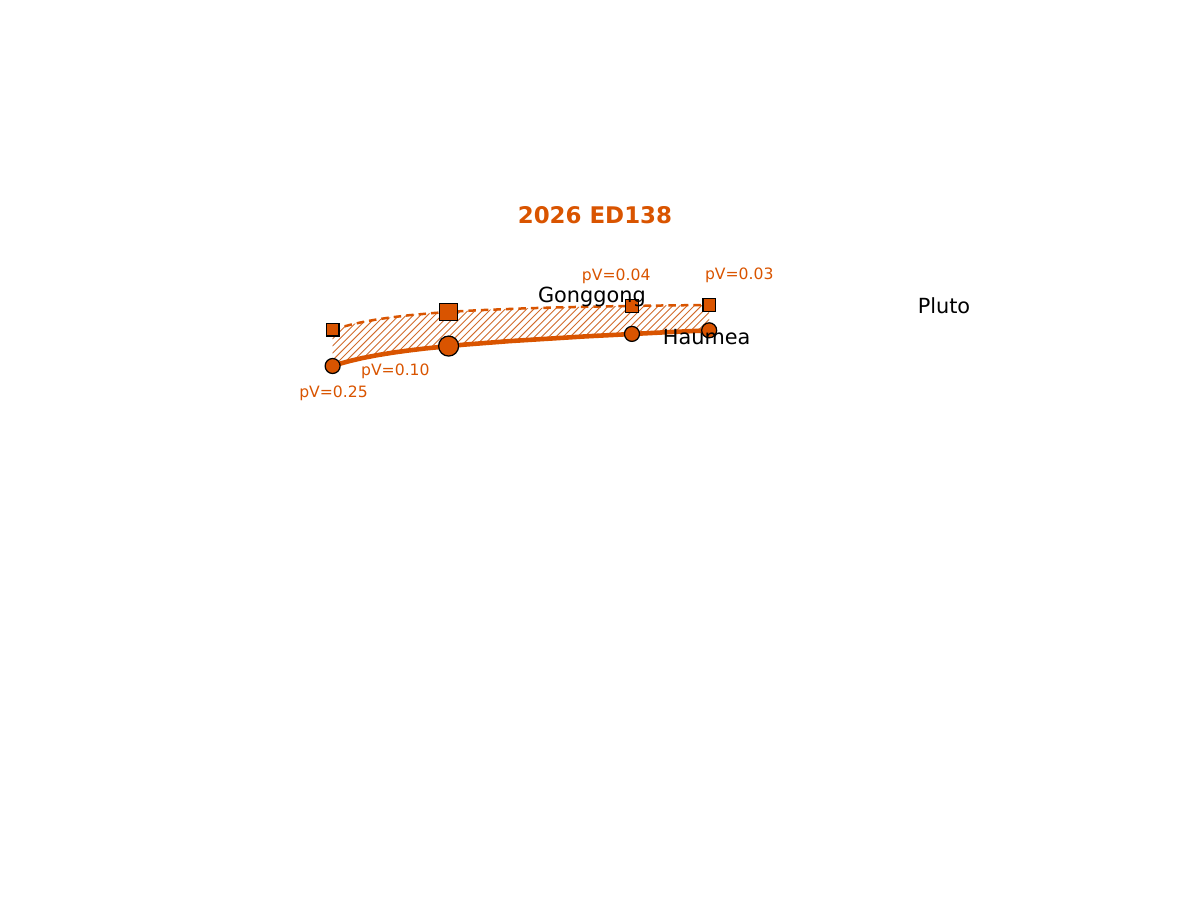}}
\caption{\objA (orange) added to the volatile-retention diagram of
\citet[their Figure~11]{Protopapa2025}, after \citet{Schaller2007}. Each colored
band spans the temperatures at which a body of the given diameter loses \fixed{1}\%
(thin line) to \fixed{100}\% (thick line) of its initial CH$_4$, N$_2$, CO, C$_2$H$_6$
or CO$_2$ over \fixed{4.6}\,Gyr by surface-bounded Jeans escape, for the vapor pressures
of \citet[solid; CH$_4$, N$_2$ and CO only]{Grundy2024vp} and \citet[dotted]{FraySchmitt2009}; a body below a
curve retains that species. Gray squares and circles are the perihelion and
equivalent temperatures of known TNOs from \citet{Protopapa2025}. For \objA, the
dashed and solid orange curves are its perihelion and equivalent (for N$_2$)
temperatures as its geometric albedo runs from $p_V=\volPvMax$ (\volDMin\kmunit) to \volPvMin
(\volDMax\kmunit); the band between them is hatched, and the markers are at
$p_V=\fixed{0.25}$, \fixed{0.10}, \fixed{0.04} and \fixed{0.03}.}
\label{fig:volatiles}
\end{figure*}

Whether \objA can have kept volatile ices such as CH$_4$, N$_2$ and CO on its surface
depends on its size and temperature. \citet{Schaller2007} estimated this with a simple
model: each ice is in equilibrium with its vapor, and that thin atmosphere escapes by
Jeans escape at a rate set by the vapor pressure and the body's gravity. They divided
each orbit into short segments, added up the loss over the age of the solar system, and
compared it with a primordial, cometary abundance of the volatile. A body retains a
volatile only if it is large and cold enough. To place a TNO on the diagram, they
converted its total loss into an equivalent temperature: the temperature of an
identical body on a circular orbit that would lose the same amount. Most of the loss
happens near perihelion, so the equivalent temperature is close to the perihelion
temperature, and lower for more eccentric orbits. \citet{Protopapa2025} updated this
simple model of Jeans escape for surface-bounded atmospheres over \fixed{4.6}\,Gyr,
with new vapor pressures \citep{FraySchmitt2009,Grundy2024vp} and revised sizes and
albedos of the known TNOs. Figure~\ref{fig:volatiles} shows their result, with \objA added.

We place \objA on the diagram in the same way. We take it to be an isothermal body,
$T=\fixed{278.6}\,\mathrm{K}\,(\fixed{1}-A)^{1/4}\,(r/\fixed{1}\,\mathrm{au})^{-1/2}$, with
Bond albedo $A=p_Vq$ and phase integral $q=\fixed{0.336}\,p_V+\fixed{0.479}$
\citep{Brucker2009}, and a bulk density of \volRho\,g\,cm$^{-3}$. For each of CH$_4$,
N$_2$ and CO we average its Jeans escape rate over the orbit, with the vapor pressures
of \citet{Grundy2024vp}, and find the equivalent temperature. For geometric albedos from
\volPvMax to \volPvMin ($D=\volDMin$--\volDMax\kmunit), the perihelion temperature of
\objA is \volTperiLo--\volTperiHi\,K and its equivalent temperature
\volTeqLo--\volTeqHi\,K. The three species give equivalent temperatures within
\volTeqSpread\,K of each other.

At a typical albedo, \objA should have lost these volatiles. If its geometric albedo is
\albedoHi or higher ($D\le\pnADiamHi$\kmunit), it lies above the curves at which Jeans
escape removes all of its CH$_4$, N$_2$ and CO (Figure~\ref{fig:volatiles}), as do the
other TNOs of its size and temperature \citep[see also][]{Emery2024}. Only a very dark,
and so larger, \objA could have kept some. Below $p_V=\volPvCOAll$
($D>\volDCOAll$\kmunit) it would keep some CO, and below \volPvCOOne
($D>\volDCOOne$\kmunit) it would lose less than \fixed{1}\% of it. Below
$p_V=\volPvNtwoAll$ ($D>\volDNtwoAll$\kmunit) it would also keep some N$_2$. Albedos
this low would be unusual for so large a TNO; even Salacia is brighter. CH$_4$ is
lighter and escapes more easily, and \objA loses all of it down to $p_V=\volPvMin$, the
darkest albedo we consider. The less volatile C$_2$H$_6$ should remain.

This model assumes simple Jeans escape of a primordial, cometary abundance of
volatiles. On large TNOs the loss and processing of volatiles are likely more complex.
Hydrodynamic escape or a giant impact would have removed more than Jeans escape alone
\citep{Protopapa2025}. On the other hand, volatiles can be resupplied from the interior:
the light hydrocarbons on Sedna, Gonggong and Quaoar suggest that their CH$_4$ is
resupplied \citep{Emery2024}. A near-infrared spectrum of \objA would
test this. If \objA is dark, it lies in the transition region of
Figure~\ref{fig:volatiles}, where it should keep some volatiles but not others. Finding
which ices are present would show whether processes other than escape shape its surface.

\section{Why \objA was missed} \label{sec:missed}

At $V\approx\fixed{20}$ ($i\approx\pnAPkgMedi$), \objA is about two magnitudes
brighter than the depth of the wide-field searches of the past two decades,
$R\approx\fixed{21.4}$--\fixed{21.6} \citep{Sheppard2011,Rabinowitz2012}, which together
are thought to have found nearly all TNOs this bright over most of the sky
\citep{Sheppard2011,Schwamb2014}. Why, then, was it still unknown?

\subsection{Earlier surveys} \label{sec:earlier}

It is at declination about $-\fixed{43}^\circ$, ecliptic latitude about
$-\fixed{20}^\circ$ and Galactic latitude about $-\fixed{12}^\circ$.
It was even closer to the plane during the earlier searches: it crossed the
plane in \fixed{2011} and lay within \fixed{10}\degsym of it from \fixed{2000}
to \fixed{2023}. The searches from the northern hemisphere, Spacewatch
\citep{Larsen2001,Larsen2007} and the Palomar surveys
\citep{Trujillo2003,Brown2008,Schwamb2010}, did not reach this far south, and the
survey of \citet{Schwamb2010} also avoided fields within \fixed{15}\degsym of the
Galactic plane. The southern and Galactic-plane survey of \citet{Sheppard2011}
reached $R=\fixed{21.6}$, deep enough, but in \fixed{2010} \objA lay at the southern
edge of the area it covered, within about a degree of its fields (their Figure~\fixed{1}).
The La Silla--QUEST survey \citep{Rabinowitz2012} searched about \fixed{7,500}
square degrees of the southern sky to $R=\fixed{21.4}$, but during that survey \objA
lay within a few degrees of the Galactic plane. The all-sky search of
\citet{Brown2015} covered only Galactic latitudes above \fixed{20}\degsym, and in the
south it was nearly complete only for $V\lesssim\fixed{18.6}$. The Dark Energy
Survey, the largest earlier search of the southern sky for TNOs
\citep{Bernardinelli2022}, stayed about \fixed{19}\degsym from \objA throughout
\citep{HolmanRubin}; Pan-STARRS cannot observe this far south, and the solar-system
objects in Gaia's data releases are objects already known \citep{Tanga2023}. Even the
Rubin search that found \objA excluded the region within \fixed{10}\degsym of the
Galactic plane, and \objA lies only about \fixed{2}\degsym outside it
\citep{HolmanRubin}.

\subsection{Rubin's single-visit catalogs} \label{sec:deblend}

In early operations, Rubin's pipelines were configured and validated for typical star
densities; crowded fields near the Galactic plane were processed on a best-effort basis.
Because of this, as \objA crosses a dense star field at low Galactic latitude Rubin's single-visit
catalogs have no entry for it in \pnAAccImagedUncat of the \pnAAccImaged visits
whose images show it (Table~\ref{tab:visits}). In \pnADbNoCatalogue of them,
Daytime AP produced no catalog for the detector at all (the nightly-validation
catalog \nvs\ did, and Table~\ref{tab:visits} classifies those visits by it). In
the other \pnADbUncatalogued, the single-frame pipeline had merged the light of
neighboring stars into detection regions larger than its deblender accepts, and
dropped every source inside them: in \pnADbUncatInTooBig of the
\pnADbUncatalogued, \objA lies inside such a region, flagged
\code{deblend\_parentTooBig}.
In the other three, it is blended with a star or lies at a detector edge.

These measurements describe the catalogs of direct images that our search read,
as Daytime AP made them when these data were taken, along the track of \objA. In LSST
operations, once templates exist, alerts come from difference images \citep{LSE163}, where subtracting a template
removes the static stars that formed the blended regions in the direct images; the completeness of
difference-image catalogs for slow-moving objects near the plane has not yet been measured.
For searches that rely on direct-imaging catalogs, a bright object can be imaged
on many nights and still be missing from them near the Galactic plane, where
many of the remaining large objects probably are (Section~\ref{sec:howmany}).
Specialized pipelines for processing crowded-field direct images are in development.

\subsection{The DECam and ATLAS catalogs} \label{sec:nscgap}

DECam did image \objA in other programs, on \dcAstNNights nights from
\dcFirstAcc to \dcLastAcc (Section~\ref{sec:decam}), but the NSC DR2 catalog
lists it in only \dcNAccInNsc of our \dcNAccepted detections, on
\dcNAccInNscNights nights.
Those nights fall in different seasons, never more than \fixed{two} in one,
and on each of them the detections span only minutes, too short to show the
object's motion. A search that links detections across nights needs several
nights within one season, and a search for motion within a night needs a longer
sequence, so NSC DR2 provided neither. Of our other detections, \dcNAccNotNsc
were made before NSC DR2 ended in \fixed{2019}. In \dcNAccNotNscDqSeven of these,
the Community Pipeline, NOIRLab's standard DECam processing, flagged the pixels
of \objA in its data-quality mask with the value it gives to transients,
which may be why NSC DR2 has no measurement there; for the rest we do not
know the reason.

ATLAS also recorded \objA: ADAM returned \adamNAtlas
ATLAS detections on \adamNAtlasNights nights, all consistent with the adopted
orbit (Appendix~\ref{sec:decamvalA}). Because these are mostly single detections on
separate nights, and near ATLAS's detection limit, they were unlikely to improve orbit or physical property measurements and were not used in this work.

\section{How many more remain?} \label{sec:howmany}

As of October \fixed{2026}, only \beBrighterKnown known TNOs have lower absolute
magnitudes than \objA in the MPC catalog (whose $H$ values come from many sources and are only
approximately in the $V$ band): Eris, Pluto, Makemake, Haumea, Sedna, Gonggong, Orcus,
Quaoar and \fixed{(532037)} Chiminigagua. Chiminigagua's catalog $H$ (\fixed{3.09}) is only about
\fixed{0.1}\,mag below our $H_V$, and the MPC's own $H$ for \objA (\fixed{2.93}; MPEC~\fixed{2026-U85}) is below
Chiminigagua's, so whether \objA ranks ninth or tenth depends on which $H$ values are compared. The Rubin
search of these early data gives a
first estimate of how many objects with $H_V<\fixed{4}$ remain. Of the \beKnownHvFour
known TNOs with $H_V<\fixed{4}$, \beDiscoverable were \emph{discoverable}: the
visits searched held at least \fixed{ten} of their detections on at least
\fixed{six} nights within one search region and time window, enough for an entry
in the catalog of the search \citep{HolmanRubin}. The search recovered
\beDiscRecovered of these \beDiscoverable. The \fixed{two} it missed were close to
its limits in different ways: \fixed{(55565)} Aya reached exactly \fixed{ten} detections,
and vetting removed one of them; \fixed{(90482)} Orcus had more but was not linked, for a
reason we have not found (\fixed{(90568)} Goibniu, with as many detections, was recovered).
It also found one new object with $H_V<\fixed{4}$, \objA. If the objects not yet
discovered were as easy to find as the known ones, about \beBefore were
undiscovered before the search (\fixed{90}\% range \beBeforeLo--\beBeforeHi),
\objA among them, so about \beRemain remain. The number before the search is
\beKnownHvFour$\times$\beNew/\beRecovered, and its range is the \fixed{90}\%
Poisson interval on the single new object. The method does not assume that the
known objects are a complete list: the ratio of new to recovered objects measures
how incomplete it is.

This estimate counts only what the Rubin search could have found, and it is probably
an underestimate. First, still undiscovered objects are likely to be harder to find
than known ones: they are the objects earlier surveys missed, because they are fainter
in the sky, more distant, or in crowded fields, as \objA was, and the same traits make
them harder for Rubin. Second, the estimate applies only away from the Galactic plane.
The known objects were found mostly away from the plane, because the earlier searches
avoided it, and the Rubin search measured its efficiency only there. The band within
\fixed{10}\degsym of the plane covers \bePlaneArea square degrees,
\bePlanePct\% of the sky within \fixed{30}\degsym of the ecliptic, and neither the
Rubin search nor the all-sky search of \citet{Brown2015} covered it. Third, it misses
the more distant objects described below.

The result agrees with earlier estimates once the quantities compared are the same.
\citet{Schwamb2014} estimated that about \fixed{12} excited Kuiper belt objects are
brighter than $R\approx\fixed{19.5}$, of which \fixed{9} were known, so that the
inventory of objects that are bright in the sky is nearly complete, with perhaps one or
two left in the Galactic plane or the southern sky; \objA, near that limit, is such an
object. \citet{Sheppard2011} similarly found the Kuiper belt nearly complete to
$R\approx\fixed{21}$. Our estimate counts objects with $H_V<\fixed{4}$ to much fainter
apparent magnitudes. Most of those that remain are more distant and fainter in the sky
than \objA, so they are harder both to find and to study.

The known objects with $H_V<\fixed{4}$ were found by surveys that could reach them only
when they were near the Sun. At the depth of the earlier wide searches, about
$V\approx\fixed{21.7}$, an object with $H_V=\fixed{4}$ is visible only within about
\fixed{60}\au. Several of the largest known TNOs are on eccentric orbits that take them
much farther: Eris, Sedna and Gonggong are now at about \fixed{95}, \fixed{83} and
\fixed{90}\au, and they were found because their absolute magnitudes are lower still,
which makes them bright enough even at those distances. An object with
$H_V<\fixed{4}$ on a similar orbit, now beyond about \fixed{60}\au, was out of reach of
every earlier wide search. \citet{Sheppard2011} estimated in this way that about
$\fixed{35}\pm\fixed{15}$ scattered-disk objects with $H\lesssim\fixed{4.1}$ exist, most
of them too far from the Sun to have been seen (their Table~\fixed{4}). At the \fixed{50}\%
depth of the Rubin search \citep{HolmanRubin}, an object with $H_V=\fixed{4}$ can be
found out to about \fixed{90}\au. Our estimate does not count these distant objects;
they add to it, and LSST, which reaches this depth over the whole southern sky, should
find many of them.

Objects move into and out of the Galactic plane over about two decades. At \fixed{40}--\fixed{50}\au a TNO
moves about \fixed{1.0}--\fixed{1.4}\degsym a year along its orbit, and the ecliptic
crosses the band within \fixed{10}\degsym of the plane over about \fixed{23}\degsym of
longitude, so an object takes about two decades to cross it. \objA lay within
\fixed{10}\degsym of the plane from \fixed{2000} to \fixed{2023} and was found just
outside it, and the \fixed{L5} cloud of Neptune Trojans, hidden by the plane for more than
\fixed{15} years, has likewise moved into Rubin's view \citep{Schwamb2026}.
Near the plane, even where Rubin observes, the current catalogs drop every source
inside large blended regions (Section~\ref{sec:deblend}). Local PSF fits at the predicted positions recover those detections,
but they need an orbit to predict from, so they can extend the arc of a known object
but cannot discover new ones. A search near the plane therefore needs one of three things:
catalogs made with crowded-field settings, catalogs of
difference images, or a search of the images along trial orbits. With that, a search
of the plane through the first years of LSST should find most of the remaining
objects.

\section{Conclusions} \label{sec:conclusions}

\objA is a new TNO with $H_V=\pnAHV$ and a well-determined orbit at the edge of the
\fixed{12:7} resonance with Neptune, like Haumea; its color shows that it is not a
member of the Haumea family. It is a dwarf-planet candidate, pending an albedo
measurement.

\objA was missed by earlier searches because it lies far south and near the
Galactic plane, a region those searches mostly avoided (Section~\ref{sec:missed}). Rubin found
it, but its single-visit catalogs have no entry for it in \pnAAccImagedUncat of
the \pnAAccImaged visits whose images show it, mostly because the deblender isn't yet tuned for crowded fields and for performance reasons skips
processing of oversized blends. Local PSF fits at the predicted positions
do recover these detections once the orbit is known. If the
undiscovered objects were as easy to find as the known ones, about \beRemain
with $H_V<\fixed{4}$ would remain away from the Galactic plane
(Section~\ref{sec:howmany}). More probably they lie within \fixed{10}\degsym of the
plane, which a search through the first years of LSST can cover with image differencing or better tuned crowded-field region processing.

At $i\approx\pnAPkgMedi$, \objA is bright enough to be studied in detail. A
near-infrared spectrum would identify the ices on its surface, and lightcurves would
give its rotation period. Thermal measurements would give its size and albedo, and a
satellite, if one is found, its mass. It lies toward the Galactic bulge, so it will
often pass in front of stars, and occultations can measure its size and shape directly.

Later data can also test our results. If its albedo is like those of other large
TNOs, an occultation or thermal measurement should give \objA a diameter of
\pnADiamLo--\pnADiamMid\kmunit, and a near-infrared spectrum should not show the
deep water-ice bands of the Haumea family. At that size and temperature it should
also have lost its CH$_4$, N$_2$ and CO (Figure~\ref{fig:volatiles}), so finding any
of them on its surface would test the models of volatile retention. Away from the Galactic plane, about
\beRemain more objects with $H_V<\fixed{4}$ should be found; finding many more
would mean that the search found new objects less readily than it recovered known
ones. More should turn up within \fixed{10}\degsym of the plane.

\begin{acknowledgments}
M.J.H.\ and K.J.N.\ acknowledge support from the NSF (grant
No.\ \fixed{AST-2206194}) and the NASA YORPD Program (grant
No.\ \fixed{80NSSC22K0239}).  M.J.\ acknowledges support from the NSF (grant
No.\ \fixed{AST-2606453}). M.J., J.A.K., A.H., J.M., P.H.B., I.S., E.C.B., C.O.C., I.C.,
S.G., C.L.\ and D.S.\ acknowledge the support of the University
of Washington College of Arts and Sciences, Department of Astronomy, and the
DiRAC (Data-intensive Research in Astrophysics and Cosmology) Institute.
I.C. acknowledges the support of the Natural Sciences and Engineering Research Council of Canada (NSERC) and is partially funded by an NSERC Postgraduate Scholarship-Doctoral.
The work of S.G. is supported by NOIRLab, which is managed by the Association of Universities for Research in Astronomy (AURA) under a cooperative agreement with the U.S. National Science Foundation.
M.E.S.\ acknowledges support from UK Science and Technology Facilities Council (STFC) grant \fixed{ST/X001253/1}.
This material is based upon work supported in part by the National Science
Foundation through Cooperative Agreements \fixed{AST-1258333} and
\fixed{AST-2241526} and Cooperative Support Agreements \fixed{AST-1202910} and
\fixed{AST-2211468} managed by the Association of Universities for Research in
Astronomy (AURA), and the Department of Energy under Contract
No.\ \fixed{DE-AC02-76SF00515} with the SLAC National Accelerator Laboratory
managed by Stanford University. Additional Rubin Observatory funding comes from
private donations, grants to universities, and in-kind support from LSST-DA
Institutional Members.  This work uses data from the NSF--DOE Vera
C.\ Rubin Observatory, obtained during its early operations, and data and
services provided by the International Astronomical Union's Minor Planet
Center.  We thank the Rubin Observatory builders.
This publication is based in part on proprietary Rubin Observatory Legacy
Survey of Space and Time (LSST) data, and was prepared in accordance with the
Rubin Observatory data rights and access policies. All authors of this
publication meet the requirements for co-authorship of proprietary LSST data.
This research uses services or data provided by the Rubin Science Platform at
NSF--DOE Vera C.\ Rubin Observatory, which is jointly funded by the U.S.\ National
Science Foundation and the U.S.\ Department of Energy, Office of Science.
This paper makes use of LSST Science Pipelines software developed by the
Vera C.\ Rubin Observatory. We thank the Rubin Observatory for making their
code available as free software at \url{https://pipelines.lsst.io/}.
We thank the Asteroid Institute, a program of B612 Foundation, for the
precovery search on its ADAM platform (\url{https://b612.ai}; run
\adamJobId).
This research uses services or data provided by the Astro Data Lab, which is
part of the Community Science and Data Center (CSDC) Program of NSF NOIRLab,
and the NOIRLab Source Catalog.
This research draws upon DECam data as distributed by the Astro Data Archive at
NSF NOIRLab. NOIRLab is managed by the Association of Universities for Research
in Astronomy (AURA) under a cooperative agreement with the U.S.\ National Science
Foundation. This project used data obtained with the Dark Energy Camera (DECam),
which was constructed by the Dark Energy Survey (DES) collaboration. Funding for
the DES Projects has been provided by the US Department of Energy, the US
National Science Foundation, the Ministry of Science and Education of Spain, the
Science and Technology Facilities Council of the United Kingdom, the Higher
Education Funding Council for England, the National Center for Supercomputing
Applications at the University of Illinois at Urbana-Champaign, the Kavli
Institute for Cosmological Physics at the University of Chicago, Center for
Cosmology and Astro-Particle Physics at the Ohio State University, the Mitchell
Institute for Fundamental Physics and Astronomy at Texas A\&M University,
Financiadora de Estudos e Projetos, Funda\c{c}\~ao Carlos Chagas Filho de Amparo
\`a Pesquisa do Estado do Rio de Janeiro, Conselho Nacional de Desenvolvimento
Cient\'{\i}fico e Tecnol\'ogico and the Minist\'erio da Ci\^encia, Tecnologia e
Inova\c{c}\~ao, the Deutsche Forschungsgemeinschaft and the Collaborating
Institutions in the Dark Energy Survey. The Collaborating Institutions are
Argonne National Laboratory, the University of California at Santa Cruz, the
University of Cambridge, Centro de Investigaciones En\'ergeticas,
Medioambientales y Tecnol\'ogicas--Madrid, the University of Chicago, University
College London, the DES-Brazil Consortium, the University of Edinburgh, the
Eidgen\"ossische Technische Hochschule (ETH) Z\"urich, Fermi National
Accelerator Laboratory, the University of Illinois at Urbana-Champaign, the
Institut de Ci\`encies de l'Espai (IEEC/CSIC), the Institut de F\'{\i}sica d'Altes
Energies, Lawrence Berkeley National Laboratory, the Ludwig-Maximilians
Universit\"at M\"unchen and the associated Excellence Cluster Universe, the
University of Michigan, NSF NOIRLab, the University of Nottingham, the Ohio State
University, the OzDES Membership Consortium, the University of Pennsylvania, the
University of Portsmouth, SLAC National Accelerator Laboratory, Stanford
University, the University of Sussex, and Texas A\&M University.
This work is based in part on observations at NSF Cerro Tololo Inter-American
Observatory, a program of NSF NOIRLab (NOIRLab Prop. IDs \noirlabPrograms), which
is managed by the Association of Universities for Research in Astronomy (AURA)
under a cooperative agreement with the U.S. National Science Foundation.
This work has made use of data from the European Space Agency (ESA) mission
Gaia (\url{https://www.cosmos.esa.int/gaia}), processed by the Gaia Data
Processing and Analysis Consortium (DPAC); funding for the DPAC has been
provided by national institutions, in particular the institutions
participating in the Gaia Multilateral Agreement. We used the JPL Horizons
ephemeris service. This research has made use of SciX, a scientific literature search and discovery platform.

We used large language models (Claude, Anthropic; and Grok, SpaceXAI) throughout this work: to
write and run the analysis pipeline, including the archive queries, the
validation, the local PSF fits and the orbit fits; to
make the figures and tables; to check the numbers in the text against the
analysis; and to draft and edit the text of the manuscript from the authors' analysis and notes. The authors designed the
analysis and made the scientific decisions, reviewed and verified the code,
results and text, and take full responsibility for them. M.E.S.\ and I.C.\ did not make use of generative AI.
\end{acknowledgments}

\begin{contribution}

M.J.H.\ wrote the \texttt{pointdexter} linking code, building on an initial
version by K.J.N., built the stationary-source and transient catalogs it
searches, and ran the search that found \objA; and carried out the
resonance, Haumea-family and dynamical-classification analyses.
M.J.H.\ is responsible for the outer solar system search algorithms on the
Rubin Solar System Operations Team.
M.J.\ led the Rubin and DECam precovery and arc extension work; carried out
crowded-field astrometry and photometry, and the orbit fits and size
estimates; and built the reproducible pipeline. M.J.\ leads the Rubin Solar
System Operations Team and also led development of the Solar System
infrastructure this work relies on (the River catalog database, the MPC
submission system, and the ephemeris and known-object matching services).
K.J.N.\ wrote the initial version of \texttt{pointdexter}, developed the
density-based thinning of the transient catalogs, and checked the
discovery.
J.A.K.\ experimented with the transient-catalog production for low-coverage regions,
including the lower minimum numbers of detections and nights per source that
the production now uses generally.
A.H.\ vetted the TNO submissions, performed independent orbit fits, defined
the MPC deduplication rules for overlapping linkages, and leads \code{HelioLinC}
development.
J.M.\ made the ADAM precovery search possible and built the MPC submission
tracking software.
M.E.S.\ contributed to the interpretation of the results, including the
dwarf-planet size threshold and the expected population of large TNOs at
low Galactic latitude.
S.P.\ contributed the volatile-retention analysis in Section~\ref{sec:volatiles}, placing \objA on the
retention diagram of \citet{Protopapa2025}, and its interpretation.
P.H.B.\ contributed TNO-survey expertise, including the comparison with the
Dark Energy Survey search, and contributed to the MPC submissions.
I.S.\ and E.C.B.\ lead the Rubin Alert Production pipeline, which produces
some of the catalogs searched in this work and the images used for the local PSF fits.
E.N.\ contributed to the Solar System pipeline tasks in the LSST Science
Pipelines, including the Solar System tables and the \code{HelioLinC} inputs, and
contributed to the MPC submissions.
C.O.C., I.C., E.C., S.C., S.E., S.G., C.L., J.E.R., D.S.\ and D.Z.\ are
members of the Rubin Solar System Operations Team, contributed to its
commissioning-era work, and commented on the manuscript.
\end{contribution}

\facilities{Rubin:Simonyi (LSSTCam; \citealt{LSSTCam}), Rubin:USDAC, Blanco (DECam), Astro Data Lab, Astro Data Archive, Gaia}

\software{pointdexter (Holman \& Napier, in preparation), ASSIST \citep{Holman2023}, REBOUND \citep{Rein2012},
LSST Science Pipelines \citep{PSTN019}, Data Butler \citep{Jenness2022}, Astropy, NumPy, SciPy, Polars,
Matplotlib, SkyBoT \citep{Berthier2006}, SEP \citep{Barbary2016}, GaiaXPy \citep{RuzMieres2026}, ADAM \citep{Kiker2023}}

\section*{Data availability}
The astrometry of \objA, Rubin and DECam, is listed in
Table~\ref{tab:astrometry}, which is also available in machine-readable form, and provided in ADES format
(pipe-separated values).
The Rubin data we use are in the \objA data release on the Rubin Science
Platform (\url{https://data.lsst.cloud}), in \code{\fixed{/rubin/shared/datasets/2026-ED138}}: the
calibrated single-visit images, the \aps and \nvs catalogs, and the visit metadata,
as a Data Butler export \citep{Jenness2022} and tables. These are proprietary data,
available to Rubin data rights holders. Every Rubin number, table and figure in this
paper is produced from this release by one scripted pipeline; the other inputs come
from read-only queries of the MPC, NOIRLab and Gaia archives and of the ADAM service.

\clearpage
\appendix
\twocolumngrid  %
\section{Rubin data and checks} \label{sec:rubindetail}

\subsection{Products} \label{sec:products}

All Rubin data products utilized in this work come from the Rubin \objA data release. By searching this repository we find that the only exposures from \fixed{2025} that cover \objA are
\pnAPcNTwentyFiveAcq acquisition frames on the path of \objA, for which no
catalogs exist (Appendix~\ref{sec:precoverydetail}). During the period these images were taken, Rubin processed each exposure in several
ways \citep{Juric2017,LSE163}, and we use three of the products:
\begin{itemize}
  \item \emph{Prompt Processing} \citep{RubinAP,RubinPromptProducts} runs within minutes of each exposure. We
        use its calibrated single-visit images
        (\code{preliminary\_visit\_image}), made with the LSST Science
        Pipelines \citep{PSTN019}. Prompt Processing itself
        detects sources only above \fixed{20}$\sigma$; we use its images,
        not its sources.
  \item \emph{Daytime Alert Production} (Daytime AP; \citealt{RubinAP}) reprocesses each
        night's exposures during the following day. Its single-frame task,
        \code{singleFrameDetectAndMeasure}, measures sources to \fixed{5}$\sigma$ on each direct
        image (not on a difference image). We call these catalogs \aps. They
        give a position error for each source.
  \item \emph{Nightly validation}, a processing run that checked each
        night's data in early operations, produces a second source catalog for each visit
        (\code{single\_visit\_star}). We call it \nvs. It gives positions but
        no position errors.
\end{itemize}
We read the \aps and \nvs catalogs, and the images, from the \objA data release
(Data availability); the positions we use are in Table~\ref{tab:astrometry}.
Astrometry is calibrated to Gaia DR3 \citep{GaiaDR3}. The catalog error of each \aps
measurement plus \rmsQuadArcsec\arcsecsym in quadrature reproduces every
uncertainty in the delivered observations
(Section~\ref{sec:package}; all \verifyRmsWithErr rows that carry one). The files we
submit to the MPC, in the format of the Astrometry Data Exchange Standard (ADES), add
\submitQuadArcsec\arcsecsym instead, as Rubin's own submissions do, and give the
correlation between the RA and Dec errors; these are the uncertainties of
Table~\ref{tab:astrometry} and of our orbit fits. The files of the delivered and
cataloged data sets (Table~\ref{tab:orbits}) carry the \rmsQuadArcsec\arcsecsym term; their fits
replace it with \submitQuadArcsec\arcsecsym. Where a catalog provides no error,
the uncertainty columns are left empty.

\subsection{Weighting of the orbit fits} \label{sec:weights}

We weight the observations by their reported uncertainties as they are.
With them, the Rubin-only fit has $\chi^2=\pnAChisqFinalRep$ for
\pnADofFinalRep degrees of freedom. For the \pnAWtNRep observations that have a reported
uncertainty, the residual rms per axis is \pnAWtResRmsMas\mas, compared with a
median reported uncertainty of \pnAWtSigMedMas\mas; the rms is dominated by the few faint
observations with larger uncertainties, and the $\chi^2$, close to the number of degrees of
freedom, shows that the reported uncertainties describe the residuals. As a check, we repeat every
fit with a \floorArcsec\arcsecsym floor added in quadrature to the reported
uncertainties. The floor inflates them by a median factor of
\pnAWtFloorInflation and gives
$\chi^2=\pnAChisqFinalFl$; Table~\ref{tab:orbits}
lists both weightings. The adopted fit, to the Rubin and DECam
observations together, has $\chi^2=\pnAChisqPrecRep$ for \pnADofPrecRep
degrees of freedom (\pnAChisqPrecFl with the floor). A $\chi^2$ this far below
the number of degrees of freedom means that the DECam uncertainties, which
include a \dcAstSys\arcsecsym systematic term (Appendix~\ref{sec:decamastrometry}),
are conservative; we report them unchanged to the MPC.

One of the Rubin observations, from \nvs, has no reported uncertainty. In both weightings, the
fit weights it as if its uncertainty were
\emptySigArcsec\arcsecsym per axis. This weight is not an uncertainty estimate: its ADES uncertainty
columns stay empty,
and it is left out of the residual statistics above.

\subsection{Was it previously known?} \label{sec:known}

We queried a replica of the MPC observation database (\code{obs\_sbn}) for
any observation within \knownBoxRadiusArcmin\arcmin and
$\pm$\knownBoxWindowHours\,h of each reported detection, and found
none.

We then selected the \knownOrbitsSelected MPC orbits that satisfy
$q>\fixed{25}$\au, $a>\fixed{25}$\au or $H<\fixed{6}$, and propagated them with \code{ASSIST}
\citep{Holman2023} to \knownEpochsPerCandidate epochs on the arc. The
nearest TNO was \pnANearestTno ($H=\pnANearestTnoH$), at
\pnANearestTnoSepMin--\pnANearestTnoSepMax\degsym.
A cone search with \code{SkyBoT} \citep{Berthier2006}, a service that lists the known
small bodies in a field, of radius
\skybotRadiusArcmin\arcmin at the same epochs returned \pnASkybotObjects
objects, none farther than \pnASkybotMaxHelioDist\au from the Sun and none
closer than \pnASkybotMinSep\arcsecsym to \objA.

Finally, we looked for tracklets in the ITF within
\knownWideArcmin\arcmin of the predicted track outside the observed
arc, from \fixed{2025} to \knownWideUntil.
We found none.

\subsection{Are the detections real?} \label{sec:real}

Every reported row matches an \aps source in the same visit
(\verifyMatchedAPS rows). The worst
matches are within \verifyMaxSepMas\mas and \verifyMaxDtMs\,ms. None of the
matched sources has any of the pixel, point-spread-function (PSF) flux or centroid flags set.
Signal-to-noise ratios (S/N) are \pnASnrMin--\pnASnrMax.

Two rows needed attention:
\begin{enumerate}
  \item The \objA row of \parentObsTime is a deblend `parent' (the combined
        footprint of overlapping sources, before the deblender splits it) flagged as
        extended, because it includes an equally bright neighbor
        \pnAOvrRepNbSep\arcsecsym away. We do
        not use extended parents, so we first replaced it with its point-like
        `child', \parentChildSepMas\mas away. The local PSF fit
        (Section~\ref{sec:remeasure}), which models the neighbor, lies
        \pnAOvrRepFitParentMas\mas from the parent and \pnAOvrRepFitCatMas\mas
        from the child, and we use it instead of both.
  \item The \objA \blendBand-band row of \blendObsTime is blended with a
        static $\blendBand\approx\blendMag$ star \blendSep\arcsecsym away.
\end{enumerate}
We kept the blended row, which fits the orbit (Section~\ref{sec:orbits}).

Figure~\ref{fig:detections} shows \objA in single-visit images,
together with the same sky on another night. At fixed sky positions we
also measured aperture photometry in every covering visit:
\pnAFixedVisitsTotal visits at \pnAFixedPositions positions. Flux appears only when
the object is present, except at \pnAFixedNeighbourPositions of the
positions, where a known static neighbor lies within the aperture.

\subsection{Additional cataloged detections} \label{sec:precoverydetail}

We predicted the position of \objA, with its covariance, from a fit
to the delivered observations with a \floorArcsec\arcsecsym floor added
to their uncertainties (Section~\ref{sec:orbitmethod}); the floor gives
wider search windows. We did this for every LSSTCam exposure within
\pcSearchDeg\degsym of the track. There were \pcNExp such exposures
(Table~\ref{tab:visits}). Using detector outlines from the export of the Rubin consolidated
database provided in the release, we found that the predicted position fell on a detector in
\pnAAccOnDetector of \pnAAccExposures visits;
in the other \pnAAccOffDetector, it falls off every detector.
In each of these visits, we searched the \aps and \nvs catalogs within
$\max(\pcMinRadius\arcsecsym, 3\sigma)$ of the prediction, and accepted
sources within \pcMagTol\,mag of the expected magnitude in the visit's band:
the median of the delivered observations in that band, or, for a band with
none, the value of the neighboring band. Where \aps had a
deblended extended parent, the point-like child was preferred.

This search added \pnAPcNew cataloged detections on \pnAPcNewNights nights that were not among the
delivered observations.
Of these, \pcNNewNVS are \nvs detections without position errors. With
the new detections, the arc grows to \pnAArcCombined days.

We classified each other covering visit by why it has no detection
(Table~\ref{tab:visits}). Most of the remaining visits fall in
one of two classes:
\begin{itemize}
  \item \pnAPcIncompleteCatalog have an \emph{incomplete catalog}: fewer
        than half of the stars brighter than the object were recovered
        nearby;
  \item \pnAPcEdge fall at detector edges.
\end{itemize}
The incomplete catalogs are the problem described in Section~\ref{sec:deblend}.

A second pass with the refitted orbit changed no classification and found no
new detections.

There is no \fixed{2025} science coverage of the track. \pnAPcNTwentyFiveAcq
LSSTCam acquisition frames from \fixed{2025} cover \objA's predicted path,
but no catalogs exist for them.

\tabletypesize{\footnotesize}
\begin{deluxetable*}{llrrrr}
\tablecaption{Every LSSTCam exposure searched, by class\label{tab:visits}}
\tablehead{\colhead{Class} & \colhead{Meaning} & \multicolumn{4}{c}{2026 ED138}\\
 \colhead{} & \colhead{} & \colhead{$N$} & \colhead{Img.} & \colhead{Add.} & \colhead{Rej.}}
\startdata
Delivered & delivered observation & 10 & 10 & 0+1r & 0+1d \\
Detected & cataloged detection found by the search & 5 & 4 & 0 & 0 \\
\multicolumn{6}{l}{\textit{On a detector, no cataloged detection:}} \\
Incomplete catalog & catalog incomplete around the position & 18 & 14 & 12 & 2 \\
Edge & near a detector edge & 9 & 5 & 1 & 4 \\
Masked & hole in the catalog at the position & 3 & 3 & 3 & 0 \\
Blended & cataloged source of the wrong brightness & 2 & 2 & 0 & 2 \\
Shallow & image too shallow for the object & 0 & 0 & 0 & 0 \\
Missed & not cataloged; no other explanation & 0 & 0 & 0 & 0 \\
No catalog & neither catalog covers the position & 4 & 0 & 0 & 0 \\
\multicolumn{6}{l}{\textit{Off every detector:}} \\
Not covered & predicted position off every detector & 12 & 0 & 0 & 0 \\
\hline
Total &  & 63 & 38 & 16+1r & 8+1d \\
Final observations & delivered + detected + added $-$ dropped & 30 &  &  & 
\enddata
\tablecomments{$N$: exposures within the search radius of the predicted track (Section~\ref{sec:precovery}), by the class the catalog search assigns. Img.: a single-visit image exists at the predicted position. Add.: local PSF fits (Section~\ref{sec:remeasure}) that pass the cuts and add an observation; ``r'' marks fits that instead replace a cataloged row: an \nvs row without position errors, or a row overridden after inspection (Section~\ref{sec:remeasure}). Rej.: local fits that fail the $S/N$ or $\chi^2$ cut; ``d'' marks cataloged rows dropped after inspection. The validation set of the local fit (Section~\ref{sec:remeasure}) is the imaged reported and detected visits with an \aps\ detection whose fit also passes the $\chi^2$ cut. Final observations count rows of the final data set (Table~\ref{tab:astrometry}).}
\end{deluxetable*}

\subsection{Local PSF fits} \label{sec:remeasuredetail}

Rubin's catalogs miss \objA in most of the visits whose images show it
(Table~\ref{tab:visits}; Section~\ref{sec:deblend} explains why).
Instead of reprocessing whole detectors, we measured \objA directly
at its predicted positions with a local PSF fit. The predictions came
from a fit to the cataloged detections, with the same floor as in
Appendix~\ref{sec:precoverydetail}. We used the same Prompt Processing images, with the PSF models, world
coordinate systems (WCS) and photometric calibration that the pipeline stored with each image.

In each visit, we cut a \remStampPx-pixel (\remStampArcsec\arcsecsym)
stamp and fitted a model with these components:
\begin{itemize}
  \item a PSF at the object, with free position and flux;
  \item a PSF at every neighbor detected above \remNbSnr$\sigma$, with
        free flux, and also free position for neighbors brighter than
        \remNbFreeSnr$\sigma$;
  \item a background that is a plane: a constant plus a gradient in each
        direction across the stamp.
\end{itemize}
We weight the fit by each image's variance plane and mask saturated, bad,
cosmic-ray, suspect and edge pixels. Positions and their covariances are
converted to RA and Dec through the image's WCS. Magnitudes come from the
PSF flux, the image's aperture correction and its photometric calibration.
The method uses no catalog subtraction and no deblending step.

We validate this approach by first applying the fit
to the \pnAValN visits where \aps already has a clean detection.
The local and catalog positions differ by a median
of $\pnAValMedRaMas$ and $\pnAValMedDecMas$\mas in RA and Dec, with
an rms of \pnAValRmsMas\mas. This is consistent with the fit's own
errors of $\approx$\pnAValSigMas\mas. The magnitudes agree to
\pnAValMagRms\,mag rms.
There is no systematic offset between the image WCS and the catalogs, so
we do not inflate the fit errors; as for the \aps\ rows, we add
\submitQuadArcsec\arcsecsym in quadrature when we report them
(Appendix~\ref{sec:products}).

We keep a measurement if it has
$S/N\geq\remSnrMin$ and a reduced $\chi^2<\remChiMax$. The $\chi^2$ cut
rejects fits whose residuals show an unmodeled neighbor or blend
(Figure~\ref{fig:remeasure}).
The fit yields \pnARemNew new measurements, with
$S/N=\pnARemNewSnrMin$--\pnARemNewSnrMax. \pnARemDropChi visits
fail the $\chi^2$ cut and \pnARemDropSnr fails the $S/N$ cut.
The earliest new points, from \pnARemNewFirst, were never in any
catalog. Daytime AP produced no catalog for their detectors, and \nvs has
no source at the object's position. They lie within a
fraction of an arcsecond of the position predicted without them.

The complete Rubin data set has \pnAFinalNobs observations of \objA on
\pnAFinalNights nights, from \pnAFinalFirst to \pnAFinalLast. These are the
\pnAAccRep delivered, \pnAAccDet additional cataloged and \pnAAccAdded remeasured
observations, less the \pnARemOvrDropped dropped after inspection (Table~\ref{tab:visits}).

\subsection{Orbit fits} \label{sec:orbitdetail}

\begin{figure*}
\centering
\includegraphics[width=\textwidth]{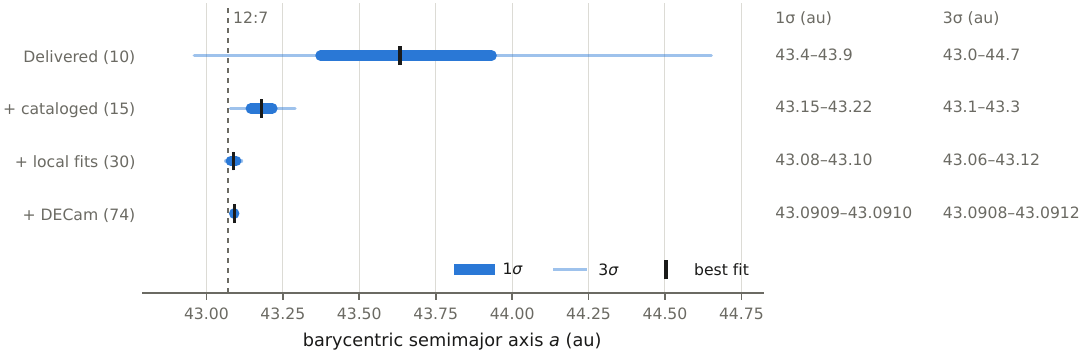}
\caption{Barycentric semimajor axis of \objA, with
its $1\sigma$ and $3\sigma$ profile ranges, from the delivered observations,
after adding cataloged detections, after adding the local PSF fits and with the DECam
precoveries (reported uncertainties).
The number of observations in each fit is given in parentheses. The dashed
line marks the nominal \fixed{12:7} resonance with Neptune.}
\label{fig:arange}
\end{figure*}

\begin{figure*}
\centering
\includegraphics[width=\textwidth]{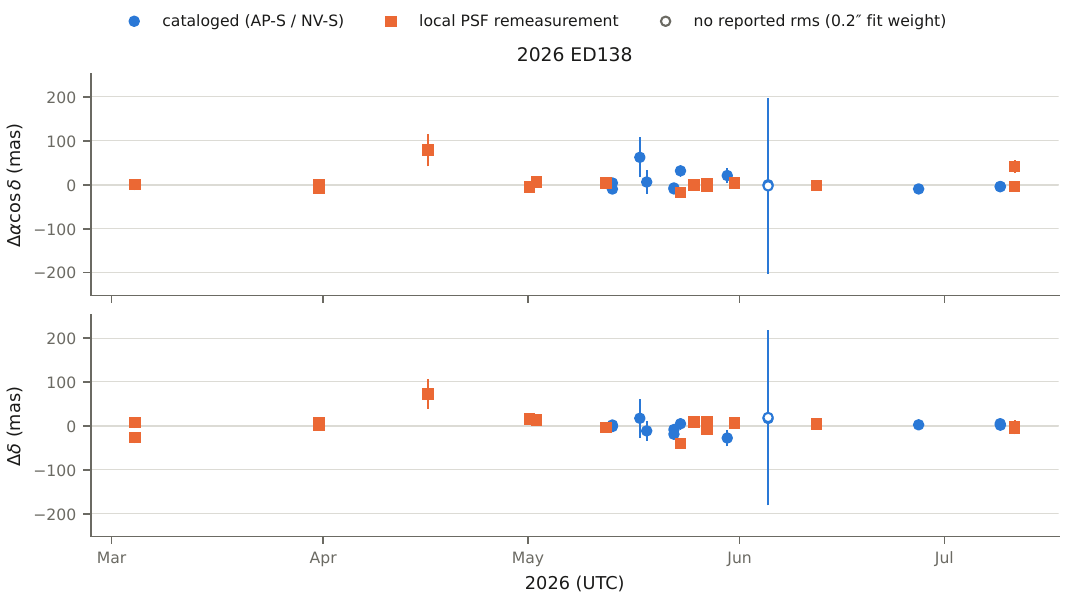}
\caption{Astrometric residuals of the Rubin observations from the Rubin-only fit (reported uncertainties; the DECam residuals are in
Table~\ref{tab:decamval}). Circles: cataloged detections (\aps or \nvs); squares:
local PSF fits; open circle: the row without a reported uncertainty,
weighted at \emptySigArcsec\arcsecsym.}
\label{fig:residuals}
\end{figure*}

Table~\ref{tab:orbits} and Figure~\ref{fig:arange} show how the orbit evolved from
the delivered observations to the complete Rubin set and with the DECam
precoveries; Figure~\ref{fig:residuals} shows the residuals of the Rubin
observations.

\tabletypesize{\scriptsize}
\setlength{\tabcolsep}{3pt}
\centerwidetable  %
\begin{deluxetable*}{llrrrrccrrr}
\tablecaption{Orbit fits for each observation set and weighting.\label{tab:orbits}}
\tablehead{
\colhead{Data} & \colhead{Weights} & \colhead{$N$} & \colhead{Arc} & \colhead{$\chi^2$/dof} &
\colhead{$a$} & \colhead{$a$ (1$\sigma$)} & \colhead{$a$ (3$\sigma$)} & \colhead{$e$} & \colhead{$q$} & \colhead{$i$} \\
\colhead{} & \colhead{} & \colhead{} & \colhead{(d)} & \colhead{} &
\colhead{(au)} & \colhead{(au)} & \colhead{(au)} & \colhead{} & \colhead{(au)} & \colhead{(deg)}}
\startdata
Delivered & reported & 10 & 35 & 17.8/14 & 43.63 & 43.37--43.93 & 43.0--44.7 & 0.167 & 36.34 & 24.416 \\
Delivered & floor & 10 & 35 & 0.5/14 & 43.51 & 42.55--46.54 & 42.5--64.3 & 0.159 & 36.61 & 24.419 \\
+ cataloged & reported & 15 & 69 & 21.7/24 & 43.18 & 43.15--43.22 & 43.1--43.3 & 0.142 & 37.06 & 24.427 \\
+ cataloged & floor & 15 & 69 & 0.7/24 & 43.16 & 42.91--43.48 & 42.6--44.3 & 0.141 & 37.10 & 24.427 \\
+ local fits & reported & 30 & 129 & 65.5/54 & 43.09 & 43.08--43.10 & 43.06--43.12 & 0.139 & 37.11 & 24.434 \\
+ local fits & floor & 30 & 129 & 2.2/54 & 43.09 & 43.02--43.17 & 42.9--43.3 & 0.139 & 37.11 & 24.434 \\
+ DECam & reported & 74 & 3789 & 84.7/142 & 43.090975 & 43.090915--43.091032 & 43.09080--43.09115 & 0.1387280 & 37.11305 & 24.4338512 \\
+ DECam & floor & 74 & 3789 & 6.2/142 & 43.09139 & 43.09066--43.09211 & 43.0892--43.0936 & 0.138729 & 37.113350 & 24.4338493 \\
\enddata
\tablecomments{Barycentric ecliptic J2000 osculating elements at each fit's mid-arc epoch. The epochs differ
between fits, but barycentric elements change little between them, unlike heliocentric ones, so the rows can be
compared directly. The adopted fits are in Table~\ref{tab:elements}. ``floor'': reported
astrometric uncertainties with $\floorArcsec''$ added in quadrature (robustness check); ``reported'': as reported
(adopted). Rows without a
reported uncertainty are weighted at $\emptySigArcsec''$ in both. Ranges are $\Delta\chi^2 = 1$ and $9$ profiles in the
radial velocity, with the other five parameters re-fitted. ``+ cataloged'' adds the additional
cataloged Rubin detections (Section~\ref{sec:precovery}); ``+ local fits'' adds the local PSF fits
(Section~\ref{sec:remeasure}); ``+ DECam'' adds the DECam precoveries (Section~\ref{sec:decam}).}
\end{deluxetable*}

\section{Astrometry} \label{sec:astrometry}

After inspection, two cataloged Rubin rows of \objA were overridden
(Appendix~\ref{sec:remeasuredetail}). The decisions
and their reasons are listed in \code{remeasure\_overrides.csv}.
\begin{itemize}
  \item \pnAOvrRepDate (\pnAOvrRepBand): the catalog row is the deblend
        child that replaced the delivered parent (Appendix~\ref{sec:real}). The local PSF fit models the equally
        bright neighbor \pnAOvrRepNbSep\arcsecsym away (flux ratio
        \pnAOvrRepNbRatio). The fitted position lies \pnAOvrRepFitParentMas\mas from the
        deblend parent and \pnAOvrRepFitCatMas\mas from the child. Its
        magnitude, \pnAOvrRepFitMag, matches the parent's
        \pnAOvrRepParentMag, not the child's \pnAOvrRepCatMag. The fit
        replaces the child.
  \item \pnAOvrDropDate (\pnAOvrDropBand): the catalog position lies
        \pnAOvrDropFitCatMas\mas from the local PSF fit. A static
        $\approx$\pnAOvrDropStaticMag\,mag star \pnAOvrDropStaticSep\arcsecsym
        away shifts the catalog position. The fit itself fails the $\chi^2$ cut
        (\pnAOvrDropChi), so there is no clean measurement and we drop the
        row. An exposure \pnAOvrDropTwinMin\,min apart on the same night
        is clean.
\end{itemize}

\section{DECam photometry and checks} \label{sec:photchecks}

\subsection{Rubin calibration}

Rubin's calibration is tied to The Monster, Rubin's reference catalog for
photometric calibration \citep{Ferguson2025}. Stars in the
Rubin visits of \objA agree with The Monster to within
\pcRubMonMaxAbs\,mag in every band (Table~\ref{tab:photchecks}). Against
Gaia XP synthetic photometry in the LSST bands, the same visits' stars differ
by \pcRubXpg\,mag (Rubin $-$ XP) in $g$ and agree to \pcRubXpMaxAbsRIZ\,mag in $riz$.
We treat the $g$ offset as a systematic uncertainty of Rubin's $g$ calibration
in this field and do not correct for it.

\subsection{DECam calibration} \label{sec:photdecamcal}

We measured \objA in all \dcPhotN DECam exposures by PSF photometry on difference
images. For each exposure we took \dcDiTplMin--\dcDiTplMax templates: the same
sky in the same band on other nights, when \objA was elsewhere. Each template is
resampled onto the exposure's pixels, and the sharper of the two images is
convolved with a kernel \citep{Alard1998}, Gaussians of three widths times
polynomials, fitted together with a linear background to all the pixels of a
\fixed{79}\arcsec\ box except those near \objA and saturated stars. \objA is
then fitted on the difference with the exposure's PSF (convolved with the
kernel when the exposure was the sharper image), so its flux is on the scale of
the stars measured in the exposure itself. The uncertainty is empirical: the
scatter of the same fit at \fixed{300} blank positions of the difference, which
includes the template's noise and the residuals of subtracted stars. The
templates are combined by inverse variance, and the combined uncertainty is the
scatter of the same combination at the same blank positions, because all the
differences share the exposure's noise. Copies of \objA injected into each exposure
\fixed{7}--\fixed{12}\arcsec\ from it are recovered with a median bias of
\dcDiFakeBiasMed\,mag (at most \dcDiFakeBiasMaxAbs\,mag in one exposure). We would
use the direct fit instead where no template is usable, or where the difference
fails a check (an uncertainty more than \fixed{3} times the direct fit's, or
injected copies recovered off by more than \fixed{0.05}\,mag plus three times
the uncertainty of their median); for \objA none does.

The direct PSF fit of Appendix~\ref{sec:decamastrometry}, with every cataloged
neighbor held at its cataloged position, differs from the difference-image
photometry by \dcDiMinusDirectSd\,mag (robust scatter) and by up to
\dcDiMinusDirectMaxAbs\,mag in single exposures. Injections elsewhere in the
image do not reveal this: it is specific to the position of \objA in each
exposure, where it is blended with uncataloged stars or with cataloged ones
close enough to share its light. The Gaia DR3 stars in each cutout are measured
with the direct fit. We tie each exposure to the Gaia DR3 XP synthetic DECam magnitudes
\citep{Montegriffo2023} of the \dcPhotCNStars (median) stars in the cutout that
have XP spectra (\dcPhotXpStars stars in all). \code{GaiaXPy} \citep{RuzMieres2026} computes the $grizY$
magnitudes; the broad DECam VR band, which it lacks, we integrated from the calibrated XP spectra
over the CTIO DECam VR throughput. The same integration reproduces \code{GaiaXPy}'s
$griz$ to \dcPhotVRValid\,mag. Per exposure, a robust linear fit of
$m_{\rm XP}-m_{\rm inst}$ against the stars' $g-i$, evaluated at the color of
\objA, gives the zero point. The stars scatter about the fit by
\dcPhotCZpScatter\,mag (median), and repeated exposures in the same band and
night agree to \dcPhotCRep\,mag. The XP spectra end at
\fixed{1020}\,nm, inside the $Y$ band, so we flag $Y$ and add
\dcPhotYExtraC\,mag, the robust exposure-to-exposure scatter of its difference
from the NSC-tied zero points, to its uncertainties. The robust scatter keeps one
exposure from dominating: in \dcPhotYNRed exposure the reference stars are
heavily reddened (median $g-i>\dcPhotYRedGi$), and there we use
\dcPhotYRedExtra\,mag, which includes that exposure's own difference. Table~\ref{tab:decamphot}
gives the results. Our ADES file uses these magnitudes.

\begin{deluxetable*}{lrrrrrrrrr}
\tablecaption{DECam photometry of \objA by band.\label{tab:decamphot}}
\tablehead{\colhead{Band} & \colhead{$N$} & \colhead{Nights} & \colhead{$m_{\rm XP}$} & \colhead{$m_{\rm NSC}$} &
\colhead{XP $-$ NSC} & \colhead{$H_{\rm XP}$} & \colhead{$H_{\rm XP}^{\rm LSST}$} & \colhead{$H_{\rm Rubin}$} &
\colhead{$\Delta H$}}
\startdata
g & 7 & 5 & 20.24 & 20.33 & $-$0.09 & 3.48 & 3.50 & 3.50 & +0.00 \\
r & 6 & 4 & 19.62 & 19.79 & $-$0.17 & 2.85 & 2.90 & 2.87 & +0.03 \\
i & 6 & 5 & 19.33 & 19.42 & $-$0.09 & 2.58 & 2.61 & 2.61 & +0.00 \\
z & 7 & 6 & 19.25 & 19.37 & $-$0.12 & 2.48 & 2.53 & 2.49 & +0.04 \\
Y\tablenotemark{a} & 6 & 5 & 19.14 & 19.20 & $-$0.08 & 2.38 & 2.40 & 2.40 & +0.00 \\
VR & 12 & 3 & 19.61 & 19.63 & $-$0.02 & 2.92 & \nodata & \nodata & \nodata
\enddata
\tablecomments{Medians over the exposures in each band. XP (adopted): zero points from Gaia DR3 XP synthetic DECam
magnitudes of stars in the same image, at the object's color; NSC: from their NSC DR2 magnitudes
(Section~\ref{sec:photdecam}). $H = m - 5\log_{10}(r\Delta) - \beta\alpha$, with the slope $\beta$ measured from the
Rubin data. $H_{\rm XP}^{\rm LSST}$: converted to the LSST bandpass at the object's color; $H_{\rm Rubin}$: the
Rubin value in that band (none for VR); $\Delta H = H_{\rm XP}^{\rm LSST} - H_{\rm Rubin}$.}
\tablenotetext{a}{The XP spectra end inside the $Y$ band; flagged, with a larger uncertainty.}
\end{deluxetable*}

We do not use the NSC DR2 mean magnitudes \citep{Nidever2021}, which are tied
to \fixed{ATLAS-Refcat2} \citep{Tonry2018} without color terms. In the
DECam fields, at Galactic latitudes \pcDecamGalBMin\ to \pcDecamGalBMax\degsym,
most reference stars are reddened.  The NSC zero points differ from the XP
ones by up to \dcPhotNscBiasMax\,mag (in $r$), and their stars scatter about
them by \dcPhotAZpScatter\,mag. Along the Rubin track of \objA ($b=\pcpnAGalB\degsym$),
NSC agrees with The Monster to \pcMonNscMaxAbs\,mag
(Table~\ref{tab:photchecks}).

\subsection{DECam versus Rubin} \label{sec:photcompare}

Table~\ref{tab:photbudget} follows the difference between the DECam and Rubin
reduced magnitudes of \objA. With NSC-tied DECam magnitudes at the observed
phases, DECam is \pcBudNscObsMin--\pcBudNscObsMax\,mag fainter. Reducing both
to zero phase removes part of that: the DECam phase angles,
\pcDecamAlphaMin--\pcDecamAlphaMax\degsym, are on average larger than the Rubin ones.
Tying DECam to XP and converting it to the LSST bandpasses at the object's
color removes most of the rest, and measuring \objA on difference images
removes the light of its neighbors. What remains is \pcBudXpg, \pcBudXpr,
\pcBudXpi, \pcBudXpz and \pcBudXpY\,mag in $g$, $r$, $i$, $z$ and $Y$; with the
direct PSF fit instead it would be \dcPhotDHCDirectg, \dcPhotDHCDirectr,
\dcPhotDHCDirecti, \dcPhotDHCDirectz and \dcPhotDHCDirectY\,mag. A
joint fit of the Rubin and XP-tied DECam magnitudes with one phase slope and
one DECam offset gives $\beta=\pcBetaJoint\pm\pcBetaJointErr$, consistent
with the Rubin value, and an offset of $\pcJointOffset\pm\pcJointOffsetErr$\,mag.

\begin{deluxetable}{lrrrrrr}
\tablecaption{DECam $-$ Rubin reduced magnitudes of \objA, step by step.\label{tab:photbudget}}
\tablehead{\colhead{Band} & \colhead{$N$} & \colhead{NSC, $\alpha_{\rm obs}$} & \colhead{Phase} &
\colhead{NSC, $\alpha=0$} & \colhead{Calib.} & \colhead{XP, $\alpha=0$}}
\startdata
g & 7 & +0.11 & $-$0.07 & +0.04 & $-$0.04 & +0.00 \\
r & 6 & +0.21 & $-$0.05 & +0.16 & $-$0.13 & +0.03 \\
i & 6 & +0.09 & $-$0.03 & +0.06 & $-$0.06 & +0.00 \\
z & 7 & +0.12 & +0.00 & +0.12 & $-$0.08 & +0.04 \\
Y\tablenotemark{a} & 6 & +0.04 & $-$0.02 & +0.03 & $-$0.03 & +0.00
\enddata
\tablecomments{Medians over the DECam exposures in each band, minus the Rubin value, in mag. NSC, $\alpha_{\rm obs}$:
NSC-tied, both at their observed phase angles; Phase: the change on reducing both to zero phase with the slope measured
from the Rubin data; Calib.: the change on tying DECam to Gaia XP and converting it to the LSST bandpass at the
object's color. The last column is the remaining difference (Section~\ref{sec:photcompare}).}
\tablenotetext{a}{Flagged (the XP spectra end inside the $Y$ band).}
\end{deluxetable}

Our direct PSF fit, applied to the Rubin images with zero points from the Rubin
catalog fluxes of the same stars, reproduces Rubin's magnitudes of \objA to
\pcSameMax\,mag in every band (Table~\ref{tab:photchecks}), so the two cameras'
photometry agree where crowding does not intervene. On two consecutive nights
\objA changed by \pcNighti\,mag in $i$ and \pcNightz\,mag in $z$, comparable to
the \dcPhotCRep\,mag repeatability within a night.

\subsection{Checks} \label{sec:photchecklist}

Table~\ref{tab:photchecks} collects the checks behind Section~\ref{sec:phot}.
\emph{The Monster $-$ NSC}: stars along the Rubin track of \objA with
$\fixed{0.6}<g-i<\fixed{1.2}$, with the \fixed{PS1}-to-ComCam color terms of \citet{Porter2025}
applied to NSC (whose southern calibration is \fixed{ATLAS-Refcat2}, itself tied to
\fixed{PS1}). \emph{Rubin $-$ Monster}: Rubin's calibrated single-visit fluxes of the
stars in the visits and detectors of \objA; the per-visit medians lie within
\pcRubMonVisitMaxAbs\,mag of zero in every band. \emph{Rubin $-$ XP}: the same
stars with XP spectra (\pcRubXpStars stars; up to \fixed{20} bright, well-measured stars per
visit), against their \code{GaiaXPy} LSST magnitudes. \emph{XP $-$ NSC} and
\emph{XP $-$ Monster}: the zero points of the DECam exposures from XP, from
NSC, and from The Monster (the last converted to the DECam system with the XP
LSST $-$ DECam relation), all evaluated at the color of \objA; The Monster
is limited to the brighter stars, and gives a usable fit in only a few
exposures. \emph{Same method}: \objA measured on \pcSameN Rubin
\texttt{preliminary\_visit\_image} cutouts with the DECam method (empirical PSF,
static neighbors, Gaia reference stars, zero point from the Rubin catalog
fluxes of the same stars), minus Rubin's magnitude.

\begin{deluxetable*}{lccccc}
\tablecaption{Checks of the photometric calibrations.\label{tab:photchecks}}
\tablehead{\colhead{Comparison} & \colhead{$g$} & \colhead{$r$} & \colhead{$i$} & \colhead{$z$} & \colhead{$Y$/$y$}}
\startdata
Monster $-$ NSC, Rubin fields & $-$0.030 (1072) & $-$0.030 (893) & $-$0.032 (796) & $-$0.005 (778) & \nodata \\
Rubin $-$ Monster, stars & $-$0.004 (2423) & $-$0.002 (5817) & +0.007 (6618) & +0.007 (12156) & \nodata \\
Rubin $-$ XP, stars & $-$0.020 (113) & +0.001 (237) & +0.009 (177) & +0.016 (289) & \nodata \\
XP $-$ NSC, DECam zero points & $-$0.087 (7) & $-$0.174 (6) & $-$0.092 (6) & $-$0.116 (7) & $-$0.082 (6) \\
XP $-$ Monster, DECam zero points & +0.035 (3) & +0.009 (3) & +0.010 (3) & $-$0.003 (4) & \nodata \\
Same method $-$ Rubin, \objA & +0.005 (3) & $-$0.001 (4) & $-$0.004 (6) & +0.005 (9) & \nodata
\enddata
\tablecomments{Median differences in mag; in parentheses the number of stars, or of exposures for the zero points
and the same-method test. Monster $-$ NSC includes the RTN-099 PS1-to-ComCam color terms. The DECam zero points are
evaluated at the color of \objA (Section~\ref{sec:photchecks}).}
\end{deluxetable*}

\section{DECam precovery} \label{sec:decamdetail}

\subsection{Candidates from ADAM} \label{sec:adam}

We first searched for archival detections with the precovery service of the
Asteroid Institute's ADAM platform \citep{Kiker2023}. Given an orbit, it predicts the object's
position, with its uncertainty, in every exposure of the surveys it indexes,
and returns the cataloged sources that lie within the uncertainty region
and have a plausible brightness. We gave it the Rubin-only orbit
(Section~\ref{sec:orbitmethod}), and it
searched from \adamSearchStart to \adamSearchEnd (run of \adamRunDate). It
examined \adamNFrames exposures and returned \adamNDet candidate detections:
\begin{itemize}
  \item \adamNDecam in \adamNDecamExp DECam exposures on \adamNDecamNights
        nights (\adamDecamFirst to \adamDecamLast), from the second data
        release of the NOIRLab Source Catalog \citep[NSC DR2;][]{Nidever2021};
  \item \adamNSkymapper from SkyMapper on \adamNSkymapperNights nights
        (\adamSkymapperFirst and \adamSkymapperLast);
  \item \adamNAtlas from ATLAS on \adamNAtlasNights nights
        (\adamAtlasFirst to \adamAtlasLast).
\end{itemize}
We followed up the DECam candidates. DECam \citep{Flaugher2015}, on the
\fixed{4}-m Blanco telescope at Cerro Tololo, reaches several magnitudes
deeper than SkyMapper and ATLAS, so it gives the highest signal-to-noise
ratio and the most precise positions. We compare the ATLAS and SkyMapper
candidates with our adopted orbit in Appendix~\ref{sec:decamvalA}, but do not use
them in it.

At these epochs the Rubin-only prediction is uncertain by
\adamDecamPredSigMin--\adamDecamPredSigMax\arcsecsym, and in this crowded
field such a region holds many cataloged sources. We therefore checked each
DECam candidate in the images. For each one we cut out its DECam CCD from the
NOIRLab Astro Data Lab \citep{Fitzpatrick2014,Nikutta2020}, measured
aperture photometry and the brightest peak at the candidate's position in
NSC DR2, and
measured the same aperture in an image of the same sky taken on another
night, when \objA was far away (at least \seedMinTplDays days earlier or later
for every candidate we accepted). We accepted a candidate as a moving point
source if $S/N\geq\seedMinSnr$, its peak lies within \seedMaxCen\arcsecsym of
that position, and the other night's image holds less than
\seedMaxTplRatio of its flux. \seedNVetted of the \seedNCand candidates pass.
Of the rest, \seedNStatic are static sources and \seedNRejectedOther show no
significant source at their NSC DR2 positions; all of these are among the
\adamBusiestNightN candidates of \adamBusiestNight.

We then added the nights one at a time to the Rubin orbit, starting from the
night closest in time to the Rubin data. On each night we kept the candidates
that agree with the current prediction (at $3\sigma$, with
\seedRms\arcsecsym position errors), refitted, and kept the night only if
every new point fitted the new orbit at $3\sigma$. All \seedNLinked
vetted candidates, on \seedNLinkedNights nights from \seedFirstNight on,
link.

\subsection{Search of the DECam archive} \label{sec:decamsearch}

The orbit from the ADAM detections predicts the position of \objA in the
DECam era to within \dcPredSigMax\arcsecsym, so we measured every image at the
predicted position instead of relying on catalogs. We listed every DECam
exposure in NSC DR2 (\dcNBoxNsc in a box around the track) and every DECam
image in the NOIRLab Astro Data Archive that NSC DR2 does not contain
(\dcNBoxArchive, mostly from after \fixed{2019}, when NSC DR2 ends). The
prediction falls within \dcFieldDeg\degsym of the pointing in \dcNField
exposures on \dcNFieldNights nights, from \dcFirstYear to \dcLastYear
(Table~\ref{tab:decam}). Using the CCD outlines, it falls on a CCD in
\dcNOnChip; the other \dcNNoChip fall in gaps between CCDs or off the array.

In each of the \dcNOnChip, we measured the image at the predicted position as in
Appendix~\ref{sec:adam}, with the data-quality mask of NOIRLab's DECam Community Pipeline
\citep{Valdes2014}, and a template from another night, at least \fixed{1}~day earlier or later (median
\fixed{271}~days), when \objA had moved well away. We accepted the detection if:
\begin{itemize}
  \item $S/N\geq\dcMinSnr$ in an aperture at the prediction;
  \item the brightest peak is within \dcMaxCen\arcsecsym of the prediction;
  \item the position is at least \dcMinEdge\arcsecsym from the CCD edge;
  \item the template holds less than \dcMaxTplRatio of the flux.
\end{itemize}
\objA is detected in \dcNAccepted exposures on \dcNAcceptedNights nights,
from \dcFirstAcc to \dcLastAcc, with $S/N=\dcSnrMin$--\dcSnrMax
(Figure~\ref{fig:decam}). Of the rejected exposures, \dcNRejEdge lie at a CCD
edge and \dcNRejOffsetPeak have their peak on a brighter star next to
\objA.

We then repeated the image check at the positions predicted by the orbit
fitted to these detections, as a check only (Appendix~\ref{sec:secondpass}).
For \objA, it found no new detections and lost none. Only \dcNAccInNsc of the \dcNAccepted detections are in
the NSC DR2 catalog (Section~\ref{sec:nscgap}).

\begin{deluxetable*}{lrrrrlcrrl}
\tablecaption{DECam exposures of \objA by night.\label{tab:decam}}
\tablehead{\colhead{Night (UTC)} & \colhead{Field} & \colhead{CCD} & \colhead{Acc.} & \colhead{Used} &
\colhead{Bands} & \colhead{$S/N$} & \colhead{NSC} & \colhead{rms} & \colhead{Not used} \\
\colhead{} & \colhead{} & \colhead{} & \colhead{} & \colhead{} & \colhead{} & \colhead{} & \colhead{} &
\colhead{(mas)} & \colhead{}}
\startdata
2016-02-25 & 4 & 4 & 4 & 4 & VR & 54--59 & 4 & 24 & -- \\
2016-08-11 & 3 & 3 & 3 & 3 & izY & 17--48 & 3 & 18 & -- \\
2016-08-16 & 9 & 3 & 0 & 0 & \nodata & \nodata & 0 & \nodata & offset peak \\
2016-08-24 & 4 & 4 & 4 & 4 & gr & 55--87 & 0 & 35 & -- \\
2016-08-25 & 4 & 2 & 2 & 1 & gr & 70--85 & 0 & 52 & seeing \\
2016-08-26 & 4 & 2 & 2 & 2 & gr & 71--86 & 0 & 42 & -- \\
2017-05-04 & 20 & 10 & 10 & 10 & grizY & 13--46 & 0 & 24 & -- \\
2017-05-05 & 10 & 4 & 0 & 0 & \nodata & \nodata & 0 & \nodata & edge \\
2017-07-02 & 4 & 4 & 4 & 4 & VR & 60--65 & 0 & 18 & -- \\
2017-07-03 & 4 & 4 & 4 & 4 & VR & 51--59 & 0 & 15 & -- \\
2018-05-10 & 2 & 0 & 0 & 0 & \nodata & \nodata & 0 & \nodata & -- \\
2018-05-12 & 4 & 2 & 2 & 2 & iz & 23--28 & 2 & 16 & -- \\
2018-05-19 & 2 & 2 & 2 & 2 & gr & 29--36 & 0 & 13 & -- \\
2018-05-21 & 4 & 0 & 0 & 0 & \nodata & \nodata & 0 & \nodata & -- \\
2019-04-28 & 2 & 1 & 1 & 1 & Y & 12--12 & 1 & 20 & -- \\
2019-04-30 & 3 & 3 & 3 & 3 & izY & 19--51 & 0 & 18 & -- \\
2019-05-01 & 3 & 3 & 3 & 3 & izY & 15--32 & 1 & 55 & -- \\
2023-04-07 & 2 & 1 & 1 & 1 & z & 34--34 & 0 & 34 & -- \\
2026-06-25 & 1 & 0 & 0 & 0 & \nodata & \nodata & 0 & \nodata & --
\enddata
\tablecomments{Field: exposures with the prediction within the DECam field; CCD: with the prediction on a CCD
(the others fall in chip gaps or off the array); Acc.: passing the image checks (Section~\ref{sec:decamsearch});
Used: in the orbit fit after the astrometric checks; NSC: accepted exposures in which NSC DR2 catalogs the
object; rms: residual of the used exposures per axis (reported uncertainties). Not used: reasons, from the
image checks (edge, offset peak, low S/N, static) and the astrometry (seeing, orbit outlier).}
\end{deluxetable*}

\subsection{Astrometry} \label{sec:decamastrometry}

We measured each detection on a \fixed{3}\arcmin cutout of its CCD. We do not
rely on the image's WCS beyond the cutout: we tie it locally to the
Gaia DR3 stars \citep{GaiaDR3} in the cutout, propagated to the exposure's
epoch with their proper motions and measured in the same way as \objA;
\dcAstNGaiaMin--\dcAstNGaiaMax stars per exposure are used. The tie is a constant offset; an affine
term never predicted the stars significantly better in a leave-one-out test.
\objA is measured by an empirical PSF, built from bright isolated stars in
the cutout and fitted together with every known static neighbor (Gaia DR3
sources, those without a proper motion taken as stationary, and NSC DR2 objects
seen on several nights). The neighbors stay at their cataloged positions and
only their fluxes are fitted; with free positions, overlapping sources trade
flux and the fit becomes degenerate.

In this field even a PSF fit is biased by stars too faint to be cataloged.
We measured this bias directly: we injected \objA, with its measured flux
and the template's PSF and noise, into two templates at its position, on a
$\dcAstInjGrid\times\dcAstInjGrid$ grid of sub-pixel offsets, and measured
it as in the data. Where the two templates agree to \dcAstAgree\mas
(\dcAstNBiasAgree exposures, and \dcAstNBiasSingle with one usable
template), we subtract the mean bias, at most \dcAstBiasMax\mas; where they
disagree (\dcAstNBiasNone), we leave the position uncorrected and add half
their difference to its uncertainty. The uncertainty of each position combines
the PSF fit, the tie, the bias correction and a \dcAstSys\arcsecsym
systematic term; its median is \dcAstErrMed\mas per axis. The times are the
exposure midpoints: the start time in the image header (DATE-OBS) plus half of the exposure time and of the
\valShutterTravel\,s shutter travel \citep{Flaugher2015,BandaHuarca2019}.

We kept \dcAstN of the \dcAstNIn detections, on \dcAstNNights nights. We
dropped the one with seeing worse than \dcAstMaxFwhm\arcsecsym. No detection failed
the other test, a fit with the Rubin data at $3\sigma$.

The positions are in
Table~\ref{tab:astrometry} and in ADES format in the accompanying repository.

\section{Checks of the DECam astrometry} \label{sec:decamchecks}

This appendix gives the details behind Section~\ref{sec:decamval}.

\subsection{The second pass} \label{sec:secondpass}

The second pass of the DECam search (Appendix~\ref{sec:decamsearch}) repeats
the image check at the positions predicted by the orbit fitted to the first
pass's detections. We use it only as a check, not as a cut: near the threshold,
measuring again at a better position lowers the $S/N$ of real faint detections.
A detection that the second pass loses would be reviewed, and a new one measured
independently (Appendix~\ref{sec:decamastrometry}).

\subsection{Validation} \label{sec:decamvalA}

Against Gaia DR3, the image WCS is offset by up
to \dcAstHdrOffMax\mas, with \dcAstHdrRmsMin--\dcAstHdrRmsMax\mas scatter
per star; our tie removes the offset. NSC DR2's own positions of the same
stars agree with Gaia to \dcAstNscOffMed\mas (median offset) and scatter about
them by \dcAstNscRmsMin--\dcAstNscRmsMax\mas per CCD (Table~\ref{tab:decamval}),
so NSC DR2's frame is sound; our tie uses Gaia directly.

The NSC DR2 catalog has \objA in \valNscNExp
exposures on \valNscNNights nights. On \seedFirstNight, NSC DR2 and our
positions differ by $(\valFirstNscMinusFinalRa, \valFirstNscMinusFinalDec)$\mas
in (RA, Dec). A \code{SExtractor}-equivalent windowed centroid
\citep{Bertin1996,Barbary2016}, the method behind NSC DR2, reproduces
NSC DR2's positions there to \valFirstNscMinusSepMax\mas. The injections show
that every method is pulled by the crowding at that position: a Gaussian
centroid by $(\valFirstGaussBiasRa, \valFirstGaussBiasDec)$\mas, the PSF fit by
$(\valFirstPsfBiasRa, \valFirstPsfBiasDec)$\mas and the windowed centroid by
$(\valFirstSepBiasRa, \valFirstSepBiasDec)$\mas, which accounts for the
difference. On \valNscWorstNight, NSC DR2's positions are
\valNscWorstMin--\valNscWorstMax\mas from ours, more than any bias the
injections show there; on the other nights they agree to
\valNscOtherMax\mas.

Our main test refits without each night and predicts it (Table~\ref{tab:decamvariants}). With our
adopted positions the left-out nights are predicted with an rms of
\valPsfCorrLooRa and \valPsfCorrLooDec\mas in RA and Dec; with NSC DR2's
positions, \valNscLooRa and \valNscLooDec\mas. For the earliest night,
which has the least support from the others, our positions are predicted to
$(\valLooFirstPsfCorrRa, \valLooFirstPsfCorrDec)$\mas and NSC DR2's to
$(\valLooFirstNscRa, \valLooFirstNscDec)$\mas, against a prediction
uncertainty of \valLooFirstPredSig\mas. All the variants give the same orbit
($a_{\rm bary}=\valABaryMin$--\valABaryMax\au).

\objA moves \valObjRateMin--\valObjRateMax\arcsecsym per
hour, so a timing error of one second moves it by at most
\valObjMasPerSecMax\mas. The shutter-open times in the headers agree with
DATE-OBS to between \valShutMin and \valShutMax\,s, except for the
\valFlagDate exposure, whose header is inconsistent (OPENSHUT
\valFlagOpenshut\,s from DATE-OBS, and a dark time of \valFlagDarktime\,s
for a \valFlagExptime\,s exposure); its timing cannot be checked. To test the
timing independently, we compared the NSC DR2 positions of
\valAstNRobust numbered asteroids (\code{SkyBoT}, $V<\valAstVmax$, on
\valAstNNights nights, in our exposures and in DECam exposures of the same
nights nearer the ecliptic) with JPL Horizons ephemerides at our midpoints.
A timing error would shift every asteroid along its track in proportion to
its rate of motion (here \valAstRateMin--\valAstRateMax\arcsecsym\,h$^{-1}$).
The median along-track residual is $\valAstAlong\pm\valAstAlongErr$\mas and
the cross-track one $\valAstCross\pm\valAstCrossErr$\mas; the implied clock
offset is $\valAstDt\pm\valAstDtErr$\,s, and a robust (Theil--Sen) slope of the
along-track residual against the rate, a less precise estimate, gives
\valAstTheilSen\,s. We find no timing error, and we take the larger of the two
as the bound: any error is below about \valAstDtBound\,s, a few milliarcseconds for \objA.

The object is bluer than the typical reddened star in
this field, so differential chromatic refraction displaces it relative to
the tie stars \citep{Filippenko1982}. We estimate at most \valDcrMax\mas
(\valDcrGMax\mas in $g$ at high airmass; median \valDcrMedian\mas), and do
not correct for it.

\valAtlasNCons of the \valAtlasN ATLAS
candidates lie within \valConsistentArcsec\arcsecsym of the adopted orbit
(median \valAtlasOcMed\arcsecsym, at most \valAtlasOcMax\arcsecsym),
consistent with their astrometric precision, and so does
\valSkymapperNCons of the \valSkymapperN SkyMapper candidates
(\valSkymapperOcMed\arcsecsym). The other \valSkymapperNIncons, on
\valSkymapperInconsDate at magnitude $\approx$\valSkymapperInconsMag, lie
\valSkymapperInconsOc\arcsecsym from the orbit, whose uncertainty there is
far below an arcsecond: they are a different object. The consistent candidates
are probably detections of \objA; we leave them for future work.

With the adopted orbit, the position of \objA
since \itfStartYear is predicted to better than \itfPredSigMax\arcsecsym.
None of the \itfNRows observations in the ITF (Section~\ref{sec:validation}) lies
within \itfMatchArcsec\arcsecsym of it; the closest is \itfNearestDeg\degsym
away.

Table~\ref{tab:decamval} lists, for each DECam exposure used, the checks above, and Table~\ref{tab:decamvariants} the orbit fits
with each centroiding method. The PSF positions were tied to Gaia DR3 with
the same stars as the Gaussian centroids; the windowed (\code{SExtractor}-like)
centroids, computed with \code{SEP} and with the pixels of other detected sources
masked, are used only for the comparison with NSC DR2.
\begin{deluxetable}{lrrrrrr}
\tablecaption{Orbit fits with each DECam centroiding variant.\label{tab:decamvariants}}
\tablehead{\colhead{Positions} & \colhead{$\chi^2$/dof} & \colhead{rms$_\alpha$} & \colhead{rms$_\delta$} &
\colhead{LOO$_\alpha$} & \colhead{LOO$_\delta$} & \colhead{$a_{\rm bary}$} \\
\colhead{} & \colhead{} & \colhead{(mas)} & \colhead{(mas)} & \colhead{(mas)} & \colhead{(mas)} & \colhead{(au)}}
\startdata
PSF + neighbors, bias-corrected (adopted) & 84.7/142 & 34 & 21 & 37 & 22 & 43.0910 \\
PSF + neighbors & 95.8/142 & 37 & 21 & 43 & 22 & 43.0910 \\
Gaussian, bias-corrected & 93.4/142 & 42 & 21 & 46 & 22 & 43.0910 \\
Gaussian & 127.7/142 & 57 & 26 & 64 & 27 & 43.0911 \\
NSC DR2 catalog where available & 138.9/142 & 118 & 46 & 125 & 48 & 43.0911
\enddata
\tablecomments{Each variant is fitted with the Rubin observations (reported uncertainties). rms: residuals of
the DECam rows in RA$\cos\delta$ and Dec. LOO: rms of the DECam rows of each night predicted by a fit without
that night.}
\end{deluxetable}

\clearpage  %
\startlongtable
\tabletypesize{\footnotesize}
\begin{deluxetable*}{lrrrrrrcll}
\tablecaption{Astrometry and photometry\label{tab:astrometry}}
\tablehead{\colhead{Time (UTC)} & \colhead{R.A.} & \colhead{Decl.} & \colhead{$\sigma_\alpha$} & \colhead{$\sigma_\delta$} & \colhead{mag} & \colhead{$\sigma_m$} & \colhead{Band} & \colhead{Source} & \colhead{Origin}\\ \colhead{} & \colhead{(deg)} & \colhead{(deg)} & \colhead{($''$)} & \colhead{($''$)} & \colhead{} & \colhead{} & \colhead{} & \colhead{} & \colhead{}}
\startdata
2016-02-25 07:49:48.7 & 260.3187519 & -44.7826230 & 0.0519 & 0.0518 & 19.751 & 0.0166 & VR & DECam & D \\
2016-02-25 07:51:31.7 & 260.3187745 & -44.7826351 & 0.0537 & 0.0530 & 19.769 & 0.0161 & VR & DECam & D \\
2016-02-25 07:53:14.8 & 260.3187903 & -44.7826461 & 0.0530 & 0.0525 & 19.740 & 0.0184 & VR & DECam & D \\
2016-02-25 07:54:59.0 & 260.3187980 & -44.7826574 & 0.0526 & 0.0522 & 19.730 & 0.0144 & VR & DECam & D \\
2016-08-11 02:37:59.6 & 257.8336367 & -44.9161799 & 0.0515 & 0.0516 & 19.386 & 0.0189 & i & DECam & D \\
2016-08-11 02:38:58.5 & 257.8336257 & -44.9161732 & 0.0521 & 0.0521 & 19.261 & 0.0268 & z & DECam & D \\
2016-08-11 02:39:57.7 & 257.8336280 & -44.9161708 & 0.0571 & 0.0546 & 19.217 & 0.1875 & Y & DECam & D \\
2016-08-24 03:50:21.4 & 257.7360115 & -44.8086511 & 0.0516 & 0.0515 & 20.272 & 0.0151 & g & DECam & D \\
2016-08-24 03:51:52.3 & 257.7359814 & -44.8086388 & 0.0516 & 0.0515 & 19.649 & 0.0153 & r & DECam & D \\
2016-08-24 03:58:56.8 & 257.7359532 & -44.8085991 & 0.0520 & 0.0519 & 19.674 & 0.0165 & r & DECam & D \\
2016-08-24 04:00:29.4 & 257.7359842 & -44.8085970 & 0.0523 & 0.0519 & 20.253 & 0.0163 & g & DECam & D \\
2016-08-25 04:28:49.8 & 257.7317654 & -44.8000678 & 0.0533 & 0.0524 & 20.236 & 0.0217 & g & DECam & D \\
2016-08-26 04:37:40.3 & 257.7280846 & -44.7916420 & 0.0514 & 0.0520 & 19.636 & 0.0147 & r & DECam & D \\
2016-08-26 04:39:11.4 & 257.7281019 & -44.7916334 & 0.0507 & 0.0575 & 20.281 & 0.0166 & g & DECam & D \\
2017-05-04 05:57:10.3 & 261.5597919 & -45.0599636 & 0.1078 & 0.1780 & 19.129 & 0.0546 & Y & DECam & D \\
2017-05-04 05:58:08.8 & 261.5597833 & -45.0599611 & 0.0703 & 0.1948 & 19.207 & 0.0219 & z & DECam & D \\
2017-05-04 05:59:07.4 & 261.5597653 & -45.0599684 & 0.0523 & 0.0678 & 19.293 & 0.0152 & i & DECam & D \\
2017-05-04 06:06:01.6 & 261.5596721 & -45.0599819 & 0.0522 & 0.0656 & 19.296 & 0.0149 & i & DECam & D \\
2017-05-04 06:06:59.6 & 261.5596548 & -45.0599803 & 0.0698 & 0.1984 & 19.189 & 0.0246 & z & DECam & D \\
2017-05-04 06:07:58.4 & 261.5596465 & -45.0599880 & 0.1032 & 0.1641 & 19.037 & 0.0511 & Y & DECam & D \\
2017-05-04 07:44:47.9 & 261.5583237 & -45.0602510 & 0.0516 & 0.0516 & 19.596 & 0.0201 & r & DECam & D \\
2017-05-04 07:46:18.8 & 261.5582925 & -45.0602640 & 0.0513 & 0.0513 & 20.213 & 0.0208 & g & DECam & D \\
2017-05-04 07:54:31.1 & 261.5581905 & -45.0602605 & 0.0512 & 0.0513 & 20.211 & 0.0173 & g & DECam & D \\
2017-05-04 07:56:02.4 & 261.5581644 & -45.0602680 & 0.0518 & 0.0514 & 19.572 & 0.0206 & r & DECam & D \\
2017-07-02 01:47:44.6 & 260.1349406 & -45.0432077 & 0.0512 & 0.0513 & 19.559 & 0.0107 & VR & DECam & D \\
2017-07-02 01:49:27.8 & 260.1349088 & -45.0432022 & 0.0512 & 0.0512 & 19.562 & 0.0105 & VR & DECam & D \\
2017-07-02 01:51:11.4 & 260.1348795 & -45.0431965 & 0.0512 & 0.0512 & 19.555 & 0.0103 & VR & DECam & D \\
2017-07-02 01:52:56.0 & 260.1348521 & -45.0431889 & 0.0513 & 0.0513 & 19.575 & 0.0124 & VR & DECam & D \\
2017-07-03 02:14:18.0 & 260.1104213 & -45.0387656 & 0.0511 & 0.0511 & 19.611 & 0.0099 & VR & DECam & D \\
2017-07-03 02:16:00.5 & 260.1103923 & -45.0387557 & 0.0511 & 0.0512 & 19.611 & 0.0088 & VR & DECam & D \\
2017-07-03 02:17:44.0 & 260.1103632 & -45.0387480 & 0.0511 & 0.0511 & 19.633 & 0.0092 & VR & DECam & D \\
2017-07-03 02:19:27.9 & 260.1103360 & -45.0387467 & 0.0512 & 0.0512 & 19.614 & 0.0110 & VR & DECam & D \\
2018-05-12 09:30:18.5 & 262.9473439 & -44.9080449 & 0.0520 & 0.0519 & 19.197 & 0.0280 & z & DECam & D \\
2018-05-12 09:31:18.1 & 262.9473351 & -44.9080449 & 0.0517 & 0.0516 & 19.322 & 0.0180 & i & DECam & D \\
2018-05-19 08:16:41.7 & 262.7943405 & -44.9255037 & 0.0515 & 0.0515 & 20.221 & 0.0157 & g & DECam & D \\
2018-05-19 08:18:13.3 & 262.7943040 & -44.9255042 & 0.0526 & 0.0517 & 19.551 & 0.0193 & r & DECam & D \\
2019-04-28 10:04:53.7 & 264.7398643 & -44.6549333 & 0.0710 & 0.0883 & 19.122 & 0.0766 & Y & DECam & D \\
2019-04-30 08:19:09.2 & 264.7089166 & -44.6639005 & 0.0564 & 0.0528 & 19.150 & 0.0576 & Y & DECam & D \\
2019-04-30 08:20:07.2 & 264.7088969 & -44.6639087 & 0.0778 & 0.0516 & 19.255 & 0.0215 & z & DECam & D \\
2019-04-30 08:21:06.6 & 264.7088858 & -44.6639064 & 0.0569 & 0.0528 & 19.342 & 0.0165 & i & DECam & D \\
2019-05-01 09:13:48.8 & 264.6916101 & -44.6685837 & 0.0627 & 0.0615 & 19.149 & 0.0590 & Y & DECam & D \\
2019-05-01 09:14:47.9 & 264.6916117 & -44.6685946 & 0.2069 & 0.1194 & 19.248 & 0.0348 & z & DECam & D \\
2019-05-01 09:15:46.6 & 264.6916290 & -44.6685956 & 0.1980 & 0.0851 & 19.361 & 0.0266 & i & DECam & D \\
2023-04-07 06:11:02.9 & 270.8309540 & -43.5573462 & 0.0592 & 0.0551 & 19.414 & 0.0416 & z & DECam & D \\
2026-03-04 09:33:14.7 & 274.7906291 & -42.4543647 & 0.0131 & 0.0130 & 19.319 & 0.0218 & y & local fit & R \\
2026-03-04 09:35:07.4 & 274.7906470 & -42.4543812 & 0.0127 & 0.0134 & 19.286 & 0.0221 & y & local fit & R \\
2026-03-31 08:28:06.4 & 275.0026189 & -42.6096994 & 0.0115 & 0.0115 & 19.414 & 0.0142 & z & local fit & R \\
2026-03-31 09:00:12.6 & 275.0026441 & -42.6098348 & 0.0108 & 0.0106 & 19.411 & 0.0134 & z & local fit & R \\
2026-04-16 07:56:43.5 & 274.9631508 & -42.7055538 & 0.0368 & 0.0336 & 21.941 & 0.0565 & u & local fit & R \\
2026-05-01 06:13:30.8 & 274.8202383 & -42.7885683 & 0.0105 & 0.0104 & 19.328 & 0.0127 & z & local fit & R \\
2026-05-02 06:03:12.7 & 274.8074060 & -42.7936882 & 0.0111 & 0.0112 & 19.350 & 0.0146 & z & local fit & R \\
2026-05-12 10:00:13.8 & 274.6542322 & -42.8423077 & 0.0104 & 0.0104 & 19.450 & 0.0119 & i & local fit & R \\
2026-05-13 08:44:09.8 & 274.6380746 & -42.8464499 & 0.0104 & 0.0105 & 19.285 & 0.0121 & z & AP-S & P \\
2026-05-13 08:44:49.6 & 274.6380717 & -42.8464509 & 0.0106 & 0.0104 & 19.281 & 0.0120 & z & AP-S & P \\
2026-05-17 09:15:23.4 & 274.5661101 & -42.8631783 & 0.0453 & 0.0446 & 21.920 & 0.0571 & u & AP-S & P \\
2026-05-18 09:19:25.0 & 274.5473351 & -42.8671390 & 0.0264 & 0.0221 & 21.774 & 0.0360 & u & AP-S & P \\
2026-05-22 08:59:35.7 & 274.4698391 & -42.8819353 & 0.0109 & 0.0107 & 20.265 & 0.0120 & g & AP-S & P \\
2026-05-22 09:00:22.4 & 274.4698273 & -42.8819401 & 0.0110 & 0.0109 & 20.230 & 0.0123 & g & AP-S & P \\
2026-05-23 07:29:07.2 & 274.4509893 & -42.8852067 & 0.0112 & 0.0113 & 19.580 & 0.0123 & r & local fit & R \\
2026-05-23 08:10:34.5 & 274.4504135 & -42.8852918 & 0.0139 & 0.0140 & 19.368 & 0.0145 & i & AP-S & P \\
2026-05-25 08:11:42.9 & 274.4093250 & -42.8919531 & 0.0108 & 0.0108 & 19.432 & 0.0123 & i & local fit & R \\
2026-05-25 08:16:21.4 & 274.4092572 & -42.8919625 & 0.0110 & 0.0110 & 19.677 & 0.0125 & r & local fit & R \\
2026-05-27 05:48:28.0 & 274.3694115 & -42.8979098 & 0.0112 & 0.0110 & 19.248 & 0.0135 & z & local fit & R \\
2026-05-27 05:49:20.7 & 274.3694008 & -42.8979164 & 0.0118 & 0.0119 & 19.259 & 0.0139 & z & local fit & R \\
2026-05-30 04:31:31.1 & 274.3058211 & -42.9064128 & 0.0177 & 0.0192 & 19.059 & 0.0330 & y & AP-S & C \\
2026-05-31 05:09:15.4 & 274.2831986 & -42.9091496 & 0.0128 & 0.0131 & 19.266 & 0.0186 & z & local fit & R \\
2026-06-05 04:09:17.7 & 274.1710311 & -42.9207871 & \nodata & \nodata & 19.236 & 0.0254 & y & NV-S & C \\
2026-06-05 04:09:59.8 & 274.1710207 & -42.9207888 & 0.0157 & 0.0154 & 19.234 & 0.0200 & y & AP-S & P \\
2026-06-12 07:03:52.1 & 274.0027490 & -42.9326467 & 0.0104 & 0.0105 & 19.373 & 0.0115 & i & local fit & R \\
2026-06-27 05:23:48.3 & 273.6353771 & -42.9376950 & 0.0103 & 0.0103 & 19.310 & 0.0114 & i & AP-S & C \\
2026-07-09 04:04:54.3 & 273.3429974 & -42.9219206 & 0.0103 & 0.0102 & 19.396 & 0.0113 & i & AP-S & C \\
2026-07-09 04:49:07.0 & 273.3422495 & -42.9218574 & 0.0104 & 0.0104 & 19.635 & 0.0124 & r & AP-S & C \\
2026-07-11 06:43:24.1 & 273.2926341 & -42.9173097 & 0.0110 & 0.0112 & 19.680 & 0.0120 & r & local fit & R \\
2026-07-11 07:21:14.5 & 273.2920220 & -42.9172443 & 0.0154 & 0.0165 & 20.358 & 0.0178 & g & local fit & R
\enddata
\tablecomments{Positions are ICRF (Gaia DR3 reference frame) at the exposure midpoint, observatory code X05 unless noted. Rubin rows are as submitted to the MPC: $\sigma_\alpha$, $\sigma_\delta$ are the measurement errors with $\submitQuadArcsec''$ added in quadrature, and $\sigma_m$ with \fixed{0.01}\,mag; NV-S catalog rows report no position errors (\nodata). Rubin magnitudes are PSF magnitudes in the LSST band shown. Origin: P, a delivered observation; C, a Rubin catalog detection added by the search of Section~\ref{sec:precovery}; R, the local PSF fit of Section~\ref{sec:remeasure}, which also replaces one cataloged row (the deblend child that had replaced a delivered parent); D, a DECam precovery (Section~\ref{sec:decam}; observatory code W84, Gaia DR3 frame, uncertainties including the $\dcAstSys''$ systematic term). DECam magnitudes are PSF fits on difference images, tied to Gaia XP synthetic photometry of stars in the same field (AB, DECam bandpasses; Appendix~\ref{sec:photdecamcal}); D rows are as submitted to the MPC.}
\end{deluxetable*}

\clearpage
\startlongtable
\tabletypesize{\scriptsize}
\setlength{\tabcolsep}{3pt}
\begin{deluxetable*}{llrcrrrrrclcc}
\tablecaption{Astrometric checks of each DECam exposure.\label{tab:decamval}}
\tablehead{\colhead{Time (UTC)} & \colhead{Band} & \colhead{$N_\star$} & \colhead{Header $-$ Gaia} &
\colhead{rms} & \colhead{NSC $-$ Gaia} & \colhead{rms} & \colhead{Ours} & \colhead{Shutter} &
\colhead{Bias} & \colhead{T} & \colhead{NSC $-$ ours} & \colhead{O$-$C} \\
\colhead{} & \colhead{} & \colhead{} & \colhead{(mas)} & \colhead{(mas)} & \colhead{(mas)} & \colhead{(mas)} &
\colhead{(mas)} & \colhead{(s)} & \colhead{(mas)} & \colhead{} & \colhead{(mas)} & \colhead{(mas)}}
\startdata
2016-02-25 07:49 & VR & 137 & $-$23, 75 & 14 & 12 & 32 & 18 & 0.04 & $-$66, 56 & a & $-$123, 102 & $-$39, $-$4 \\
2016-02-25 07:51 & VR & 126 & 14, 57 & 16 & 4 & 38 & 20 & 0.04 & $-$34, 34 & a & $-$144, 144 & $-$18, $-$14 \\
2016-02-25 07:53 & VR & 107 & 8, 57 & 13 & 10 & 33 & 18 & 0.04 & $-$35, 35 & a & $-$158, 113 & $-$15, $-$20 \\
2016-02-25 07:54 & VR & 138 & $-$4, 79 & 12 & 2 & 32 & 15 & 0.05 & $-$36, 35 & a & $-$137, 115 & $-$33, $-$27 \\
2016-08-11 02:37 & i & 93 & 49, 24 & 17 & 4 & 38 & 21 & 0.06 & 6, $-$1 & a & 145, 91 & $-$25, $-$1 \\
2016-08-11 02:38 & z & 94 & 76, 43 & 12 & 1 & 41 & 15 & 0.07 & $-$2, 5 & a & 238, 135 & $-$34, 3 \\
2016-08-11 02:39 & Y & 96 & 52, 30 & 8 & 11 & 40 & 13 & 0.26 & 27, $-$6 & a & 649, 162 & $-$9, $-$9 \\
2016-08-24 03:50 & g & 120 & 24, 35 & 29 & 32 & 47 & 43 & $-$0.54 & $-$7, 0 & s & \nodata & 57, 14 \\
2016-08-24 03:51 & r & 120 & 75, 52 & 15 & 23 & 39 & 21 & $-$0.53 & 18, $-$18 & s & \nodata & $-$8, 26 \\
2016-08-24 03:58 & r & 116 & 32, 44 & 23 & 21 & 41 & 29 & $-$0.54 & 16, $-$21 & s & \nodata & $-$23, 18 \\
2016-08-24 04:00 & g & 116 & 29, 57 & 47 & 48 & 61 & 58 & $-$0.53 & $-$14, 2 & s & \nodata & 69, $-$8 \\
2016-08-25 04:28 & g & 99 & 23, 43 & 49 & 61 & 58 & 69 & $-$0.63 & $-$7, 3 & a & \nodata & 69, $-$26 \\
2016-08-26 04:37 & r & 103 & 21, 69 & 15 & 6 & 31 & 20 & $-$0.66 & $-$4, $-$15 & a & \nodata & 20, $-$24 \\
2016-08-26 04:39 & g & 104 & 27, 53 & 26 & 33 & 41 & 34 & $-$0.66 & \nodata & d & \nodata & 74, $-$25 \\
2017-05-04 05:57 & Y & 84 & $-$1, 0 & 9 & 0 & 39 & 12 & 1.02 & \nodata & d & \nodata & 1, 12 \\
2017-05-04 05:58 & z & 81 & 6, $-$2 & 10 & 11 & 33 & 15 & 1.01 & \nodata & d & \nodata & 13, 30 \\
2017-05-04 05:59 & i & 80 & 5, $-$22 & 9 & 11 & 38 & 14 & 1.01 & \nodata & d & \nodata & 1, 13 \\
2017-05-04 06:06 & i & 90 & 4, $-$1 & 13 & \nodata & \nodata & 17 & 1.01 & \nodata & d & \nodata & 2, 29 \\
2017-05-04 06:06 & z & 90 & 19, 20 & 13 & \nodata & \nodata & 21 & 1.01 & \nodata & d & \nodata & $-$8, 44 \\
2017-05-04 06:07 & Y & 90 & 15, $-$17 & 13 & \nodata & \nodata & 19 & 1.01 & \nodata & d & \nodata & 5, 25 \\
2017-05-04 07:44 & r & 80 & 17, $-$18 & 13 & 6 & 26 & 19 & 1.01 & 8, 6 & s & \nodata & $-$7, $-$24 \\
2017-05-04 07:46 & g & 79 & 0, 6 & 11 & 5 & 23 & 15 & 1.01 & 19, 15 & s & \nodata & $-$34, $-$57 \\
2017-05-04 07:54 & g & 93 & 8, $-$22 & 17 & \nodata & \nodata & 22 & 1.00 & 7, 4 & a & \nodata & $-$9, 31 \\
2017-05-04 07:56 & r & 91 & $-$3, 7 & 12 & \nodata & \nodata & 18 & 1.01 & 11, 3 & a & \nodata & $-$22, 18 \\
2017-07-02 01:47 & VR & 133 & 36, $-$35 & 12 & 16 & 53 & 22 & 0.03 & $-$3, 5 & a & \nodata & $-$18, 7 \\
2017-07-02 01:49 & VR & 133 & 30, $-$2 & 14 & 12 & 47 & 22 & 0.03 & $-$1, 5 & a & \nodata & $-$25, 9 \\
2017-07-02 01:51 & VR & 133 & 11, $-$15 & 13 & 16 & 52 & 18 & 0.03 & $-$1, 6 & a & \nodata & $-$24, 11 \\
2017-07-02 01:52 & VR & 134 & 25, $-$23 & 13 & 12 & 42 & 22 & 0.03 & $-$1, 7 & a & \nodata & $-$18, 20 \\
2017-07-03 02:14 & VR & 134 & 35, 11 & 8 & 8 & 28 & 11 & 0.03 & $-$3, $-$2 & a & \nodata & $-$17, $-$3 \\
2017-07-03 02:16 & VR & 136 & 19, 5 & 9 & 11 & 30 & 12 & 0.03 & $-$4, $-$3 & a & \nodata & $-$17, 14 \\
2017-07-03 02:17 & VR & 134 & 24, 2 & 8 & 9 & 33 & 12 & 0.02 & $-$4, $-$2 & a & \nodata & $-$17, 23 \\
2017-07-03 02:19 & VR & 134 & 33, 1 & 13 & 8 & 34 & 18 & 0.03 & $-$5, $-$2 & a & \nodata & $-$11, 9 \\
2018-05-12 09:30 & z & 135 & $-$107, 4 & 13 & 6 & 41 & 18 & 0.04 & $-$2, 10 & a & $-$6, 66 & 14, $-$4 \\
2018-05-12 09:31 & i & 134 & $-$75, 20 & 18 & 4 & 37 & 24 & 0.06 & 0, 6 & a & 8, 26 & 29, 3 \\
2018-05-19 08:16 & g & 108 & 18, 9 & 15 & 6 & 36 & 22 & 0.06 & 0, 4 & a & \nodata & 23, $-$6 \\
2018-05-19 08:18 & r & 110 & 12, 27 & 15 & 15 & 35 & 21 & 0.06 & $-$11, $-$5 & a & \nodata & $-$7, 0 \\
2019-04-28 10:04 & Y & 142 & 14, 17 & 19 & 4 & 42 & 29 & $-$0.18 & \nodata & d & 30, 19 & $-$2, $-$28 \\
2019-04-30 08:19 & Y & 96 & $-$22, 13 & 11 & 18 & 30 & 17 & $-$0.18 & \nodata & d & \nodata & 34, 16 \\
2019-04-30 08:20 & z & 95 & $-$4, 21 & 9 & 4 & 27 & 12 & $-$0.17 & \nodata & d & \nodata & 12, $-$2 \\
2019-04-30 08:21 & i & 91 & $-$12, 35 & 10 & 15 & 31 & 13 & $-$0.09 & \nodata & d & \nodata & 13, 17 \\
2019-05-01 09:13 & Y & 98 & 1, 33 & 15 & 14 & 42 & 23 & $-$0.18 & 12, $-$19 & a & 83, $-$35 & 11, $-$3 \\
2019-05-01 09:14 & z & 87 & 43, $-$3 & 21 & 18 & 49 & 29 & $-$0.16 & \nodata & d & \nodata & 45, $-$31 \\
2019-05-01 09:15 & i & 89 & $-$23, 20 & 25 & 33 & 53 & 36 & $-$0.17 & \nodata & d & \nodata & 119, $-$24 \\
2023-04-07 06:11 & z & 278 & $-$28, $-$52 & 15 & \nodata & \nodata & 21 & $-$24.72 & \nodata & d & \nodata & 44, 21
\enddata
\tablecomments{$N_\star$: Gaia DR3 stars measured in the cutout. Header $-$ Gaia: the image WCS's offset from
Gaia at the object (RA$\cos\delta$, Dec) and the scatter per star. NSC $-$ Gaia: NSC DR2's positions of the
same stars (offset, scatter). Ours: leave-one-out scatter of our PSF tie. Shutter: OPENSHUT $-$ DATE-OBS.
Bias: the injection bias removed from the PSF position; T: the two templates agree (a), only one was
usable (s), or they disagree (d; no correction, larger uncertainty).
NSC $-$ ours: NSC DR2's catalog position of \objA minus ours. O$-$C: residual from the adopted fit.}
\end{deluxetable*}

\twocolumngrid  %

\bibliography{refs}{}

\begin{thebibliography}{}
\expandafter\ifx\csname natexlab\endcsname\relax\def\natexlab#1{#1}\fi
\providecommand{\url}[1]{\href{#1}{#1}}
\providecommand{\dodoi}[1]{doi:~\href{http://doi.org/#1}{\nolinkurl{#1}}}
\providecommand{\doeprint}[1]{\href{http://ascl.net/#1}{\nolinkurl{http://ascl.net/#1}}}
\providecommand{\doarXiv}[1]{\href{https://arxiv.org/abs/#1}{\nolinkurl{https://arxiv.org/abs/#1}}}

% type= article
\bibitem[{C. Alard \& R.~H. Lupton(1998)Alard \& Lupton}]{Alard1998}
Alard, C., \& Lupton, R.~H. 1998, \bibinfo{title}{A Method for Optimal Image
  Subtraction,} \apj, 503, 325, \dodoi{10.1086/305984}

% type= article
\bibitem[{A. {Alvarez-Candal} {et~al.}(2019){Alvarez-Candal}, {Ayala-Loera},
  {Gil-Hutton}, {Ortiz}, {Santos-Sanz}, \& {Duffard}}]{AlvarezCandal2019}
{Alvarez-Candal}, A., {Ayala-Loera}, C., {Gil-Hutton}, R., {et~al.} 2019,
  \bibinfo{title}{{Absolute colours and phase coefficients of trans-Neptunian
  objects: correlations and populations},} \mnras, 488, 3035

% type= article
\bibitem[{A. {Alvarez-Candal} {et~al.}(2016){Alvarez-Candal}, {Pinilla-Alonso},
  {Ortiz}, {Duffard}, {Morales}, {Santos-Sanz}, {Thirouin}, \&
  {Silva}}]{AlvarezCandal2016}
{Alvarez-Candal}, A., {Pinilla-Alonso}, N., {Ortiz}, J.~L., {et~al.} 2016,
  \bibinfo{title}{{Absolute magnitudes and phase coefficients of
  trans-Neptunian objects},} \aap, 586, A155,
  \dodoi{10.1051/0004-6361/201527161}

% type= article
\bibitem[{C. {Ayala-Loera} {et~al.}(2018){Ayala-Loera}, {Alvarez-Candal},
  {Ortiz}, {Duffard}, {Fern{\'a}ndez-Valenzuela}, {Santos-Sanz}, \&
  {Morales}}]{AyalaLoera2018}
{Ayala-Loera}, C., {Alvarez-Candal}, A., {Ortiz}, J.~L., {et~al.} 2018,
  \bibinfo{title}{{Absolute colours and phase coefficients of trans-Neptunian
  objects: H$_V$ - H$_R$ and relative phase coefficients},} \mnras, 481, 1848

% type= article
\bibitem[{M.~V. Banda-Huarca {et~al.}(2019)Banda-Huarca, Camargo, Desmars,
  {et~al.}}]{BandaHuarca2019}
Banda-Huarca, M.~V., Camargo, J. I.~B., Desmars, J., {et~al.} 2019,
  \bibinfo{title}{Astrometry and Occultation Predictions to Trans-{N}eptunian
  and {C}entaur Objects Observed within the {D}ark {E}nergy {S}urvey,} \aj,
  157, 120, \dodoi{10.3847/1538-3881/aafb37}

% type= article
\bibitem[{M.~T. {Bannister} {et~al.}(2016){Bannister}, {Alexandersen},
  {Benecchi}, {Chen}, {Delsanti}, {Fraser}, {Gladman}, {Granvik}, {Grundy},
  {Guilbert-Lepoutre}, {Gwyn}, {Ip}, {Jakubik}, {Jones}, {Kaib}, {Kavelaars},
  {Lacerda}, {Lawler}, {Lehner}, {Lin}, {Lykawka}, {Marsset}, {Murray-Clay},
  {Noll}, {Parker}, {Petit}, {Pike}, {Rousselot}, {Schwamb}, {Shankman},
  {Veres}, {Vernazza}, {Volk}, {Wang}, \& {Weryk}}]{Bannister2016}
{Bannister}, M.~T., {Alexandersen}, M., {Benecchi}, S.~D., {et~al.} 2016,
  \bibinfo{title}{{OSSOS. IV. Discovery of a Dwarf Planet Candidate in the 9:2
  Resonance with Neptune},} \aj, 152, 212, \dodoi{10.3847/0004-6256/152/6/212}

% type= article
\bibitem[{K. Barbary(2016)Barbary}]{Barbary2016}
Barbary, K. 2016, \bibinfo{title}{{SEP}: Source Extractor as a library,} The
  Journal of Open Source Software, 1, 58, \dodoi{10.21105/joss.00058}

% type= article
\bibitem[{P.~H. Bernardinelli {et~al.}(2022)Bernardinelli, Bernstein, Sako,
  {et~al.}}]{Bernardinelli2022}
Bernardinelli, P.~H., Bernstein, G.~M., Sako, M., {et~al.} 2022,
  \bibinfo{title}{A Search of the Full Six Years of the {D}ark {E}nergy
  {S}urvey for Outer Solar System Objects,} \apjs, 258, 41,
  \dodoi{10.3847/1538-4365/ac3914}

% type= article
\bibitem[{G. Bernstein \& B. Khushalani(2000)Bernstein \&
  Khushalani}]{Bernstein2000}
Bernstein, G., \& Khushalani, B. 2000, \bibinfo{title}{Orbit Fitting and
  Uncertainties for {K}uiper Belt Objects,} \aj, 120, 3323,
  \dodoi{10.1086/316868}

% type= inproceedings
\bibitem[{J. Berthier {et~al.}(2006)Berthier, Vachier, Thuillot,
  {et~al.}}]{Berthier2006}
Berthier, J., Vachier, F., Thuillot, W., {et~al.} 2006,
  \bibinfo{title}{{SkyBoT}, a new {VO} service to identify {S}olar {S}ystem
  objects,} in ASP Conf. Ser., Vol. 351, Astronomical Data Analysis Software
  and Systems XV, 367

% type= article
\bibitem[{E. Bertin \& S. Arnouts(1996)Bertin \& Arnouts}]{Bertin1996}
Bertin, E., \& Arnouts, S. 1996, \bibinfo{title}{{SExtractor}: Software for
  source extraction,} \aaps, 117, 393, \dodoi{10.1051/aas:1996164}

% type= article
\bibitem[{F. {Braga-Ribas} {et~al.}(2013){Braga-Ribas}, {Sicardy}, {Ortiz},
  {Lellouch}, {Tancredi}, {Lecacheux}, {Vieira-Martins}, {Camargo}, {Assafin},
  {Behrend}, {Vachier}, {Colas}, {Morales}, {Maury}, {Emilio}, {Amorim},
  {Unda-Sanzana}, {Roland}, {Bruzzone}, {Almeida}, {Rodrigues}, {Jacques},
  {Gil-Hutton}, {Vanzi}, {Milone}, {Schoenell}, {Salvo}, {Almenares}, {Jehin},
  {Manfroid}, {Sposetti}, {Tanga}, {Klotz}, {Frappa}, {Cacella}, {Colque},
  {Neves}, {Alvarez}, {Gillon}, {Pimentel}, {Giacchini}, {Roques}, {Widemann},
  {Magalh{\~a}es}, {Thirouin}, {Duffard}, {Leiva}, {Toledo}, {Capeche},
  {Beisker}, {Pollock}, {Cede{\~n}o Monta{\~n}a}, {Ivarsen}, {Reichart},
  {Haislip}, \& {Lacluyze}}]{BragaRibas2013}
{Braga-Ribas}, F., {Sicardy}, B., {Ortiz}, J.~L., {et~al.} 2013,
  \bibinfo{title}{{The Size, Shape, Albedo, Density, and Atmospheric Limit of
  Transneptunian Object (50000) Quaoar from Multi-chord Stellar Occultations},}
  \apj, 773, 26, \dodoi{10.1088/0004-637X/773/1/26}

% type= incollection
\bibitem[{M.~E. Brown(2008)Brown}]{Brown2008}
Brown, M.~E. 2008, \bibinfo{title}{The Largest {K}uiper Belt Objects,} in The
  Solar System Beyond {N}eptune, ed. M.~A. Barucci, H.~Boehnhardt, D.~P.
  Cruikshank, \& A.~Morbidelli (Tucson: University of Arizona Press), 335--344

% type= article
\bibitem[{M.~E. Brown {et~al.}(2007)Brown, Barkume, Ragozzine, \&
  Schaller}]{Brown2007Family}
Brown, M.~E., Barkume, K.~M., Ragozzine, D., \& Schaller, E.~L. 2007,
  \bibinfo{title}{A collisional family of icy objects in the {K}uiper belt,}
  \nat, 446, 294, \dodoi{10.1038/nature05619}

% type= article
\bibitem[{M.~E. Brown {et~al.}(2015)Brown, Bannister, Schmidt, Drake,
  Djorgovski, Graham, Mahabal, Donalek, Larson, Christensen, Beshore, \&
  McNaught}]{Brown2015}
Brown, M.~E., Bannister, M.~T., Schmidt, B.~P., {et~al.} 2015,
  \bibinfo{title}{A Serendipitous All Sky Survey for Bright Objects in the
  Outer Solar System,} \aj, 149, 69, \dodoi{10.1088/0004-6256/149/2/69}

% type= article
\bibitem[{M.~J. Brucker {et~al.}(2009)Brucker, Grundy, Stansberry, Spencer,
  Sheppard, Chiang, \& Buie}]{Brucker2009}
Brucker, M.~J., Grundy, W.~M., Stansberry, J.~A., {et~al.} 2009,
  \bibinfo{title}{High albedos of low inclination Classical {K}uiper belt
  objects,} \icarus, 201, 284, \dodoi{10.1016/j.icarus.2008.12.040}

% type= article
\bibitem[{S. {Cheng} {et~al.}(2026){Cheng}, {Li}, \& {Yang}}]{Cheng2026}
{Cheng}, S., {Li}, J., \& {Yang}, E. 2026, \bibinfo{title}{{Discovery of a
  Dwarf Planet Candidate in an Extremely Wide Orbit: 2017 OF201},} \apjl, 998,
  L6, \dodoi{10.3847/2041-8213/ae3a75}

% type= article
\bibitem[{J.~P. Emery {et~al.}(2024)Emery, Wong, Brunetto, Cook,
  Pinilla-Alonso, Stansberry, Holler, Grundy, Protopapa, Souza-Feliciano,
  Fern{\'a}ndez-Valenzuela, Lunine, \& Hines}]{Emery2024}
Emery, J.~P., Wong, I., Brunetto, R., {et~al.} 2024, \bibinfo{title}{A tale of
  3 dwarf planets: Ices and organics on {S}edna, {G}onggong, and {Q}uaoar from
  {JWST} spectroscopy,} \icarus, 414, 116017,
  \dodoi{10.1016/j.icarus.2024.116017}

% type= techreport
\bibitem[{P.~S. Ferguson {et~al.}(2025)Ferguson, Rykoff, Carlin, Saunders, \&
  Parejko}]{Ferguson2025}
Ferguson, P.~S., Rykoff, E.~S., Carlin, J.~L., Saunders, C., \& Parejko, J.~K.
  2025, The Monster: A reference catalog with synthetic ugrizy-band fluxes for
  the {Vera C. Rubin Observatory}, Data Management Technical Note DMTN-277,
  Vera C. Rubin Observatory.
\newblock \url{https://dmtn-277.lsst.io/}

% type= article
\bibitem[{A.~V. Filippenko(1982)Filippenko}]{Filippenko1982}
Filippenko, A.~V. 1982, \bibinfo{title}{The importance of atmospheric
  differential refraction in spectrophotometry,} \pasp, 94, 715,
  \dodoi{10.1086/131052}

% type= inproceedings
\bibitem[{M.~J. Fitzpatrick {et~al.}(2014)Fitzpatrick, Olsen, Economou, Stobie,
  {et~al.}}]{Fitzpatrick2014}
Fitzpatrick, M.~J., Olsen, K., Economou, F., Stobie, E.~B., {et~al.} 2014,
  \bibinfo{title}{The {NOAO} Data Laboratory: a conceptual overview,} in Proc.
  SPIE, Vol. 9149, Software and Cyberinfrastructure for Astronomy III, 91491T,
  \dodoi{10.1117/12.2057445}

% type= article
\bibitem[{B. Flaugher {et~al.}(2015)Flaugher, Diehl, Honscheid, Abbott,
  {et~al.}}]{Flaugher2015}
Flaugher, B., Diehl, H.~T., Honscheid, K., Abbott, T. M.~C., {et~al.} 2015,
  \bibinfo{title}{The {D}ark {E}nergy {C}amera,} \aj, 150, 150,
  \dodoi{10.1088/0004-6256/150/5/150}

% type= article
\bibitem[{S. {Fornasier} {et~al.}(2013){Fornasier}, {Lellouch}, {M{\"u}ller},
  {Santos-Sanz}, {Panuzzo}, {Kiss}, {Lim}, {Mommert}, {Bockel{\'e}e-Morvan},
  {Vilenius}, {Stansberry}, {Tozzi}, {Mottola}, {Delsanti}, {Crovisier},
  {Duffard}, {Henry}, {Lacerda}, {Barucci}, \& {Gicquel}}]{Fornasier2013}
{Fornasier}, S., {Lellouch}, E., {M{\"u}ller}, T., {et~al.} 2013,
  \bibinfo{title}{{TNOs are Cool: A survey of the trans-Neptunian region. VIII.
  Combined Herschel PACS and SPIRE observations of nine bright targets at
  70-500 {\ensuremath{\mu}}m},} \aap, 555, A15,
  \dodoi{10.1051/0004-6361/201321329}

% type= article
\bibitem[{N. Fray \& B. Schmitt(2009)Fray \& Schmitt}]{FraySchmitt2009}
Fray, N., \& Schmitt, B. 2009, \bibinfo{title}{Sublimation of ices of
  astrophysical interest: A bibliographic review,} Planet. Space Sci., 57,
  2053, \dodoi{10.1016/j.pss.2009.09.011}

% type= article
\bibitem[{ {Gaia Collaboration} {et~al.}(2023{\natexlab{a}}){Gaia
  Collaboration}, Montegriffo, Bellazzini, {et~al.}}]{Montegriffo2023}
{Gaia Collaboration}, Montegriffo, P., Bellazzini, M., {et~al.}
  2023{\natexlab{a}}, \bibinfo{title}{{Gaia} {D}ata {R}elease 3: The {G}alaxy
  in your preferred colours: Synthetic photometry from {Gaia} low-resolution
  spectra,} \aap, 674, A33, \dodoi{10.1051/0004-6361/202243709}

% type= article
\bibitem[{ {Gaia Collaboration} {et~al.}(2023{\natexlab{b}}){Gaia
  Collaboration}, Vallenari, {et~al.}}]{GaiaDR3}
{Gaia Collaboration}, Vallenari, A., {et~al.} 2023{\natexlab{b}},
  \bibinfo{title}{{Gaia} Data Release 3. Summary of the content and survey
  properties,} \aap, 674, A1, \dodoi{10.1051/0004-6361/202243940}

% type= article
\bibitem[{W.~M. {Grundy} {et~al.}(2019){Grundy}, {Noll}, {Buie}, {Benecchi},
  {Ragozzine}, \& {Roe}}]{Grundy2019}
{Grundy}, W.~M., {Noll}, K.~S., {Buie}, M.~W., {et~al.} 2019,
  \bibinfo{title}{{The mutual orbit, mass, and density of transneptunian binary
  G!k{\'u}n||'h{\`o}md{\'\i}m{\`a} (229762 2007 UK$_{126}$)},} \icarus, 334,
  30, \dodoi{10.1016/j.icarus.2018.12.037}

% type= article
\bibitem[{W.~M. Grundy {et~al.}(2024)Grundy, Tegler, Steckloff, Tan, Loeffler,
  Jasko, Koga, Blakley, Raposa, Engle, Thieberger, Hanley, Lindberg, Gomez, \&
  Madden-Watson}]{Grundy2024vp}
Grundy, W.~M., Tegler, S.~C., Steckloff, J.~K., {et~al.} 2024,
  \bibinfo{title}{Laboratory measurement of volatile ice vapor pressures with a
  quartz crystal microbalance,} \icarus, 410, 115767,
  \dodoi{10.1016/j.icarus.2023.115767}

% type= article
\bibitem[{M.~J. Holman {et~al.}(2023)Holman, Akmal, Farnocchia, Rein, Payne,
  Weryk, Tamayo, \& Hernandez}]{Holman2023}
Holman, M.~J., Akmal, A., Farnocchia, D., {et~al.} 2023,
  \bibinfo{title}{{ASSIST}: An Ephemeris-quality Test-particle Integrator,}
  \psj, 4, 69, \dodoi{10.3847/PSJ/acc9a9}

% type= misc
\bibitem[{M.~J. Holman {et~al.}(2026)Holman, Napier, {et~al.}}]{HolmanRubin}
Holman, M.~J., Napier, K.~J., {et~al.} 2026, First Outer Solar System Searches
  with Rubin, in preparation,

% type= article
\bibitem[{M.~J. {Holman} {et~al.}(2018){Holman}, {Payne}, {Fraser}, {Lacerda},
  {Bannister}, {Lackner}, {Chen}, {Lin}, {Smith}, {Kokotanekova}, {Young},
  {Chambers}, {Chastel}, {Denneau}, {Fitzsimmons}, {Flewelling}, {Grav},
  {Huber}, {Induni}, {Kudritzki}, {Krolewski}, {Jedicke}, {Kaiser}, {Lilly},
  {Magnier}, {Mark}, {Meech}, {Micheli}, {Murray}, {Parker}, {Protopapas},
  {Ragozzine}, {Veres}, {Wainscoat}, {Waters}, \& {Weryk}}]{Holman2018b}
{Holman}, M.~J., {Payne}, M.~J., {Fraser}, W., {et~al.} 2018,
  \bibinfo{title}{{A Dwarf Planet Class Object in the 21:5 Resonance with
  Neptune},} \apjl, 855, L6, \dodoi{10.3847/2041-8213/aaadb3}

% type= article
\bibitem[{{\v{Z}}. {Ivezi{\'c}} {et~al.}(2019){Ivezi{\'c}}, {Kahn}, {Tyson},
  {Abel}, {Acosta}, {Allsman}, {Alonso}, {AlSayyad}, {Anderson}, {Andrew},
  {Angel}, {Angeli}, {Ansari}, {Antilogus}, {Araujo}, {Armstrong}, {Arndt},
  {Astier}, {Aubourg}, {Auza}, {Axelrod}, {Bard}, {Barr}, {Barrau}, {Bartlett},
  {Bauer}, {Bauman}, {Baumont}, {Bechtol}, {Bechtol}, {Becker}, {Becla},
  {Beldica}, {Bellavia}, {Bianco}, {Biswas}, {Blanc}, {Blazek}, {Blandford},
  {Bloom}, {Bogart}, {Bond}, {Booth}, {Borgland}, {Borne}, {Bosch}, {Boutigny},
  {Brackett}, {Bradshaw}, {Brandt}, {Brown}, {Bullock}, {Burchat}, {Burke},
  {Cagnoli}, {Calabrese}, {Callahan}, {Callen}, {Carlin}, {Carlson},
  {Chandrasekharan}, {Charles-Emerson}, {Chesley}, {Cheu}, {Chiang}, {Chiang},
  {Chirino}, {Chow}, {Ciardi}, {Claver}, {Cohen-Tanugi}, {Cockrum}, {Coles},
  {Connolly}, {Cook}, {Cooray}, {Covey}, {Cribbs}, {Cui}, {Cutri}, {Daly},
  {Daniel}, {Daruich}, {Daubard}, {Daues}, {Dawson}, {Delgado}, {Dellapenna},
  {de Peyster}, {de Val-Borro}, {Digel}, {Doherty}, {Dubois},
  {Dubois-Felsmann}, {Durech}, {Economou}, {Eifler}, {Eracleous}, {Emmons},
  {Fausti Neto}, {Ferguson}, {Figueroa}, {Fisher-Levine}, {Focke}, {Foss},
  {Frank}, {Freemon}, {Gangler}, {Gawiser}, {Geary}, {Gee}, {Geha}, {Gessner},
  {Gibson}, {Gilmore}, {Glanzman}, {Glick}, {Goldina}, {Goldstein}, {Goodenow},
  {Graham}, {Gressler}, {Gris}, {Guy}, {Guyonnet}, {Haller}, {Harris},
  {Hascall}, {Haupt}, {Hernandez}, {Herrmann}, {Hileman}, {Hoblitt}, {Hodgson},
  {Hogan}, {Howard}, {Huang}, {Huffer}, {Ingraham}, {Innes}, {Jacoby}, {Jain},
  {Jammes}, {Jee}, {Jenness}, {Jernigan}, {Jevremovi{\'c}}, {Johns}, {Johnson},
  {Johnson}, {Jones}, {Juramy-Gilles}, {Juri{\'c}}, {Kalirai}, {Kallivayalil},
  {Kalmbach}, {Kantor}, {Karst}, {Kasliwal}, {Kelly}, {Kessler}, {Kinnison},
  {Kirkby}, {Knox}, {Kotov}, {Krabbendam}, {Krughoff}, {Kub{\'a}nek},
  {Kuczewski}, {Kulkarni}, {Ku}, {Kurita}, {Lage}, {Lambert}, {Lange},
  {Langton}, {Le Guillou}, {Levine}, {Liang}, {Lim}, {Lintott}, {Long},
  {Lopez}, {Lotz}, {Lupton}, {Lust}, {MacArthur}, {Mahabal}, {Mandelbaum},
  {Markiewicz}, {Marsh}, {Marshall}, {Marshall}, {May}, {McKercher}, {McQueen},
  {Meyers}, {Migliore}, {Miller}, \& {Mills}}]{Ivezic2019}
{Ivezi{\'c}}, {\v{Z}}., {Kahn}, S.~M., {Tyson}, J.~A., {et~al.} 2019,
  \bibinfo{title}{{LSST: From Science Drivers to Reference Design and
  Anticipated Data Products},} \apj, 873, 111, \dodoi{10.3847/1538-4357/ab042c}

% type= inproceedings
\bibitem[{T. Jenness {et~al.}(2022)Jenness, Bosch, Lust, Pease, Gower, Kowalik,
  Dubois-Felsmann, Mueller, \& Schellart}]{Jenness2022}
Jenness, T., Bosch, J.~F., Lust, N.~B., {et~al.} 2022, \bibinfo{title}{The
  {Vera C. Rubin Observatory} Data Butler and Pipeline Execution System,} in
  Proc. SPIE, Vol. 12189, Software and Cyberinfrastructure for Astronomy VII,
  1218911

% type= inproceedings
\bibitem[{M. Juri{\'c} {et~al.}(2017)Juri{\'c}, Kantor, Lim, Lupton,
  Dubois-Felsmann, Jenness, Axelrod, {et~al.}}]{Juric2017}
Juri{\'c}, M., Kantor, J., Lim, K.-T., {et~al.} 2017, \bibinfo{title}{The
  {LSST} Data Management System,} in ASP Conf. Ser., Vol. 512, Astronomical
  Data Analysis Software and Systems XXV, 279

% type= techreport
\bibitem[{M. Juri{\'c} {et~al.}(2023)Juri{\'c}, Axelrod, Becker, Becla, Bellm,
  Bosch, Ciardi, Connolly, Dubois-Felsmann, Economou, Freemon, Johnson, Gill,
  Graham, Guy, Ivezi{\'c}, Jenness, Kantor, Krughoff, Lim, Lupton, Mueller,
  Nidever, O'Mullane, Patterson, Petravick, Shaw, Slater, Strauss, Swinbank,
  Tyson, Wood-Vasey, \& Wu}]{LSE163}
Juri{\'c}, M., Axelrod, T.~S., Becker, A.~C., {et~al.} 2023, {Data Products
  Definition Document}, Systems Engineering Controlled Document LSE-163,
  NSF-DOE Vera C. Rubin Observatory, \dodoi{10.71929/rubin/2587118}

% type= article
\bibitem[{S.~C. {Keller} {et~al.}(2007){Keller}, {Schmidt}, {Bessell},
  {Conroy}, {Francis}, {Granlund}, {Kowald}, {Oates}, {Martin-Jones},
  {Preston}, {Tisserand}, {Vaccarella}, \& {Waterson}}]{Keller2007}
{Keller}, S.~C., {Schmidt}, B.~P., {Bessell}, M.~S., {et~al.} 2007,
  \bibinfo{title}{{The SkyMapper Telescope and The Southern Sky Survey},}
  \pasa, 24, 1, \dodoi{10.1071/AS07001}

% type= article
\bibitem[{T. {Khain} {et~al.}(2020){Khain}, {Becker}, {Lin}, {Gerdes}, {Adams},
  {Bernardinelli}, {et~al.}}]{Khain2020}
{Khain}, T., {Becker}, J.~C., {Lin}, H.~W., {et~al.} 2020,
  \bibinfo{title}{Dynamical Classification of Trans-Neptunian Objects Detected
  by the Dark Energy Survey,} \aj, 159, 133, \dodoi{10.3847/1538-3881/ab7002}

% type= article
\bibitem[{K. Kiker {et~al.}(2023)Kiker, Tellis, Moeyens, Juri{\'c}, Nelson,
  Stetzler, Posner, Grabowski, Lu, Remy, Barerra, Whitlow, Herpich, Yang,
  Davidge, Lingat, \& Naik}]{Kiker2023}
Kiker, K., Tellis, N., Moeyens, J., {et~al.} 2023, \bibinfo{title}{{The
  Asteroid Discovery, Analysis, and Mapping (ADAM) Precovery Service},}
  Bulletin of the American Astronomical Society, 55, 105.41.
\newblock \url{https://baas.aas.org/pub/2023n2i105p41}

% type= article
\bibitem[{J.~A. {Kurlander} {et~al.}(2025){Kurlander}, {Bernardinelli},
  {Schwamb}, {Juri{\'c}}, {et~al.}}]{Kurlander2025}
{Kurlander}, J.~A., {Bernardinelli}, P.~H., {Schwamb}, M.~E., {Juri{\'c}}, M.,
  {et~al.} 2025, \bibinfo{title}{{Predictions of the LSST Solar System Yield:
  Near-Earth Objects, Main Belt Asteroids, Jupiter Trojans, and Trans-Neptunian
  Objects},} \aj, 170, 99, \dodoi{10.3847/1538-3881/add685}

% type= article
\bibitem[{J.~A. {Larsen} {et~al.}(2001){Larsen}, {Gleason}, {Danzl}, {Descour},
  {McMillan}, {Gehrels}, {Jedicke}, {Montani}, \& {Scotti}}]{Larsen2001}
{Larsen}, J.~A., {Gleason}, A.~E., {Danzl}, N.~M., {et~al.} 2001,
  \bibinfo{title}{{The Spacewatch Wide-Area Survey for Bright Centaurs and
  Trans-Neptunian Objects},} \aj, 121, 562, \dodoi{10.1086/318022}

% type= article
\bibitem[{J.~A. {Larsen} {et~al.}(2007){Larsen}, {Roe}, {Albert}, {Descour},
  {McMillan}, {Gleason}, {Jedicke}, {Block}, {Bressi}, {Cochran}, {Gehrels},
  {Montani}, {Perry}, {Read}, {Scotti}, \& {Tubbiolo}}]{Larsen2007}
{Larsen}, J.~A., {Roe}, E.~S., {Albert}, C.~E., {et~al.} 2007,
  \bibinfo{title}{{The Search for Distant Objects in the Solar System Using
  Spacewatch},} \aj, 133, 1247, \dodoi{10.1086/511155}

% type= article
\bibitem[{E. {Lellouch} {et~al.}(2013){Lellouch}, {Santos-Sanz}, {Lacerda},
  {Mommert}, {Duffard}, {Ortiz}, {M{\"u}ller}, {Fornasier}, {Stansberry},
  {Kiss}, {Vilenius}, {Mueller}, {Peixinho}, {Moreno}, {Groussin}, {Delsanti},
  \& {Harris}}]{Lellouch2013}
{Lellouch}, E., {Santos-Sanz}, P., {Lacerda}, P., {et~al.} 2013,
  \bibinfo{title}{{``TNOs are Cool'': A survey of the trans-Neptunian region.
  IX. Thermal properties of Kuiper belt objects and Centaurs from combined
  Herschel and Spitzer observations},} \aap, 557, A60,
  \dodoi{10.1051/0004-6361/201322047}

% type= article
\bibitem[{D.~L. Nidever {et~al.}(2021)Nidever, Dey, Fasbender, Juneau,
  {et~al.}}]{Nidever2021}
Nidever, D.~L., Dey, A., Fasbender, K., Juneau, S., {et~al.} 2021,
  \bibinfo{title}{Second Data Release of the All-sky {NOIRLab} Source Catalog,}
  \aj, 161, 192, \dodoi{10.3847/1538-3881/abd6e1}

% type= article
\bibitem[{R. Nikutta {et~al.}(2020)Nikutta, Fitzpatrick, Scott, \&
  Weaver}]{Nikutta2020}
Nikutta, R., Fitzpatrick, M., Scott, A., \& Weaver, B.~A. 2020,
  \bibinfo{title}{Data {L}ab---A community science platform,} Astronomy and
  Computing, 33, 100411, \dodoi{10.1016/j.ascom.2020.100411}

% type= misc
\bibitem[{ {NSF-DOE Vera C. Rubin Observatory}(2026{\natexlab{a}}){NSF-DOE Vera
  C. Rubin Observatory}}]{RubinPromptProducts}
{NSF-DOE Vera C. Rubin Observatory}. 2026{\natexlab{a}}, {Legacy Survey of
  Space and Time Prompt Products}, \dodoi{10.71929/rubin/3020138}

% type= misc
\bibitem[{ {NSF-DOE Vera C. Rubin Observatory}(2026{\natexlab{b}}){NSF-DOE Vera
  C. Rubin Observatory}}]{RubinAP}
{NSF-DOE Vera C. Rubin Observatory}. 2026{\natexlab{b}}, {Alert Production with
  the Vera C. Rubin Observatory}, \dodoi{10.71929/rubin/3019817}

% type= article
\bibitem[{A. {P{\'a}l} {et~al.}(2016){P{\'a}l}, {Kiss}, {M{\"u}ller},
  {Moln{\'a}r}, {Szab{\'o}}, {Szab{\'o}}, {S{\'a}rneczky}, \& {Kiss}}]{Pal2016}
{P{\'a}l}, A., {Kiss}, C., {M{\"u}ller}, T.~G., {et~al.} 2016,
  \bibinfo{title}{{Large Size and Slow Rotation of the Trans-Neptunian Object
  (225088) 2007 OR$_{10}$ Discovered from Herschel and K2 Observations},} \aj,
  151, 117, \dodoi{10.3847/0004-6256/151/5/117}

% type= article
\bibitem[{R.~S. Park {et~al.}(2021)Park, Folkner, Williams, \&
  Boggs}]{Park2021}
Park, R.~S., Folkner, W.~M., Williams, J.~G., \& Boggs, D.~H. 2021,
  \bibinfo{title}{The {JPL} Planetary and Lunar Ephemerides {DE440} and
  {DE441},} \aj, 161, 105, \dodoi{10.3847/1538-3881/abd414}

% type= techreport
\bibitem[{M.~N. Porter {et~al.}(2025)Porter, Tucker, Smith, \&
  Adair}]{Porter2025}
Porter, M.~N., Tucker, D.~L., Smith, J.~A., \& Adair, C.~L. 2025, Photometric
  Transformation Relations for the {LSST} Data Preview 1, Rubin Technical Note
  RTN-099, Vera C. Rubin Observatory, \dodoi{10.71929/rubin/3006074}

% type= article
\bibitem[{P. Pravec \& A.~W. Harris(2007)Pravec \& Harris}]{Pravec2007}
Pravec, P., \& Harris, A.~W. 2007, \bibinfo{title}{Binary asteroid population.
  1. Angular momentum content,} \icarus, 190, 250,
  \dodoi{10.1016/j.icarus.2007.02.023}

% type= article
\bibitem[{S. Protopapa {et~al.}(2025)Protopapa, Wong, Lellouch, Johnson,
  Grundy, Glein, M{\"u}ller, Kiss, Emery, Brunetto, Holler, Parker, Stansberry,
  Hammel, Milam, Guilbert-Lepoutre, Santos-Sanz, \&
  Pinilla-Alonso}]{Protopapa2025}
Protopapa, S., Wong, I., Lellouch, E., {et~al.} 2025, \bibinfo{title}{{JWST}
  Detection of Hydrocarbon Ices and Methane Gas on {M}akemake,} \apjl, 991,
  L34, \dodoi{10.3847/2041-8213/adfe63}

% type= article
\bibitem[{B. {Proudfoot} \& D. {Ragozzine}(2022){Proudfoot} \&
  {Ragozzine}}]{Proudfoot2022}
{Proudfoot}, B., \& {Ragozzine}, D. 2022, \bibinfo{title}{{The formation of
  Haumea and its family via binary merging},} Nature Communications, 13, 2262,
  \dodoi{10.1038/s41467-022-29901-5}

% type= article
\bibitem[{B.~C.~N. {Proudfoot} \& D. {Ragozzine}(2019){Proudfoot} \&
  {Ragozzine}}]{Proudfoot2019}
{Proudfoot}, B. C.~N., \& {Ragozzine}, D. 2019, \bibinfo{title}{{Modeling the
  Formation of the Family of the Dwarf Planet Haumea},} \aj, 157, 230,
  \dodoi{10.3847/1538-3881/ab19c4}

% type= article
\bibitem[{D. {Rabinowitz} {et~al.}(2012){Rabinowitz}, {Schwamb}, {Hadjiyska},
  \& {Tourtellotte}}]{Rabinowitz2012}
{Rabinowitz}, D., {Schwamb}, M.~E., {Hadjiyska}, E., \& {Tourtellotte}, S.
  2012, \bibinfo{title}{{The La Silla-QUEST Kuiper Belt Survey},} \aj, 144,
  140, \dodoi{10.1088/0004-6256/144/5/140}

% type= article
\bibitem[{D.~L. Rabinowitz {et~al.}(2006)Rabinowitz, Barkume, Brown, Roe,
  Schwartz, Tourtellotte, \& Trujillo}]{Rabinowitz2006}
Rabinowitz, D.~L., Barkume, K., Brown, M.~E., {et~al.} 2006,
  \bibinfo{title}{Photometric Observations Constraining the Size, Shape, and
  Albedo of 2003 {EL}61, a Rapidly Rotating, {P}luto-sized Object in the
  {K}uiper Belt,} \apj, 639, 1238, \dodoi{10.1086/499575}

% type= article
\bibitem[{D.~L. {Rabinowitz} {et~al.}(2007){Rabinowitz}, {Schaefer}, \&
  {Tourtellotte}}]{Rabinowitz2007}
{Rabinowitz}, D.~L., {Schaefer}, B.~E., \& {Tourtellotte}, S.~W. 2007,
  \bibinfo{title}{{The Diverse Solar Phase Curves of Distant Icy Bodies. I.
  Photometric Observations of 18 Trans-Neptunian Objects, 7 Centaurs, and
  Nereid},} \aj, 133, 26, \dodoi{10.1086/508931}

% type= article
\bibitem[{D. Ragozzine \& M.~E. Brown(2007)Ragozzine \& Brown}]{Ragozzine2007}
Ragozzine, D., \& Brown, M.~E. 2007, \bibinfo{title}{Candidate Members and Age
  Estimate of the Family of {K}uiper Belt Object 2003 {EL}61,} \aj, 134, 2160,
  \dodoi{10.1086/522334}

% type= article
\bibitem[{H. Rein \& S.-F. Liu(2012)Rein \& Liu}]{Rein2012}
Rein, H., \& Liu, S.-F. 2012, \bibinfo{title}{{REBOUND}: an open-source
  multi-purpose {N}-body code for collisional dynamics,} \aap, 537, A128,
  \dodoi{10.1051/0004-6361/201118085}

% type= article
\bibitem[{H. Rein \& D.~S. Spiegel(2015)Rein \& Spiegel}]{Rein2015}
Rein, H., \& Spiegel, D.~S. 2015, \bibinfo{title}{{IAS15}: a fast, adaptive,
  high-order integrator for gravitational dynamics, accurate to machine
  precision over a billion orbits,} \mnras, 446, 1424,
  \dodoi{10.1093/mnras/stu2164}

% type= article
\bibitem[{F.~L. {Rommel} {et~al.}(2023){Rommel}, {Braga-Ribas}, {Ortiz},
  {Sicardy}, {Santos-Sanz}, {Desmars}, {Camargo}, {Vieira-Martins}, {Assafin},
  {Morgado}, {Boufleur}, {Benedetti-Rossi}, {Gomes-J{\'u}nior},
  {Fern{\'a}ndez-Valenzuela}, {Holler}, {Souami}, {Duffard}, {Margoti},
  {Vara-Lubiano}, {Lecacheux}, {Plouvier}, {Morales}, {Maury}, {Fabrega},
  {Ceravolo}, {Jehin}, {Albanese}, {Mariey}, {Cikota}, {Ru{\v{z}}djak},
  {Cikota}, {Szak{\'a}ts}, {Baba Aissa}, {Gringahcene}, {Kashuba}, {Koshkin},
  {Zhukov}, {Fi{\textcommabelow s}ek}, {{\c{C}}akir}, {{\"O}zer}, {Schnabel},
  {Schnabel}, {Signoret}, {Morrone}, {Santana-Ros}, {Pereira}, {Emilio},
  {Burdanov}, {de Wit}, {Barkaoui}, {Gillon}, {Leto}, {Frasca}, {Catanzaro},
  {Sanchez}, {Tagliaferri}, {Di Sora}, {Isopi}, {Krugly}, {Slyusarev},
  {Chiorny}, {Miku{\v{z}}}, {Bacci}, {Maestripieri}, {Grazia}, {de la Cueva},
  {Yuste-Moreno}, {Ciabattari}, {Kozhukhov}, {Serra-Ricart}, {Alarcon},
  {Licandro}, {Masi}, {Bacci}, {Bosch}, {Behem}, {Prost}, {Renner}, {Conjat},
  {Bachini}, {Succi}, {Stoian}, {Juravle}, {Carosati}, {Gowe}, {Carrillo},
  {Zheleznyak}, {Montigiani}, {Foster}, {Mannucci}, {Ruocco}, {Cuevas}, {Di
  Marcantonio}, {Coretti}, {Iafrate}, {Baldini}, {Collins}, {P{\'a}l},
  {Cs{\'a}k}, {Fern{\'a}ndez-Garcia}, {Castro-Tirado}, {Hudin}, {Madiedo},
  {Anghel}, {Calvo-Fern{\'a}ndez}, {Valvasori}, {Guido}, {Gherase}, {Kamoun},
  {Fafet}, {S{\'a}nchez-Gonz{\'a}lez}, {Curelaru}, {V{\^\i}ntdevar{\u{a}}},
  {Danescu}, {Gout}, {Schmitz}, {Sota}, {Belskaya}, {Rodr{\'\i}guez-Marco},
  {Kilic}, {Frappa}, {Klotz}, {Lavayssi{\`e}re}, {Oliveira}, {Popescu},
  {Mammana}, {Fern{\'a}ndez-Laj{\'u}s}, {Schmidt}, {Hopp}, {Kom{\v{z}}{\'\i}k},
  {Pribulla}, {Tomko}, {Hus{\'a}rik}, {Erece}, {Eryilmaz}, {Buzzi},
  {G{\"a}hrken}, {Nardiello}, {Hornoch}, {Sonbas}, {Er}, {Burwitz}, {Sybilski},
  {Bykowski}, {M{\"u}ller}, {Ogloza}, {Gon{\c{c}}alves}, {Ferreira},
  {Ferreira}, {Bento}, {Meister}, {Bagiran}, {Teke{\textcommabelow s}},
  {Marciniak}, {Moravec}, {Delin{\v{c}}{\'a}k}, {Gianni}, {Casalnuovo},
  {Boutet}, {Sanchez}, {Klemt}, {Wuensche}, {Burzynski}, {Borkowski}, {Serrau},
  {Dangl}, {Kl{\"o}s}, {Weber}, {Urban{\'\i}k}, {Rousselot}, {Kub{\'a}nek},
  {Andr{\'e}}, {Colazo}, {Spagnotto}, {Sickafoose}, {Hueso},
  {S{\'a}nchez-Lavega}, {Fisher}, {Rengstorf}, {Perell{\'o}}, {Dascalu},
  {Altan}, {Gazeas}, {de Santana}, {Sfair}, {Winter}, {Kalkan},
  {Canales-Moreno}, {Trigo-Rodr{\'\i}guez}, {Tsamis}, {Tigani}, {Sioulas},
  {Lekkas}, {Bertesteanu}, {Dumitrescu}, {Wilberger}, {Barnes}, {Fieber-Beyer},
  {Swaney}, \& {Fuentes}}]{Rommel2023}
{Rommel}, F.~L., {Braga-Ribas}, F., {Ortiz}, J.~L., {et~al.} 2023,
  \bibinfo{title}{{A large topographic feature on the surface of the
  trans-Neptunian object (307261) 2002 MS$_{4}$ measured from stellar
  occultations},} \aap, 678, A167, \dodoi{10.1051/0004-6361/202346892}

% type=
\bibitem[{D. Ruz-Mieres {et~al.}(2026)Ruz-Mieres, De~Angeli,
  {et~al.}}]{RuzMieres2026}
Ruz-Mieres, D., De~Angeli, F., {et~al.} 2026, {GaiaXPy}, 2.1.5 Zenodo,
  \dodoi{10.5281/zenodo.23400597}

% type= article
\bibitem[{E.~L. Schaller \& M.~E. Brown(2007)Schaller \& Brown}]{Schaller2007}
Schaller, E.~L., \& Brown, M.~E. 2007, \bibinfo{title}{Volatile Loss and
  Retention on {K}uiper Belt Objects,} \apjl, 659, L61, \dodoi{10.1086/516709}

% type= article
\bibitem[{E.~L. {Schaller} \& M.~E. {Brown}(2008){Schaller} \&
  {Brown}}]{Schaller2008}
{Schaller}, E.~L., \& {Brown}, M.~E. 2008, \bibinfo{title}{{Detection of
  Additional Members of the 2003 EL61 Collisional Family via Near-Infrared
  Spectroscopy},} \apjl, 684, L107, \dodoi{10.1086/592232}

% type= article
\bibitem[{M.~E. {Schwamb} {et~al.}(2014){Schwamb}, {Brown}, \&
  {Fraser}}]{Schwamb2014}
{Schwamb}, M.~E., {Brown}, M.~E., \& {Fraser}, W.~C. 2014, \bibinfo{title}{{The
  Small Numbers of Large Kuiper Belt Objects},} \aj, 147, 2,
  \dodoi{10.1088/0004-6256/147/1/2}

% type= article
\bibitem[{M.~E. Schwamb {et~al.}(2010)Schwamb, Brown, Rabinowitz, \&
  Ragozzine}]{Schwamb2010}
Schwamb, M.~E., Brown, M.~E., Rabinowitz, D.~L., \& Ragozzine, D. 2010,
  \bibinfo{title}{Properties of the Distant {K}uiper Belt: Results from the
  {P}alomar Distant Solar System Survey,} \apj, 720, 1691,
  \dodoi{10.1088/0004-637X/720/2/1691}

% type= article
\bibitem[{M.~E. {Schwamb} {et~al.}(2023){Schwamb}, {Jones}, {Yoachim}, {Volk},
  {Dorsey}, {Opitom}, {Greenstreet}, {Lister}, {Snodgrass}, {Bolin}, {Inno},
  {Bannister}, {Eggl}, {Solontoi}, {Kelley}, {Juri{\'c}}, {Lin}, {Ragozzine},
  {Bernardinelli}, {Chesley}, {Daylan}, {{\v{D}}urech}, {Fraser}, {Granvik},
  {Knight}, {Lisse}, {Malhotra}, {Oldroyd}, {Thirouin}, \& {Ye}}]{Schwamb2023}
{Schwamb}, M.~E., {Jones}, R.~L., {Yoachim}, P., {et~al.} 2023,
  \bibinfo{title}{{Tuning the Legacy Survey of Space and Time (LSST) Observing
  Strategy for Solar System Science},} \apjs, 266, 22,
  \dodoi{10.3847/1538-4365/acc173}

% type= article
\bibitem[{M.~E. {Schwamb} {et~al.}(2026){Schwamb}, {Bernardinelli}, {Volk},
  {Murtagh}, {Lin}, {Rivkin}, {Chandler}, {Jones}, {Benecchi}, {Juri{\'c}},
  {Kurlander}, {Greenstreet}, {Wilson}, {West}, \& {Lyttle}}]{Schwamb2026}
{Schwamb}, M.~E., {Bernardinelli}, P.~H., {Volk}, K., {et~al.} 2026,
  \bibinfo{title}{{NSF-DOE Vera C. Rubin Observatory's First View of the
  Neptune Trojan L5 (Trailing) Cloud: Two New Long-term Stable Trojans},}
  \apjl, 1010, L36, \dodoi{10.3847/2041-8213/aead9a}

% type= article
\bibitem[{S.~S. {Sheppard} {et~al.}(2018){Sheppard}, {Fernandez}, \&
  {Moullet}}]{Sheppard2018}
{Sheppard}, S.~S., {Fernandez}, Y.~R., \& {Moullet}, A. 2018,
  \bibinfo{title}{{The Albedos, Sizes, Colors, and Satellites of Dwarf Planets
  Compared with Newly Measured Dwarf Planet 2013 FY27},} \aj, 156, 270,
  \dodoi{10.3847/1538-3881/aae92a}

% type= article
\bibitem[{S.~S. Sheppard {et~al.}(2011)Sheppard, Udalski, Trujillo, Kubiak,
  Pietrzynski, Poleski, Soszynski, Szyma{\'n}ski, \& Ulaczyk}]{Sheppard2011}
Sheppard, S.~S., Udalski, A., Trujillo, C., {et~al.} 2011, \bibinfo{title}{A
  Southern Sky and Galactic Plane Survey for Bright {K}uiper Belt Objects,}
  \aj, 142, 98, \dodoi{10.1088/0004-6256/142/4/98}

% type= misc
\bibitem[{ {SLAC National Accelerator Laboratory} \&  {NSF-DOE Vera C. Rubin
  Observatory}(2025){SLAC National Accelerator Laboratory} \& {NSF-DOE Vera C.
  Rubin Observatory}}]{LSSTCam}
{SLAC National Accelerator Laboratory}, \& {NSF-DOE Vera C. Rubin Observatory}.
  2025, {The LSST Camera (LSSTCam)}, SLAC National Accelerator Laboratory,
  \dodoi{10.71929/rubin/2571927}

% type= article
\bibitem[{C. Snodgrass {et~al.}(2010)Snodgrass, Carry, Dumas, \&
  Hainaut}]{Snodgrass2010}
Snodgrass, C., Carry, B., Dumas, C., \& Hainaut, O. 2010,
  \bibinfo{title}{Characterisation of candidate members of (136108) {H}aumea's
  family,} \aap, 511, A72, \dodoi{10.1051/0004-6361/200913031}

% type= article
\bibitem[{G. {Tancredi} \& S. {Favre}(2008){Tancredi} \&
  {Favre}}]{Tancredi2008}
{Tancredi}, G., \& {Favre}, S. 2008, \bibinfo{title}{{Which are the dwarfs in
  the Solar System?},} \icarus, 195, 851, \dodoi{10.1016/j.icarus.2007.12.020}

% type= article
\bibitem[{P. {Tanga} {et~al.}(2023){Tanga}, {Pauwels}, {Mignard}, {Muinonen},
  {Cellino}, {David}, {Hestroffer}, {Spoto}, {Berthier}, {Guiraud}, {Roux},
  {Carry}, {Delbo}, {Dell'Oro}, {Fouron}, {Galluccio}, {Jonckheere}, {Klioner},
  {Lefustec}, {Liberato}, {Ord{\'e}novic}, {Oreshina-Slezak}, {Penttil{\"a}},
  {Pailler}, {Panem}, {Petit}, {Portell}, {Poujoulet}, {Thuillot}, {Van
  Hemelryck}, {Burlacu}, {Lasne}, \& {Managau}}]{Tanga2023}
{Tanga}, P., {Pauwels}, T., {Mignard}, F., {et~al.} 2023, \bibinfo{title}{{Gaia
  Data Release 3. The Solar System survey},} \aap, 674, A12,
  \dodoi{10.1051/0004-6361/202243796}

% type= article
\bibitem[{J.~L. Tonry {et~al.}(2018)Tonry, Denneau, Flewelling,
  {et~al.}}]{Tonry2018}
Tonry, J.~L., Denneau, L., Flewelling, H., {et~al.} 2018, \bibinfo{title}{The
  {ATLAS} All-Sky Stellar Reference Catalog,} \apj, 867, 105,
  \dodoi{10.3847/1538-4357/aae386}

% type= article
\bibitem[{J.~L. {Tonry} {et~al.}(2018){Tonry}, {Denneau}, {Heinze}, {Stalder},
  {Smith}, {Smartt}, {Stubbs}, {Weiland}, \& {Rest}}]{Tonry2018ATLAS}
{Tonry}, J.~L., {Denneau}, L., {Heinze}, A.~N., {et~al.} 2018,
  \bibinfo{title}{{ATLAS: A High-cadence All-sky Survey System},} \pasp, 130,
  064505, \dodoi{10.1088/1538-3873/aabadf}

% type= article
\bibitem[{C.~A. {Trujillo} \& M.~E. {Brown}(2003){Trujillo} \&
  {Brown}}]{Trujillo2003}
{Trujillo}, C.~A., \& {Brown}, M.~E. 2003, \bibinfo{title}{{The Caltech Wide
  Area Sky Survey},} Earth Moon and Planets, 92, 99,
  \dodoi{10.1023/B:MOON.0000031929.19729.a1}

% type= inproceedings
\bibitem[{F. Valdes {et~al.}(2014)Valdes, Gruendl, \& {DES
  Project}}]{Valdes2014}
Valdes, F., Gruendl, R., \& {DES Project}. 2014, \bibinfo{title}{The {DECam}
  Community Pipeline,} in Astronomical Society of the Pacific Conference
  Series, Vol. 485, Astronomical Data Analysis Software and Systems XXIII, ed.
  N.~Manset \& P.~Forshay, 379

% type= techreport
\bibitem[{ {Vera C. Rubin Observatory Science Pipelines Developers}(2025){Vera
  C. Rubin Observatory Science Pipelines Developers}}]{PSTN019}
{Vera C. Rubin Observatory Science Pipelines Developers}. 2025, {The LSST
  Science Pipelines Software: Optical Survey Pipeline Reduction and Analysis
  Environment}, Project Science Technical Note PSTN-019, Vera C. Rubin
  Observatory, \dodoi{10.71929/rubin/2570545}

% type= article
\bibitem[{J. Virtanen {et~al.}(2001)Virtanen, Muinonen, \&
  Bowell}]{Virtanen2001}
Virtanen, J., Muinonen, K., \& Bowell, E. 2001, \bibinfo{title}{Statistical
  Ranging of Asteroid Orbits,} \icarus, 154, 412,
  \dodoi{10.1006/icar.2001.6592}

\end{thebibliography}
\bibliographystyle{aasjournalv7}

\end{document}